\documentclass[a4paper,10pt,showpacs,notitlepage,nofootinbib,floatfix,superscriptaddress,prd]{revtex4-1}
\usepackage{graphicx}
\usepackage{amsfonts}
\usepackage{amsmath}
\usepackage{float}
\usepackage{amssymb}

\begin{document}

\title{On the relative importance of second-order terms in relativistic
dissipative fluid dynamics}

\author{E.\ Moln\'ar}
\affiliation{MTA-DE Particle Physics Research Group, H-4010 Debrecen, P.O.\ Box 105,
Hungary}
\affiliation{Frankfurt Institute for Advanced Studies, Ruth-Moufang-Str.\ 1, D-60438
Frankfurt am Main, Germany}

\author{H.\ Niemi}
\affiliation{Department of Physics, P.O.\ Box 35 (YFL), FI-40014 University of
Jyv\"askyl\"a, Finland }
\affiliation{Helsinki Institute of Physics, P.O.Box 64, FI-00014 University of Helsinki, Finland}

\author{G.S.\ Denicol}
\affiliation{Department of Physics, McGill University, 3600 rue University, Montr\'{e}al,
Canada}

\author{D.H.\ Rischke}
\affiliation{Institut f\"ur Theoretische Physik, Johann Wolfgang Goethe-Universit\"at,
Max-von-Laue-Str.\ 1, D-60438 Frankfurt am Main, Germany}

\begin{abstract}
In Denicol {\it et al.}, Phys.\ Rev.\ D \textbf{%
85}, 114047 (2012), the equations of motion of relativistic dissipative
fluid dynamics were derived from the relativistic Boltzmann equation.
These equations contain a multitude of terms of second order in Knudsen number, 
in inverse Reynolds number, or their product. Terms of second order in Knudsen 
number give rise to non-hyperbolic (and thus acausal) behavior and must be neglected in
(numerical) solutions of relativistic dissipative fluid dynamics. The
coefficients of the terms which are of the order of the product of Knudsen
and inverse Reynolds numbers have been explicitly computed in the above reference,
in the limit of a massless Boltzmann gas. Terms of second order in
inverse Reynolds number arise from the collision
term in the Boltzmann equation, upon expansion to second order in deviations from
the single-particle distribution function in local thermodynamical
equilibrium. In this work, we compute these second-order terms for a massless Boltzmann gas
with constant scattering cross section. Consequently, we assess their
relative importance in comparison to the terms which are of the order of the
product of Knudsen and inverse Reynolds numbers.
\end{abstract}


\maketitle


\section{Introduction and Conclusions}

Relativistic fluid dynamics has found widespread applications in heavy-ion
physics, in modelling nuclear collisions at ultrarelativistic bombarding
energies \cite{Heinz:2013th,Gale:2013da}, in astrophysics, for instance in
modelling binary mergers of compact stellar objects, see e.g.\ Ref.\ \cite%
{Baiotti:2008ra}, as well as in cosmology \cite%
{Andersson_Comer,Maartens:1995wt,Maartens:1996vi,Zimdahl:1996tg,Chen:2011ia}%
. In the past, in order to solve the equations of motion of relativistic
fluid dynamics, one has often made the assumption that the fluid is ideal,
i.e., one demands instantaneous local thermodynamical equilibrium, which in
turn allows to neglect all dissipative effects. However, there are no ideal
fluids in nature, as can be seen for instance from the fact that the shear
viscosity coefficient may attain a lower limit, but never vanishes \cite%
{Danielewicz:1984ww,Kovtun:2004de,Csernai:2006zz}.

A more realistic modelling of the dynamics of relativistic fluids thus
demands that one uses the equations of relativistic \emph{dissipative} fluid
dynamics. First attempts to formulate such equations were made by Eckart 
\cite{Eckart:1940te} and Landau and Lifshitz \cite{Landau_book} based on a
relativistic generalization of the non-relativistic Navier-Stokes equations.
However, their equations suffer from instabilities and acausal signal
propagation \cite{his}. The reason for this is the (erroneous) assumption
that the dissipative quantities, like bulk viscous pressure $\Pi $, particle
diffusion current $n^{\mu }$, and shear-stress tensor $\pi ^{\mu \nu }$,
react \emph{instantaneously} to the thermodynamic forces, like gradients of
the fluid velocity or temperature and chemical potential. If one relaxes
this assumption by introducing certain time scales $\tau _{\Pi },\,\tau
_{n}, $ and $\tau _{\pi }$, on which the dissipative quantities are allowed
to approach the values determined by the corresponding thermodynamic forces, 
these problems can be
cured [provided the relaxation times fulfill certain conditions \cite%
{Pu:2009fj}]. Similarly to earlier works by Grad \cite{Grad} and M\"{u}ller 
\cite{Muller_67,Muller_1986,Muller_99} in the non-relativistic context,
Israel and Stewart (IS) were among the first to suggest equations of motion
for relativistic dissipative fluids that were stable and causal \cite%
{Stewart:1972hg,Stewart:1977,Israel:1979wp}.

In recent years, relativistic dissipative fluid dynamics based on the IS
formulation was extensively applied to describe the dynamics of nuclear
collisions. At the same time, the theoretical foundations of this theory
were further explored both from kinetic theory 
\cite{Betz:2008me,Betz:2010cx,Muronga:2006zw,Tsumura:2007ji,Tsumura:2009vm,Tsumura:2012gq,York:2008rr,Monnai:2009ad,Monnai:2010qp,Denicol:2010xn,Denicol:2012cn,Denicol:2012es,Florkowski:2010cf,Martinez:2010sc,Jaiswal:2012qm,Jaiswal:2013npa,Jaiswal:2013vta,Teaney:2013gca}
and from irreversible thermodynamics \cite{Gariel:1994nw,ext_irr_thermo,
ext_irr_thermo_2,Koide:2006ef, Van:2007pw,
Biro:2008be,Van:2011yn,El:2009vj,Muronga:2010zz,Osada:2011gx,Jaiswal:2013fc}. 
In particular, in Refs.\ \cite%
{Denicol:2010xn,Denicol:2012cn,Denicol:2012es} it has been investigated how
to derive the equations of motion for the dissipative quantities using the
Boltzmann equation as underlying microscopic theory.

In Ref.\ \cite{Denicol:2012cn} a derivation of the equations of motion of
relativistic dissipative fluid dynamics was presented, which is based on a
systematic power-counting scheme in Knudsen and inverse Reynolds number. The
Knudsen number, Kn=$\lambda /L$, is the ratio between a characteristic
microscopic time/length scale, $\lambda $, e.g., the mean-free path between
collisions, and a characteristic macroscopic scale of the fluid $L$. In this
context, the inverse Reynolds numbers are the ratios of dissipative
quantities and (local) equilibrium values of macroscopic fields, e.g.\ like R%
$_{\Pi }^{-1}=|\Pi |/P_{0}$, R$_{n}^{-1}=|n^{\mu }|/n_{0}$, or R$_\pi^{-1} =
|\pi^{\mu \nu}|/P_0$, where $P_{0}$ is the thermodynamic pressure and $n_{0}$
is the particle density in equilibrium. The time scales $\tau_\Pi, \tau_n$,
and $\tau_\pi$ are identified with the slowest \emph{microscopic} time
scales of the Boltzmann equation.

The physical picture that emerges is that microscopic processes (i.e., in
the case of the Boltzmann equation, binary collisions) occur on time scales
smaller than (or at most as large as) $\tau_\Pi, \tau_n,$ and $\tau_\pi$.
These processes affect that the dissipative quantities $\Pi, n^\mu,$ and $%
\pi^{\mu \nu}$ approach the values given by the (relativistic generalization
of the) Navier-Stokes equations on the time scales $\tau_\Pi, \tau_n$, and $%
\tau_\pi$. Since microscopic physics influences the motion of the fluid only
on short time scales, the term ``transient fluid dynamics'' was coined for
such theories of relativistic dissipative fluid dynamics. The fact that the
microscopic dynamics of the Boltzmann equation gives rise to relaxation-type
equations of motion for the dissipative quantities, i.e., where these
quantities exponentially decay towards the values given by the Navier-Stokes
equations, was confirmed in Ref.\ \cite{Denicol:2011fa}. It was also shown
in that paper that approaches based on the AdS/CFT correspondence lead to
equations of motion which are of the type encountered for an underdamped
harmonic oscillator. The relaxation towards the Navier-Stokes values is then
not exponential but oscillatory.

Let us recall the relaxation-type equations for the dissipative quantities
derived in Ref.\ \cite{Denicol:2012cn}, 
\begin{align}
\tau _{\Pi }\dot{\Pi}+\Pi & =-\zeta \theta +\mathcal{J}+\mathcal{K}+\mathcal{%
R},  \label{Final1} \\
\tau _{n}\dot{n}^{\left\langle \mu \right\rangle }+n^{\mu }& =\kappa I^{\mu
}+\mathcal{J}^{\mu }+\mathcal{K}^{\mu }+\mathcal{R}^{\mu },  \label{Final2}
\\
\tau _{\pi }\dot{\pi}^{\left\langle \mu \nu \right\rangle }+\pi ^{\mu \nu }&
=2\eta \sigma ^{\mu \nu }+\mathcal{J}^{\mu \nu }+\mathcal{K}^{\mu \nu }+%
\mathcal{R}^{\mu \nu },  \label{Final3}
\end{align}%
where the over-dot denotes the proper time derivative, $\dot{A}\equiv
DA=u^{\mu }\partial _{\mu }A$. Here, $\zeta $ is the coefficient of the bulk
viscosity, $\kappa $ the coefficient of particle diffusion (which is related
to that of heat conduction) and $\eta $ the coefficient of shear viscosity.
Furthermore, with the fluid 4-velocity $u^{\mu }$ (chosen in the Landau
frame), where $u^{\mu }u_{\mu }=1$, with $\Delta ^{\mu \nu }=g^{\mu \nu
}-u^{\mu }u^{\nu }$ being the 3-projector onto the subspace orthogonal to $%
u^{\mu }$, and with $\nabla ^{\mu }=\Delta ^{\mu \nu }\partial _{\nu }$
being the 3-gradient, $\theta =\nabla _{\mu }u^{\mu }$ is the expansion
scalar, $\sigma ^{\mu \nu }\equiv \nabla ^{\langle \mu }u^{\nu \rangle }=%
\frac{1}{2}\left( \nabla ^{\mu }u^{\nu }+\nabla ^{\nu }u^{\mu }\right) -%
\frac{1}{3}\theta \Delta ^{\mu \nu }$ is the shear tensor, while $I^{\mu
}=\nabla ^{\mu }\alpha _{0}$ is the gradient of $\alpha _{0}=\mu /T$, the
ratio of chemical potential $\mu $ and temperature $T$.

In the above equations, the tensors $\mathcal{J}$, $\mathcal{J}^{\mu }$, and 
$\mathcal{J}^{\mu \nu }$ contain all terms of first order in the product of
Knudsen and inverse Reynolds number,%
\begin{align}
\mathcal{J}& =-\ell _{\Pi n}\nabla \cdot n-\tau _{\Pi n}n\cdot F-\delta
_{\Pi \Pi }\Pi \theta -\lambda _{\Pi n}n\cdot I+\lambda _{\Pi \pi }\pi ^{\mu
\nu }\sigma _{\mu \nu },  \notag \\
\mathcal{J}^{\mu }& =-\tau _{n}n_{\nu }\omega ^{\nu \mu }-\delta _{nn}n^{\mu
}\theta -\ell _{n\Pi }\nabla ^{\mu }\Pi +\ell _{n\pi }\Delta ^{\mu \nu
}\nabla _{\lambda }\pi _{\nu }^{\lambda }+\tau _{n\Pi }\Pi F^{\mu }-\tau
_{n\pi }\pi ^{\mu \nu }F_{\nu }  
 -\lambda _{nn}n_{\nu }\sigma ^{\mu \nu }+\lambda _{n\Pi }\Pi I^{\mu
}-\lambda _{n\pi }\pi ^{\mu \nu }I_{\nu },  \notag \\
\mathcal{J}^{\mu \nu }& =2\tau _{\pi }\pi _{\lambda }^{\left\langle \mu
\right. }\omega ^{\left. \nu \right\rangle \lambda }-\delta _{\pi \pi }\pi
^{\mu \nu }\theta -\tau _{\pi \pi }\pi ^{\lambda \left\langle \mu \right.
}\sigma _{\lambda }^{\left. \nu \right\rangle }+\lambda _{\pi \Pi }\Pi
\sigma ^{\mu \nu }-\tau _{\pi n}n^{\left\langle \mu \right. }F^{\left. \nu
\right\rangle }  
 +\ell _{\pi n}\nabla ^{\left\langle \mu \right. }n^{\left. \nu
\right\rangle }+\lambda _{\pi n}n^{\left\langle \mu \right. }I^{\left. \nu
\right\rangle },  \label{14_moment_terms}
\end{align}%
where $\omega ^{\mu \nu }=\frac{1}{2}\left( \nabla ^{\mu }u^{\nu }-\nabla
^{\nu }u^{\mu }\right) $ denotes the vorticity and we defined $F^{\mu
}=\nabla ^{\mu }P_{0}$ as the gradient of the thermodynamic pressure. The
tensors $\mathcal{K}$, $\mathcal{K}^{\mu }$, and $\mathcal{K}^{\mu \nu }$
contain all terms of second order in Knudsen number,%
\begin{align}
\mathcal{K}& =\tilde{\zeta}_{1}\omega _{\mu \nu }\omega ^{\mu \nu }+\tilde{%
\zeta}_{2}\sigma _{\mu \nu }\sigma ^{\mu \nu }+\tilde{\zeta}_{3}\theta ^{2}+%
\tilde{\zeta}_{4}\,I\cdot I+\tilde{\zeta}_{5}F\cdot F+\tilde{\zeta}%
_{6}I\cdot F+\tilde{\zeta}_{7}\nabla \cdot I+\tilde{\zeta}_{8}\nabla \cdot F,
\notag \\
\mathcal{K}^{\mu }& =\tilde{\kappa}_{1}\sigma ^{\mu \nu }I_{\nu }+\tilde{%
\kappa}_{2}\sigma ^{\mu \nu }F_{\nu }+\tilde{\kappa}_{3}I^{\mu }\theta +%
\tilde{\kappa}_{4}F^{\mu }\theta +\tilde{\kappa}_{5}\omega ^{\mu \nu }I_{\nu
}+\tilde{\kappa}_{6}\Delta _{\lambda }^{\mu }\nabla_{\nu }\sigma ^{\lambda
\nu }+\tilde{\kappa}_{7}\nabla ^{\mu }\theta ,  \notag \\
\mathcal{K}^{\mu \nu }& =\tilde{\eta}_{1}\omega _{\lambda }^{\left.
{}\right. \left\langle \mu \right. }\omega ^{\left. \nu \right\rangle
\lambda }+\tilde{\eta}_{2}\theta \sigma ^{\mu \nu }+\tilde{\eta}_{3}\sigma
^{\lambda \left\langle \mu \right. }\sigma _{\lambda }^{\left. \nu
\right\rangle }+\tilde{\eta}_{4}\sigma _{\lambda }^{\left\langle \mu \right.
}\omega ^{\left. \nu \right\rangle \lambda }+\tilde{\eta}_{5}I^{\left\langle
\mu \right. }I^{\left. \nu \right\rangle }+\tilde{\eta}_{6}F^{\left\langle
\mu \right. }F^{\left. \nu \right\rangle }  \notag \\
& +\tilde{\eta}_{7}I^{\left\langle \mu \right. }F^{\left. \nu \right\rangle
}+\tilde{\eta}_{8}\nabla ^{\left\langle \mu \right. }I^{\left. \nu
\right\rangle }+\tilde{\eta}_{9}\nabla ^{\left\langle \mu \right. }F^{\left.
\nu \right\rangle }.  \label{Ks}
\end{align}
Note that, in contrast to Ref.\ \cite{Denicol:2012cn}, we now write the term
proportional to $\tilde{\kappa}_6$ with a 3-gradient operator $\nabla_\nu$
instead of a partial derivative $\partial_\nu$. Finally, the tensors $%
\mathcal{R}$, $\mathcal{R}^{\mu }$, and $\mathcal{R}^{\mu \nu }$ contain all
terms of second order in inverse Reynolds number,%
\begin{align}
\mathcal{R}& =\varphi _{1}\Pi ^{2}+\varphi _{2}n\cdot n+\varphi _{3}\pi
_{\mu \nu }\pi ^{\mu \nu },  \label{R_scalar} \\
\mathcal{R}^{\mu }& =\varphi _{4}n_{\nu }\pi ^{\mu \nu }+\varphi _{5}\Pi
n^{\mu },  \label{R_vector} \\
\mathcal{R}^{\mu \nu }& =\varphi _{6}\Pi \pi ^{\mu \nu }+\varphi _{7}\pi
^{\lambda \left\langle \mu \right. }\pi _{\lambda }^{\left. \nu
\right\rangle }+\varphi _{8}n^{\left\langle \mu \right. }n^{\left. \nu
\right\rangle }.  \label{R_tensor}
\end{align}%
These second-order terms follow from computing the collision integral beyond
linear order in the dissipative quantities.

Observing the plethora of transport coefficients occurring in Eqs.\ (\ref%
{14_moment_terms}) -- (\ref{R_tensor}), a natural question to ask is whether
all of them are of the same order of magnitude or whether some coefficients
are larger and thus more important than others. The goal of this paper is to
answer this question. The coefficients in Eqs.\ (\ref{14_moment_terms}) were
explicitly computed in Ref.\ \cite{Denicol:2012cn} for a massless Boltzmann
gas with constant scattering cross section. We now supplement these results
by computing the coefficients in Eqs.\ (\ref{Ks}) -- (\ref{R_tensor}). We
restrict ourselves to the 14-moment approximation. In this case, the
coefficients in Eqs.\ (\ref{Ks}) vanish identically (cf.\ Appendix \ref%
{Kcoeff}), and we only need to focus on the coefficients $\varphi_1, \ldots,
\varphi_8$ in Eqs.\ (\ref{R_scalar}) -- (\ref{R_tensor}). The derivation and
calculation of these coefficients is quite demanding and is presented in
detail in the remainder of this paper. For the rest of this introductory
section, we simply quote the results and draw our conclusions.

For massless particles, the bulk viscous pressure vanishes identically, $\Pi
= 0$, and we do not have to solve Eq.\ (\ref{Final1}). Also, terms in Eqs.\ (%
\ref{14_moment_terms}) -- (\ref{R_tensor}) proportional to $\Pi$ can be
neglected, such that we do not need to compute the corresponding
coefficients. Moreover, as mentioned above, in the 14-moment approximation
all coefficients in Eqs.\ (\ref{Ks}) vanish. Dividing Eq.\ (\ref{Final2}) by 
$n_0$, and Eq.\ (\ref{Final3}) by $P_0$, these two equations can be written
in the following form, 
\begin{align}
\tau_n \frac{\dot{n}^{\left\langle \mu \right\rangle }}{n_0} +\frac{n^{\mu }%
}{n_0} & =\frac{\kappa}{n_0} \nabla^{\mu }\alpha_0  \notag \\
& + \left( \tau_n \omega^{\mu \nu} - \lambda_{nn} \sigma^{\mu \nu} -
\delta_{nn} \theta g^{\mu \nu} \right) \frac{n_\nu}{n_0}  \notag \\
& + \frac{\ell_{n \pi}}{\beta_0}\frac{ \Delta^{\mu \nu} \nabla_\lambda
\pi^\lambda_{\nu} }{P_0} - \left( \frac{\tau_{n \pi} P_0}{\beta_0} \frac{%
\nabla_\nu P_0}{P_0} + \frac{\lambda_{n \pi}}{\beta_0} \nabla_\nu \alpha_0 -
\varphi_4 P_0 \frac{n_\nu}{n_0} \right) \frac{\pi^{\mu \nu}}{P_0},
\label{neq2} \\
\tau _{\pi } \frac{\dot{\pi}^{\left\langle \mu \nu \right\rangle }}{P_0} +%
\frac{\pi ^{\mu \nu }}{P_0}& =\frac{2\eta}{P_0} \sigma ^{\mu \nu }  \notag \\
& + \frac{\pi_\lambda^{\langle \mu}}{P_0} \left( 2 \tau_{\pi} \omega^{\nu
\rangle \lambda} - \tau_{\pi\pi} \sigma^{\nu \rangle \lambda} - \delta_{\pi
\pi} \theta g^{\nu \rangle \lambda} + \varphi_7 P_0 \frac{ \pi^{\nu \rangle
\lambda}}{P_0}\right)  \notag \\
& + \ell_{\pi n}\beta_0 \frac{\nabla^{\langle \mu} n^{\nu \rangle}}{n_0} + 
\frac{n^{\langle \mu}}{n_0} \left( \lambda_{\pi n} \beta_0 \nabla^{\nu
\rangle} \alpha_0 - \tau_{\pi n} n_0 \frac{\nabla^{\nu \rangle} P_0}{P_0} +
\varphi_8 \beta_0^2 P_0 \frac{n^{\nu \rangle}}{n_0} \right).  \label{pieq2}
\end{align}
Here, we have made use of the equation of state of the massless Boltzmann
gas, $P_0=n_0/ \beta_0$, with $\beta_0 = 1/T$. 
By dividing the dissipative quantities by $n_0$ or $P_0$,
respectively, we immediately identify terms which are proportional to
inverse Reynolds number. Furthermore, the coefficients of terms involving
gradients (or time derivatives) all have dimension of time (or mean-free
path) and are thus proportional to the Knudsen number. In this form, it is
easy to apply power-counting arguments to estimate the order of magnitude of
the various terms. The Navier-Stokes terms appearing in the first lines are
of first order in Knudsen number. The second lines contain terms
proportional to the dissipative quantity that is evolved in the respective
equation ($n^\mu$ in the first and $\pi^{\mu \nu}$ in the second equation),
while the third lines contain cross terms proportional to the other
dissipative quantity that is not evolved ($\pi^{\mu \nu}$ in the first and $%
n^\mu$ in the second equation). Terms in the second and third lines are of
first order in the product of Knudsen and inverse Reynolds number as well as
of second order in inverse Reynolds number.

At this point, one cannot draw any further conclusion without making
assumptions about the relative magnitude of Knudsen and inverse Reynolds
numbers. In the following, let us assume that all of them are of the \emph{%
same\/} order of magnitude, $\mathrm{Kn} \sim \mathrm{R}_n^{-1} \sim \mathrm{%
R}_\pi^{-1}$ [situations where this is no longer the case were studied,
e.g., in Ref.\ \cite{Denicol:2012vq}]. 
At least for asymptotically long times, when the values of the dissipative quantities approach their respective Navier-Stokes limits, this assumption is fulfilled.

Then we can simply assess the order
of magnitude (and thus the importance) of the various terms by comparing the
values of the coefficients accompanying them.
These are listed in Tables \ref%
{diff_massless} [for Eq.\ (\ref{neq2})] and \ref{shear_massless} [for Eq.\ (%
\ref{pieq2})]. Note that the coefficients of terms containing a power of
Knudsen number are given in units of the relaxation times $\tau_n$ and $%
\tau_\pi$, respectively.

\begin{table}[h]
\begin{center}
\begin{tabular}{|c|c|c|c|c|c|c|c|}
\hline
$\displaystyle \frac{\kappa}{n_0} [\lambda_{\mathrm{mfp}}]$ & $\tau
_{n}[\lambda _{\mathrm{mfp}}]$ & ${\lambda }_{nn}[\tau _{n}]$ & $\delta
_{nn}[\tau _{n}]$  & $\displaystyle \frac{\ell_{n\pi }}{%
\beta_0} [\tau _{n}]$ & $\displaystyle \frac{\tau _{n\pi }P_0}{\beta_0}
[\tau _{n}]$ & $\displaystyle \frac{{\lambda }_{n\pi }}{\beta_0}[\tau _{n}]$
& $\varphi_4 P_0$ \\ \hline
$\displaystyle \frac{3}{16}$ & $\displaystyle\frac{9}{4}$ & $\displaystyle%
\frac{3}{5}$ & $1$  & $\displaystyle\frac{1}{20}
$ & $\displaystyle\frac{1}{80}$ & $\displaystyle\frac{1}{20}$ & $\displaystyle\frac{1}{25}$
\\ \hline
\end{tabular}%
\end{center}
\caption{{\protect\small The coefficients in the particle diffusion equation
for a massless Boltzmann gas with constant cross section $\protect\sigma _{T}
$ in the 14-moment approximation [from Ref.\ \protect\cite{Denicol:2012cn}]. 
$\protect\lambda _{\mathrm{mfp}}=1/(n_{0}\protect\sigma _{T})$ is the
mean-free path.}}
\label{diff_massless}
\end{table}
\begin{table}[h]
\begin{center}
\begin{tabular}{|c|c|c|c|c|c|c|c|c|}
\hline
$\displaystyle \frac{2\eta}{P_0}[\lambda_{\mathrm{mfp}}] $ & $\tau _{\pi
}[\lambda _{\mathrm{mfp}}]$ & ${\tau }_{\pi \pi }[\tau_{\pi}]$ & $\delta
_{\pi \pi }[\tau_{\pi}] $ & $\varphi_7 P_0$ & $\ell _{\pi n} \beta_0 [\tau
_{\pi}]$ & ${\lambda }_{\pi n} \beta_0[\tau _{\pi}]$ & $\tau _{\pi n}
n_0[\tau_{\pi}]$ & $\varphi_8 \beta_0^2 P_0$ \\ \hline
$\displaystyle \frac{8}{3}$ & $\displaystyle\frac{5}{3}$ & $\displaystyle 
\frac{10}{7}$ & $\displaystyle \frac{4}{3}$ & $\displaystyle \frac{9}{70}$ & 
$0$ & $0$ & $0$ & $\displaystyle \frac{8}{5}$ \\ \hline
\end{tabular}%
\end{center}
\caption{{\protect\small The coefficients in the shear-stress tensor
equation for a massless Boltzmann gas with constant cross section $\protect%
\sigma _{T}$ in the 14-moment approximation [from Ref.\ \protect\cite%
{Denicol:2012cn}]. $\protect\lambda _{\mathrm{mfp}}=1/(n_{0}\protect\sigma %
_{T})$ is the mean-free path.}}
\label{shear_massless}
\end{table}
One immediately observes that the coefficients in the third line of Eq.\ (%
\ref{neq2}) are (at least) an order of magnitude smaller than the ones in
the second line. If Knudsen and inverse Reynolds numbers are of the same
order of magnitude, it is then a reasonable assumption to drop these terms
altogether, including the last term in Eq.\ (\ref{neq2}) that arises from
nonlinear terms in the collision integral.

In the 14-moment approximation, all terms in the third line of Eq.\ (\ref%
{pieq2}), except for the last one, vanish identically. We remark, however,
that this is accidental; going beyond the 14-moment approximation \cite%
{Denicol:2012cn}, all terms are of the same order of magnitude. From the
terms in the second line, the last one (arising from the nonlinear terms in
the collision integral) is an order of magnitude smaller than the other
second-order terms and hence can safely be neglected.

To summarize our results, we conclude that, in the 14-moment approximation,
for a massless Boltzmann gas, and for situations where Knudsen and inverse
Reynolds numbers are of the same order of magnitude, the equations of motion
for the dissipative quantities read, to good approximation, as follows 
\begin{align}
\tau_n \dot{n}^{\left\langle \mu \right\rangle } +n^{\mu } & \simeq\kappa
\nabla^{\mu }\alpha_0 + \left( \tau_n \omega^{\mu \nu} - \lambda_{nn}
\sigma^{\mu \nu}\right) n_\nu - \delta_{nn} \theta n^\mu, \\
\tau _{\pi } \dot{\pi}^{\left\langle \mu \nu \right\rangle } +\pi ^{\mu \nu
}& \simeq 2\eta \sigma ^{\mu \nu } + \pi_\lambda^{\langle \mu} \left( 2
\tau_{\pi} \omega^{\nu \rangle \lambda} - \tau_{\pi\pi} \sigma^{\nu \rangle
\lambda}\right) - \delta_{\pi \pi} \theta \pi^{\mu \nu} +\varphi_8
n^{\langle \mu} n^{\nu \rangle}.
\end{align}
In the application of relativistic dissipative fluid dynamics to describe
the dynamics of nuclear collisions, this form of the equations should be
applicable (far) above the QCD transition, i.e., (deep) in the quark-gluon
plasma phase, where the velocity of sound is close to that of an
ultrarelativistic gas, $c_s^2= 1/3$, and the mass of the quasi-particles can
be neglected. Nevertheless, the values of the transport coefficients may
change from the values given in Tables \ref{diff_massless} and \ref%
{shear_massless} when accounting for proper quantum statistics, the correct
number of degrees of freedom, as well as a more realistic
(angular-dependent) scattering cross section. Closer to the QCD phase
transition, however, the breaking of conformality and the effect from
non-vanishing bulk viscous pressure $\Pi$ can no longer be neglected.

The remainder of this paper is organized as follows. In Secs.\ \ref{BTE_FD}
and \ref{IIB} we briefly recapitulate the derivation of the equations of
relativistic dissipative fluid dynamics from the Boltzmann equation using
the method of moments \cite{Denicol:2012cn}. In Sec.\ \ref{BTE_coll_int} we
give a general derivation of the second-order corrections to the collision
integral. Explicit results will be derived in Sec.\ \ref{BTE_14moment} by
reducing the number of moments to 14. Details of our calculations are
relegated to several Appendices.

We use natural units throughout this work, $\hbar =k_{B}=c=1$. Covariant, $%
x_{\mu }=(t,-x,-y,-z)$, and contravariant, $x^{\mu }=(t,x,y,z)$, 4-vectors
are related through the metric of flat space-time, $g^{\mu \nu }\equiv $diag$%
(1,-1,-1,-1)\equiv g_{\mu \nu }$, by $x_{\mu }=g_{\mu \nu }x^{\nu }$ and $%
x^{\mu }=g^{\mu \nu }x_{\nu }$. Furthermore, the symmetrization and
anti-symmetrization operations are denoted by parentheses or brackets around
indices, $A^{(\mu \nu )}=\left( A^{\mu \nu }+A^{\nu \mu }\right) /2$ and $%
A^{[\mu \nu ]}=\left( A^{\mu \nu }-A^{\nu \mu }\right) /2$, respectively. 

\section{Relativistic fluid dynamics from kinetic theory}

\subsection{General variables}

\label{BTE_FD}

In relativistic kinetic theory of dilute single-component gases, an ensemble
of particles with mass $m$ and 4-momenta $k^{\mu }=(k^0,\mathbf{k})$, 
with $k^{0}=\sqrt{\mathbf{k}^{2}+m^{2}}$, at a given space-time point $x^{\mu }=(t,\mathbf{x}%
)$, is characterized by the invariant single-particle distribution function $%
f_{\mathbf{k}}(t,\mathbf{x}) \equiv f_{\mathbf{k}}$. In the absence of external
forces or fields, the space-time evolution of $f_{\mathbf{k}}$ is described
by the relativistic Boltzmann equation \cite{deGroot,Cercignani_book}, 
\begin{equation}
k^{\mu }\partial _{\mu }f_{\mathbf{k}}=C\left[ f\right] ,  \label{BTE}
\end{equation}%
where $C\left[ f\right] $ is the collision term. In the case of binary
elastic collisions it is given by 
\begin{equation}
C\left[ f\right] =\frac{1}{\nu }\int dK^{\prime }dPdP^{\prime }\,W_{\mathbf{%
kk}\prime \rightarrow \mathbf{pp}\prime }\left( f_{\mathbf{p}}f_{\mathbf{p}%
^{\prime }}\tilde{f}_{\mathbf{k}}\tilde{f}_{\mathbf{k}^{\prime }}-f_{\mathbf{%
k}}f_{\mathbf{k}^{\prime }}\tilde{f}_{\mathbf{p}}\tilde{f}_{\mathbf{p}%
^{\prime }}\right) .  \label{COLL_INT}
\end{equation}%
Here $\tilde{f}_{\mathbf{k}}=1-af_{\mathbf{k}}$, with $a=1\left( -1\right) $
for fermions (bosons) and $a=0$ for classical particles, $dK=g\,d^{3}\mathbf{%
k/}\left[ (2\pi)^{3}k^{0}\right] $ is the Lorentz-invariant
momentum-space volume, where $g$ denotes the number of internal degrees of
freedom, and $W_{\mathbf{kk}\prime \rightarrow \mathbf{pp}\prime }$ is the
Lorentz-invariant transition rate. The factor $\nu =2$ takes into account
that the particles are indistinguishable, while the transition rate
satisfies detailed balance, i.e., it is symmetric for
time-reversed states $W_{\mathbf{kk}\prime \rightarrow \mathbf{pp}\prime
}=W_{\mathbf{pp}\prime \rightarrow \mathbf{kk}\prime }$.

The conservation of particle number and of energy and momentum in individual
collisions leads to the following continuity equations for the particle
4-current, $N^{\mu }$, and energy-momentum tensor, $T^{\mu \nu }$ \cite%
{deGroot,Cercignani_book}, 
\begin{align}
\partial _{\mu }N^{\mu }& \equiv \int dKC\left[ f\right] =0,
\label{kinetic:cons_N_mu} \\
\partial _{\mu }T^{\mu \nu }& \equiv \int dKk^{\nu }C\left[ f\right] =0.
\label{kinetic:cons_T_mu_nu}
\end{align}%
The currents $N^{\mu }=N^{\mu }\left( t,\mathbf{x}\right) $ and $T^{\mu \nu
}=T^{\mu \nu }\left( t,\mathbf{x}\right) $ are identified as the first and
second moments of the single-particle distribution function, respectively.
Without loss of generality, they can be tensor-decomposed in terms of the
fluid 4-velocity as 
\begin{align}
N^{\mu }& \equiv \left\langle k^{\mu }\right\rangle =\left\langle E_{\mathbf{%
k}}\right\rangle u^{\mu }+\left\langle k^{\left\langle \mu \right\rangle
}\right\rangle ,  \label{kinetic:N_mu} \\
T^{\mu \nu }& \equiv \left\langle k^{\mu }k^{\nu }\right\rangle
=\left\langle E_{\mathbf{k}}^{2}\right\rangle u^{\mu }u^{\nu }+\frac{1}{3}%
\Delta ^{\mu \nu }\left\langle \Delta ^{\alpha \beta }k_{\alpha }k_{\beta
}\right\rangle +\left\langle k^{\left\langle \mu \right. }k^{\left. \nu
\right\rangle }\right\rangle ,  \label{kinetic:T_mu_nu}
\end{align}%
where $\left\langle \cdots \right\rangle =\int dK\left( \cdots \right) f_{%
\mathbf{k}}$.
Here, $E_{\mathbf{k}}=k^{\mu }u_{\mu }$ and $k^{\left\langle \mu \right\rangle }=\Delta ^{\mu
\nu }k_{\nu }$ correspond to the energy and the 3-momentum, respectively, of the particle in
the local rest frame of the fluid, such that $k^{\mu }=E_{%
\mathbf{k}}u^{\mu }+k^{\left\langle \mu \right\rangle }$ \cite%
{Denicol:2012cn, Anderson_1, Denicol:2010xn, Struchtrup, Struchtrup_2}.
Moreover, we denoted the orthogonal projection of a first-rank tensor as $%
A^{\left\langle \mu \right\rangle }\equiv \Delta ^{\mu \nu }A_{\nu }$, while
the symmetric, traceless, and orthogonal projection of second-rank tensors $%
A^{\mu \nu }$ is defined as $A^{\left\langle \mu \nu \right\rangle }\equiv
\Delta _{\alpha \beta }^{\mu \nu }A^{\alpha \beta }$, with $\Delta ^{\mu \nu
\alpha \beta }=\left( \Delta ^{\mu \alpha }\Delta ^{\beta \nu }+\Delta ^{\nu
\alpha }\Delta ^{\beta \mu }\right) /2-\Delta ^{\mu \nu }\Delta ^{\alpha
\beta }/3$. In this work, the flow velocity $u^{\mu }$ is defined according
to the Landau prescription \cite{Landau_book} as the eigenvector of the
energy-momentum tensor, i.e., $T^{\mu \nu }u_{\nu }=\left\langle E_{\mathbf{k%
}}^{2}\right\rangle u^{\mu }$. As a consequence, the energy-momentum
diffusion current vanishes, $W^{\mu }\equiv \left\langle E_{\mathbf{k}%
}k^{\left\langle \mu \right\rangle }\right\rangle =0$.

Using Eqs.\ (\ref{kinetic:N_mu}) -- (\ref{kinetic:T_mu_nu}) we are able to
identify the fundamental fluid-dynamical quantities as
\begin{align}
n& \equiv N^{\mu }u_{\mu }=\left\langle E_{\mathbf{k}}\right\rangle ,
\label{def_n} \\
\varepsilon & \equiv T^{\mu \nu }u_{\mu }u_{\nu }=\left\langle E_{\mathbf{k}%
}^{2}\right\rangle ,  \label{def_e} \\
P& \equiv -\frac{1}{3}T^{\mu \nu }\Delta _{\mu \nu }=-\frac{1}{3}%
\left\langle \Delta ^{\alpha \beta }k_{\alpha }k_{\beta }\right\rangle ,
\label{def_P} \\
n^{\mu }& \equiv N^{\nu }\Delta _{\nu }^{\mu }=\left\langle k^{\left\langle
\mu \right\rangle }\right\rangle ,  \label{def_n_mu} \\
\pi ^{\mu \nu }& \equiv T^{\alpha \beta }\Delta _{\alpha \beta }^{\mu \nu
}=\left\langle k^{\left\langle \mu \right. }k^{\left. \nu \right\rangle
}\right\rangle ,  \label{def_pi_mu_nu}
\end{align}%
where $n$ is the particle number density, $\varepsilon $ is the energy
density, $P$ is the local isotropic pressure, $n^{\mu }$ is the particle
diffusion current, and $\pi ^{\mu \nu }$ is the shear-stress tensor.

It is customary to separate the isotropic pressure into two components, $%
P=P_{0}+\Pi $, with $P_{0}$ being the thermodynamic pressure and $\Pi $ the
bulk viscous pressure. The thermodynamic and bulk viscous pressures are
defined with respect to the local equilibrium distribution function, 
\begin{equation}
f_{0\mathbf{k}}=\left[ \exp (\beta _{0}E_{\mathbf{k}}-\alpha _{0})+a\right]
^{-1},  \label{Equilibrium_dist}
\end{equation}%
and 
\begin{eqnarray}
P_{0} &=&-\frac{1}{3}\left\langle \Delta ^{\alpha \beta }k_{\alpha }k_{\beta
}\right\rangle _{0},  \label{def_eq_P} \\
\Pi &=&-\frac{1}{3}\left\langle \Delta ^{\alpha \beta }k_{\alpha }k_{\beta
}\right\rangle _{\delta },  \label{def_Pi}
\end{eqnarray}%
where $\left\langle \cdots \right\rangle _{0}=\int dK\left( \cdots \right)
f_{0\mathbf{k}}$ and $\left\langle \cdots \right\rangle _{\delta }=\int
dK\left( \cdots \right) \delta f_{\mathbf{k}}$, with
$\delta f_{\mathbf{k}}\equiv f_{\mathbf{k}}-f_{0\mathbf{k}}$. The
temperature and chemical potential, $\mu =T\alpha _{0}$, introduced in $f_{0%
\mathbf{k}}$ are defined by the so-called matching conditions which impose
that the particle number density and energy density are given by their
respective values in a fictitious local thermodynamic equilibrium state,
i.e., $n_{0}\equiv \left\langle E_{\mathbf{k}}\right\rangle _{0}=n$ and $%
\varepsilon _{0}\equiv \left\langle E_{\mathbf{k}}^{2}\right\rangle
_{0}=\varepsilon $. In this state, there is an equation of state of the form 
$P_{0}(T,\mu )$, such that $n_{0}=\partial P_{0}/\partial \mu $, $%
s_{0}=\partial P_{0}/\partial T$, and the fundamental thermodynamical
relation $\varepsilon _{0}=Ts_{0}+\mu n_{0}-P_{0}$ is fulfilled.

It is also convenient to introduce the irreducible moments of $\delta f_{%
\mathbf{k}}$, 
\begin{equation}
\rho _{r}^{\mu _{1}\cdots \mu _{\ell }}=\left\langle E_{\mathbf{k}%
}^{r}k^{\left\langle \mu _{1}\right. }\cdots k^{\left. \mu _{\ell
}\right\rangle }\right\rangle _{\delta }.  \label{rho_moment}
\end{equation}%
Such irreducible moments are constructed to be symmetric, traceless, and
orthogonal to the 4-velocity, with the symmetrized, traceless, and
orthogonal projections being defined as 
\begin{equation}
k^{\left\langle \mu _{1}\right. }\cdots k^{\left. \mu _{\ell }\right\rangle
}=\Delta _{\nu _{1}\cdots \nu _{\ell }}^{\mu _{1}\cdots \mu _{\ell }}k^{\nu
_{1}}\cdots k^{\nu _{\ell }}.  \label{k_irreducible}
\end{equation}%
The details of construction and the properties of such tensors can be found
in Appendix \ref{irreducible_tensors}. Note that the bulk viscous pressure,
particle diffusion current, and shear-stress tensor are also irreducible
moments of $\delta f_{\mathbf{k}}$, 
\begin{equation}
\Pi =-\frac{m^{2}}{3}\rho _{0},\ n^{\mu }=\rho _{0}^{\mu },\ \pi ^{\mu \nu
}=\rho _{0}^{\mu \nu }.  \label{matching}
\end{equation}%
Furthermore, the matching conditions and the definition of the local rest
frame can also be expressed using irreducible moments. The matching
conditions correspond to 
\begin{eqnarray}
\rho _{1} &\equiv &\left\langle E_{\mathbf{k}}\right\rangle _{\delta }=0,\ 
\label{Landau_matching} \\
\rho _{2} &\equiv &\left\langle E_{\mathbf{k}}^{2}\right\rangle _{\delta }=0,
\end{eqnarray}%
while the Landau definition of the fluid 4-velocity leads to 
\begin{equation}
\rho _{1}^{\mu }\equiv \left\langle k^{\left\langle \mu \right\rangle }E_{%
\mathbf{k}}\right\rangle _{\delta }=0.  \label{Landau_flow}
\end{equation}

\subsection{Moment expansion of $f_{\mathbf{k}}$ and the equations of motion
for the irreducible moments}

\label{IIB}

Following Ref.\ \cite{Denicol:2012cn}, $\delta f_{\mathbf{k}}$ is
parametrized as 
\begin{equation}
\delta f_{\mathbf{k}}=f_{0\mathbf{k}}\tilde{f}_{0\mathbf{k}}\phi _{\mathbf{k}%
}.
\end{equation}%
The function $\phi _{\mathbf{k}}$ is then expanded in terms of a series in
the irreducible tensors given in Eq.\ (\ref{k_irreducible}), 
\begin{equation}
\phi _{\mathbf{k}}=\sum_{\ell =0}^{\infty }\lambda _{\mathbf{k}%
}^{\left\langle \mu _{1}\right. \cdots \left. \mu _{\ell }\right\rangle
}k_{\left\langle \mu _{1}\right. }...k_{\left. \mu _{\ell }\right\rangle }.
\label{expansion1}
\end{equation}%
By expanding the tensor $\lambda _{\mathbf{k}}^{\left\langle \mu _{1}\right.
\cdots \left. \mu _{\ell }\right\rangle }$ using a set of orthogonal
polynomials, it is straightforward to prove that 
\begin{equation}
\lambda _{\mathbf{k}}^{\left\langle \mu _{1}\cdots \mu _{\ell }\right\rangle
}=\sum_{n=0}^{N_{\ell }}\mathcal{H}_{\mathbf{k}n}^{\left( \ell \right) }\rho
_{n}^{\mu _{1}\cdots \mu _{\ell }},  \label{lambda_k}
\end{equation}%
where $N_{\ell }$ denotes the order at which the expansion is truncated (for
the coefficient of rank $\ell $) and $\rho _{n}^{\mu _{1}\cdots \mu _{\ell
}} $ is the irreducible moment defined in Eq.\ (\ref{rho_moment}). The
coefficients $\mathcal{H}_{\mathbf{k}n}^{(\ell )}$ are found to be \cite%
{Denicol:2012cn} 
\begin{equation}
\mathcal{H}_{\mathbf{k}n}^{\left( \ell \right) }=\frac{W^{\left( \ell
\right) }}{\ell !}\sum_{m=n}^{N_{\ell }}a_{mn}^{(\ell )}P_{\mathbf{k}%
m}^{\left( \ell \right) },  \label{H_kn}
\end{equation}%
with $P_{\mathbf{k}m}^{\left( \ell \right) }$ being orthogonal polynomials
in $E_{\mathbf{k}}$, 
\begin{equation}
P_{\mathbf{k}m}^{\left( \ell \right) }=\sum_{r=0}^{m}a_{mr}^{(\ell )}E_{%
\mathbf{k}}^{r}.  \label{P_kn}
\end{equation}%
The coefficients $a_{mr}^{(\ell )}$ are determined from the orthonormality
condition 
\begin{equation}
\int dK\,\omega ^{\left( \ell \right) }\,P_{\mathbf{k}m}^{\left( \ell
\right) }P_{\mathbf{k}n}^{\left( \ell \right) }=\delta _{mn},
\label{P_normalization}
\end{equation}%
using Gram-Schmidt orthogonalization. The measure $\omega ^{\left( \ell
\right) }$ depends on the rank $\ell $ of the tensor being expanded and
reads 
\begin{equation}
\omega ^{\left( \ell \right) }=\frac{W^{\left( \ell \right) }}{\left( 2\ell
+1\right) !!}\left( \Delta ^{\alpha \beta }k_{\alpha }k_{\beta }\right)
^{\ell }f_{0\mathbf{k}}\tilde{f}_{0\mathbf{k}},  \label{w_measure}
\end{equation}%
where $W^{\left( \ell \right) }$ is a normalization constant defined as%
\begin{equation}
W^{\left( \ell \right) }=\left( -1\right) ^{\ell }\left( J_{2\ell ,\ell
}\right) ^{-1}.  \label{W_normalization}
\end{equation}%
For more details, see Refs.\ \cite{Denicol:2012cn,Denicol:2012es}.

Using the Boltzmann equation, one can derive the general equations of motion
satisfied by $\rho _{r}^{\mu _{1}\cdots \mu _{\ell }}$. This is accomplished
by explicitly taking the comoving derivative of the corresponding
irreducible moment, i.e., $\dot{\rho}_{r}^{\left\langle \mu _{1}\cdots \mu
_{\ell }\right\rangle }=\Delta _{\nu _{1}\cdots \nu _{\ell }}^{\mu
_{1}\cdots \mu _{\ell }}D\int dKE_{\mathbf{k}}^{r}k^{\left\langle \nu
_{1}\right. }\cdots k^{\left. \nu _{\ell }\right\rangle }\delta f_{\mathbf{k}%
}$, and using the Boltzmann equation to express the comoving derivative of $%
\delta f_{\mathbf{k}}$ in terms of the collision term, $f_{0\mathbf{k}}$ and
its derivatives, and spatial derivatives of $\delta f_{\mathbf{k}}$. The
details of this derivation as well as the general form of the resulting
equations of motion is contained in Refs.\ \cite%
{Denicol:2012cn,Denicol:2012es}. For the three lowest-rank moments, these
equations of motion read 
\begin{align}
\dot{\rho}_{r}-C_{r-1}& =\alpha _{r}^{(0)}\theta +\left( 
\mbox{nonlinear
terms}\right) ,  \label{rho_dot} \\
\dot{\rho}_{r}^{\left\langle \mu \right\rangle }-C_{r-1}^{\left\langle \mu
\right\rangle }& =\alpha _{r}^{(1)}I^{\mu }+\left( \mbox{nonlinear terms}%
\right) ,  \label{rho_mu_dot} \\
\dot{\rho}_{r}^{\left\langle \mu \nu \right\rangle }-C_{r-1}^{\left\langle
\mu \nu \right\rangle }& =2\alpha _{r}^{(2)}\sigma ^{\mu \nu }+\left( %
\mbox{nonlinear terms}\right) .  \label{rho_mu_nu_dot}
\end{align}%
Here we defined the following thermodynamic quantities,%
\begin{align}
\alpha _{r}^{\left( 0\right) }& =\left( 1-r\right) I_{r1}-I_{r0}-\frac{n_{0}%
}{D_{20}}\left( h_{0}G_{2r}-G_{3r}\right) ,  \label{alpha_r_0} \\
\alpha _{r}^{\left( 1\right) }& =J_{r+1,1}-h_{0}^{-1}J_{r+2,1},
\label{alpha_r_1} \\
\alpha _{r}^{\left( 2\right) }& =I_{r+2,1}+\left( r-1\right) I_{r+2,2},
\label{alpha_r_2}
\end{align}%
where $h_{0}=\left( \varepsilon _{0}+P_{0}\right) /n_{0}$ denotes the
enthalpy per particle and 
\begin{eqnarray}
G_{nm} &=&J_{n0}J_{m0}-J_{n-1,0}J_{m+1,0},  \label{Gnm} \\
D_{nq} &=&J_{n+1,q}J_{n-1,q}-\left( J_{nq}\right) ^{2}.  \label{Dnq}
\end{eqnarray}%
The variables $I_{n+r,q}\left( \alpha _{0},\beta _{0}\right) $ and $%
J_{n+r,q}\left( \alpha _{0},\beta _{0}\right) $ correspond to thermodynamic
integrals defined as 
\begin{align}
I_{r+n,q}& =\frac{\left( -1\right) ^{q}}{\left( 2q+1\right) !!}\left\langle
E_{\mathbf{k}}^{n+r-2q}\left( \Delta ^{\alpha \beta }k_{\alpha }k_{\beta
}\right) ^{q}\right\rangle _{0},  \label{I_nq} \\
\text{ }J_{r+n,q}& =\left. \frac{\partial I_{n+r,q}}{\partial \alpha _{0}}%
\right\vert _{\beta _{0}}.  \label{J_nq}
\end{align}%
More details can be found in Appendix \ref{thermo_integrals}.

We also introduced the generalized irreducible collision integral $%
C_{r}^{\mu _{1}\cdots \mu _{\ell }}$ and its symmetric, traceless, and
orthogonal projection, 
\begin{equation}
C_{r}^{\left\langle \mu _{1}\cdots \mu _{\ell }\right\rangle }\equiv \Delta
_{\nu _{1}...\nu _{\ell }}^{\mu _{1}...\mu _{\ell }}C_{r}^{\nu _{1}\cdots
\nu _{\ell }}=\int dKE_{\mathbf{k}}^{r}k^{\left\langle \mu _{1}\right.
}\cdots k^{\left. \mu _{\ell }\right\rangle }C\left[ f\right] .
\label{General_Coll_Int}
\end{equation}

As was shown in Ref.~\cite{Denicol:2012cn}, the moment equations\ (\ref%
{rho_dot}) -- (\ref{rho_mu_nu_dot}) reduce to the fluid-dynamical equations
for the dissipative variables when the fast-varying modes of the Boltzmann
equation can be neglected and, simultaneously, the Knudsen number(s) and
inverse Reynolds number(s) are small. In this case, the linear parts of the
collision integrals introduced above determine the relaxation times for the
dissipative variables, while their nonlinear parts give rise to the terms
that are of second order in inverse Reynolds number(s), i.e., the tensors $%
\mathcal{R}$, $\mathcal{R}^{\mu }$, and $\mathcal{R}^{\mu \nu }$ that appear
in Eqs.\ (\ref{R_scalar}) -- (\ref{R_tensor}). In Ref.~\cite{Denicol:2012cn}
the existence of such nonlinear terms was pointed out but the explicit
calculation of the corresponding transport coefficients was left for future
work. In the next sections we shall complete this task.


\subsection{Expansion of the collision integral}

\label{BTE_coll_int}

In this section we show how to express the collision integrals in terms of
irreducible moments of $\delta f_{\mathbf{k}}$. Substituting the expression
of the collision term for binary elastic collisions (\ref{COLL_INT}) into
the expression for the irreducible collision integral (\ref{General_Coll_Int}%
), one obtains 
\begin{equation}
C_{r-1}^{\left\langle \mu _{1}\cdots \mu _{\ell }\right\rangle }=\frac{1}{%
\nu }\int dKdK^{\prime }dPdP^{\prime }W_{\mathbf{kk}\prime \rightarrow 
\mathbf{pp}\prime }E_{\mathbf{k}}^{r-1}k^{\left\langle \mu _{1}\right.
}\cdots k^{\left. \mu _{\ell }\right\rangle }\left( f_{\mathbf{p}}f_{\mathbf{%
p}^{\prime }}\tilde{f}_{\mathbf{k}}\tilde{f}_{\mathbf{k}^{\prime }}-f_{%
\mathbf{k}}f_{\mathbf{k}^{\prime }}\tilde{f}_{\mathbf{p}}\tilde{f}_{\mathbf{p%
}^{\prime }}\right) .
\end{equation}%
Substituting the distribution function $f_{\mathbf{k}}=f_{0\mathbf{k}}+f_{0%
\mathbf{k}}\tilde{f}_{0\mathbf{k}}\phi _{\mathbf{k}}$ into the above formula
and using
\begin{align}
f_{\mathbf{p}}f_{\mathbf{p}^{\prime }}& =f_{0\mathbf{p}}f_{0\mathbf{p}%
^{\prime }}\left( 1+\tilde{f}_{0\mathbf{p}^{\prime }}\phi _{\mathbf{p}%
^{\prime }}+\tilde{f}_{0\mathbf{p}}\phi _{\mathbf{p}}\right) +f_{0\mathbf{p}%
}f_{0\mathbf{p}^{\prime }}\tilde{f}_{0\mathbf{p}}\tilde{f}_{0\mathbf{p}%
^{\prime }}\phi _{\mathbf{p}}\phi _{\mathbf{p}^{\prime }},  \label{A1} \\
\tilde{f}_{\mathbf{p}}\tilde{f}_{\mathbf{p}^{\prime }}& =\tilde{f}_{0\mathbf{%
p}}\tilde{f}_{0\mathbf{p}^{\prime }}\left( 1-af_{0\mathbf{p}^{\prime }}\phi
_{\mathbf{p}^{\prime }}-af_{0\mathbf{p}}\phi _{\mathbf{p}}\right) +a^{2}f_{0%
\mathbf{p}}f_{0\mathbf{p}^{\prime }}\tilde{f}_{0\mathbf{p}}\tilde{f}_{0%
\mathbf{p}^{\prime }}\phi _{\mathbf{p}}\phi _{\mathbf{p}^{\prime }},
\label{A2}
\end{align}%
together with the equality $f_{0\mathbf{k}}f_{0\mathbf{k}^{\prime }}\tilde{f}%
_{0\mathbf{p}}\tilde{f}_{0\mathbf{p}^{\prime }}=f_{0\mathbf{p}}f_{0\mathbf{p}%
^{\prime }}\tilde{f}_{0\mathbf{k}}\tilde{f}_{0\mathbf{k}^{\prime }}$, the
part that is linear in $\phi _{\mathbf{k}}$ reads 
\begin{equation}
L_{r-1}^{\left\langle \mu _{1}\cdots \mu _{\ell }\right\rangle }=\frac{1}{%
\nu }\int_{f}E_{\mathbf{k}}^{r-1}k^{\left\langle \mu _{1}\right. }\cdots
k^{\left. \mu _{\ell }\right\rangle }\left( \phi _{\mathbf{p}}+\phi _{%
\mathbf{p}^{\prime }}-\phi _{\mathbf{k}}-\phi _{\mathbf{k}^{\prime }}\right)
,
\end{equation}%
where we abbreviated $\int_{f}=\int dKdK^{\prime }dPdP^{\prime }W_{\mathbf{kk%
}\prime \rightarrow \mathbf{pp}\prime }f_{0\mathbf{k}}f_{0\mathbf{k}^{\prime
}}\tilde{f}_{0\mathbf{p}}\tilde{f}_{0\mathbf{p}^{\prime }}$ and used the
fact that the collision term vanishes for the local equilibrium distribution
function $C_{r}^{\mu _{1}\cdots \mu _{\ell }}\left[ f_{0}\right] =0$.
Inserting the expression for $\phi _{\mathbf{k}}$ from the moment expansion\
(\ref{expansion1}) into the previous equation, we obtain 
\begin{align}
L_{r-1}^{\left\langle \mu _{1}\cdots \mu _{\ell }\right\rangle }& =\frac{1}{%
\nu }\sum_{m=0}^{\infty }\sum_{n=0}^{N_{m}}\int_{f}E_{\mathbf{k}%
}^{r-1}k^{\left\langle \mu _{1}\right. }\cdots k^{\left. \mu _{\ell
}\right\rangle }\rho _{n}^{\nu _{1}\cdots \nu _{m}}  \notag \\
& \times \left( \mathcal{H}_{\mathbf{p}n}^{\left( m\right) }p_{\left\langle
\nu _{1}\right. }\cdots p_{\left. \nu _{m}\right\rangle }+\mathcal{H}_{%
\mathbf{p}^{\prime }n}^{\left( m\right) }p_{\left\langle \nu _{1}\right.
}^{\prime }\cdots p_{\left. \nu _{m}\right\rangle }^{\prime }-\mathcal{H}_{%
\mathbf{k}n}^{\left( m\right) }k_{\left\langle \nu _{1}\right. }\cdots
k_{\left. \nu _{m}\right\rangle }-\mathcal{H}_{\mathbf{k}^{\prime
}n}^{\left( m\right) }k_{\left\langle \nu _{1}\right. }^{\prime }\cdots
k_{\left. \nu _{m}\right\rangle }^{\prime }\right) .  \label{Lin_coll_int}
\end{align}

It was shown in Ref.\ \cite{Denicol:2012cn} that the linear part of the
collision integral simplifies to 
\begin{equation}
L_{r-1}^{\left\langle \mu _{1}\cdots \mu _{\ell }\right\rangle
}=\sum_{m=0}^{\infty }\sum_{n=0}^{N_{m }}\left( \mathcal{A}_{rn}\right)
_{\nu _{1}\cdots \nu _{m}}^{\mu _{1}\cdots \mu _{\ell }}\rho _{n}^{\nu
_{1}\cdots \nu _{m}}=\sum_{m=0}^{\infty }\sum_{n=0}^{N_{m }}\mathcal{A}%
_{rn}^{\left( \ell \right) }\rho _{n}^{\mu _{1}\cdots \mu _{\ell }},
\label{Lin_collint}
\end{equation}%
where 
\begin{align}
\left( \mathcal{A}_{rn}\right) _{\nu _{1}\cdots \nu _{m}}^{\mu _{1}\cdots
\mu _{\ell }}& = \frac{1}{\nu }\int_{f}E_{\mathbf{k}}^{r-1}k^{\left\langle
\mu _{1}\right. }\cdots k^{\left. \mu _{\ell }\right\rangle }  \notag \\
& \times \left( \mathcal{H}_{\mathbf{p}n}^{\left( m\right) }p_{\left\langle
\nu _{1}\right. }\cdots p_{\left. \nu _{m}\right\rangle }+\mathcal{H}_{%
\mathbf{p}^{\prime }n}^{\left( m\right) }p_{\left\langle \nu _{1}\right.
}^{\prime }\cdots p_{\left. \nu _{m}\right\rangle }^{\prime }-\mathcal{H}_{%
\mathbf{k}n}^{\left( m\right) }k_{\left\langle \nu _{1}\right. }\cdots
k_{\left. \nu _{m}\right\rangle }-\mathcal{H}_{\mathbf{k}^{\prime
}n}^{\left( m\right) }k_{\left\langle \nu _{1}\right. }^{\prime }\cdots
k_{\left. \nu _{m}\right\rangle }^{\prime }\right) ,  \label{Arn_tensor}
\end{align}%
while using the properties of the irreducible projection tensors one can
show that%
\begin{equation}
\mathcal{A}_{rn}^{\left( \ell \right) }=\left[ \Delta _{\alpha _{1}\cdots \alpha
_{\ell }}^{\alpha _{1}\cdots \alpha _{\ell }}\right] ^{-1}\Delta _{\mu _{1}\cdots
\mu _{\ell }}^{\nu _{1}\cdots \nu _{\ell }}\left( \mathcal{A}_{rn}\right)
_{\nu _{1}\cdots \nu _{\ell }}^{\mu _{1}\cdots \mu _{\ell }}.  \label{A_rn}
\end{equation}%
The coefficient $\mathcal{A}_{rn}^{\left( \ell \right) }$ is the $\left(
rn\right) $ element of an $\left( N_{\ell }+1\right) \times \left( N_{\ell
}+1\right) $ matrix, $\mathcal{A}^{\left( \ell \right) }$, and, in the
linearized case, contains all the information of the underlying microscopic
theory. We remark that, for $\ell =0$, the second and third rows and columns
($r,n=1,2$) and, for $\ell =1$, the second row and column ($r,n=1$) are
zero, because the moments $\rho _{1}$, $\rho _{2}$, and $\rho _{1}^{\mu }$
vanish due to the matching conditions and our choice of frame.

The computation of the nonlinear part of the collision integral is
analogous. Inspecting the previous formulas we observe that the collision
integral is a quartic function of $\phi _{\mathbf{k}}$. However, in this
paper we shall restrict our calculations to the case of Boltzmann statistics 
$\left( a=0\right) $, in which case the dependence on $\phi _{\mathbf{k}}$
becomes quadratic. The collision integral can be written as 
\begin{equation}
C_{r-1}^{\left\langle \mu _{1}\cdots \mu _{\ell }\right\rangle
}=L_{r-1}^{\left\langle \mu _{1}\cdots \mu _{\ell }\right\rangle
}+N_{r-1}^{\left\langle \mu _{1}\cdots \mu _{\ell }\right\rangle },
\end{equation}%
where the quadratic contribution to the collision integral reads 
\begin{align}
N_{r-1}^{\left\langle \mu _{1}\cdots \mu _{\ell }\right\rangle }& \equiv 
\frac{1}{\nu }\int_{f}E_{\mathbf{k}}^{r-1}k^{\left\langle \mu _{1}\right.
}\cdots k^{\left. \mu _{\ell }\right\rangle }\left( \phi _{\mathbf{p}}\phi _{%
\mathbf{p}^{\prime }}-\phi _{\mathbf{k}}\phi _{\mathbf{k}^{\prime }}\right) 
\notag \\
& =\frac{1}{\nu }\sum_{m,m^{\prime }=0}^{\infty
}\sum_{n=0}^{N_{m}}\sum_{n^{\prime }=0}^{N_{m^{\prime }}}\int_{f}E_{\mathbf{k%
}}^{r-1}k^{\left\langle \mu _{1}\right. }\cdots k^{\left. \mu _{\ell
}\right\rangle }\rho _{n}^{\alpha _{1}\cdots \alpha _{m}}\rho _{n^{\prime
}}^{\beta _{1}\cdots \beta _{m^{\prime }}}  \notag \\
& \times \left( \mathcal{H}_{\mathbf{p}n}^{\left( m\right) }\,\mathcal{H}_{%
\mathbf{p}^{\prime }n^{\prime }}^{(m^{\prime })}\,p_{\left\langle \alpha
_{1}\right. }\cdots p_{\left. \alpha _{m}\right\rangle }\,p_{\left\langle
\beta _{1}\right. }^{\prime }\cdots p_{\left. \beta _{m^{\prime
}}\right\rangle }^{\prime }-\mathcal{H}_{\mathbf{k}n}^{\left( m\right) }\,%
\mathcal{H}_{\mathbf{k}^{\prime }n^{\prime }}^{(m^{\prime
})}\,k_{\left\langle \alpha _{1}\right. }\cdots k_{\left. \alpha
_{m}\right\rangle }\,k_{\left\langle \beta _{1}\right. }^{\prime }\cdots
k_{\left. \beta _{m^{\prime }}\right\rangle }^{\prime }\right) .
\label{Non_Lin_coll_int}
\end{align}%
This nonlinear contribution can be further simplified to 
\begin{equation}
N_{r-1}^{\left\langle \mu _{1}\cdots \mu _{\ell }\right\rangle
}=\sum_{m^{\prime }=0}^{\infty }\sum_{m=0}^{m^{\prime
}}\sum_{n=0}^{N_{m}}\sum_{n^{\prime }=0}^{N_{m^{\prime }}}\left( \mathcal{N}%
_{rnn^{\prime }}\right) _{\alpha _{1}\cdots \alpha _{m}\beta _{1}\cdots
\beta _{m^{\prime }}}^{\mu _{1}\cdots \mu _{\ell }}\rho _{n}^{\alpha
_{1}\cdots \alpha _{m}}\rho _{n^{\prime }}^{\beta _{1}\cdots \beta
_{m^{\prime }}},  \label{NonLin_collint}
\end{equation}%
where we defined the following tensor of rank $\ell +m+m^{\prime }$,%
\begin{align}
\left( \mathcal{N}_{rnn^{\prime }}\right) _{\alpha _{1}\cdots \alpha
_{m}\beta _{1}\cdots \beta _{m^{\prime }}}^{\mu _{1}\cdots \mu _{\ell }}& =%
\frac{1}{\nu }\int_{f}E_{\mathbf{k}}^{r-1}k^{\left\langle \mu _{1}\right.
}\cdots k^{\left. \mu _{\ell }\right\rangle }  \notag \\
& \times \left[ \mathcal{H}_{\mathbf{p}n}^{\left( m\right) }\,\mathcal{H}_{%
\mathbf{p}^{\prime }n^{\prime }}^{(m^{\prime })}\,p_{\left\langle \alpha
_{1}\right. }\cdots p_{\left. \alpha _{m}\right\rangle }\,p_{\left\langle
\beta _{1}\right. }^{\prime }\cdots p_{\left. \beta _{m^{\prime
}}\right\rangle }^{\prime }+\left( 1-\delta _{mm^{\prime }}\right) \,%
\mathcal{H}_{\mathbf{p}^{\prime }n}^{\left( m\right) }\,\mathcal{H}_{\mathbf{%
p}n^{\prime }}^{(m^{\prime })}\,p_{\left\langle \alpha _{1}\right. }^{\prime
}\cdots p_{\left. \alpha _{m}\right\rangle }^{\prime }\,p_{\left\langle
\beta _{1}\right. }\cdots p_{\left. \beta _{m^{\prime }}\right\rangle
}\right.  \notag \\
& \left. -\mathcal{H}_{\mathbf{k}n}^{\left( m\right) }\,\mathcal{H}_{\mathbf{%
k}^{\prime }n^{\prime }}^{(m^{\prime })}\,k_{\left\langle \alpha _{1}\right.
}\cdots k_{\left. \alpha _{m}\right\rangle }\,k_{\left\langle \beta
_{1}\right. }^{\prime }\cdots k_{\left. \beta _{m^{\prime }}\right\rangle
}^{\prime }-\left( 1-\delta _{mm^{\prime }}\right) \,\mathcal{H}_{\mathbf{k}%
^{\prime }n}^{\left( m\right) }\,\mathcal{H}_{\mathbf{k}n^{\prime
}}^{(m^{\prime })}\,k_{\left\langle \alpha _{1}\right. }^{\prime }\cdots
k_{\left. \alpha _{m}\right\rangle }^{\prime }\,k_{\left\langle \beta
_{1}\right. }\cdots k_{\left. \beta _{m^{\prime }}\right\rangle }\right] .
\label{BIG_NonLin}
\end{align}%
In comparison with Eq.\ (\ref{Non_Lin_coll_int}), we have split the double
sum $\sum_{m=0}^{\infty }\sum_{m^{\prime }=0}^{\infty }$ into a double sum $%
\sum_{m^{\prime }=0}^{\infty }\sum_{m=0}^{m^{\prime }}$ and a double sum $%
\sum_{m=0}^{\infty }\sum_{m^{\prime }=0}^{m}$, and subtracted the
superfluous term $m=m^{\prime }$ in the last sum with the help of a
Kronecker delta. Then we interchanged indices $m\leftrightarrow m^{\prime
},\,n\leftrightarrow n^{\prime }$ in the second sum.

The tensor $\left( \mathcal{N}_{rnn^{\prime }}\right) _{\alpha _{1}\cdots
\alpha _{m}\beta _{1}\cdots \beta _{m^{\prime }}}^{\mu _{1}\cdots \mu _{\ell
}}$ is symmetric under
permutations of $\mu $--type, $\alpha $--type, and $\beta $--type indices,
and depends solely on equilibrium distribution functions and the
corresponding cross-section(s). The equilibrium distribution function
contains only one 4-vector, i.e., the fluid 4-velocity $u^{\mu }$.
Therefore, $\left( \mathcal{N}_{rnn^{\prime }}\right) _{\alpha _{1}\cdots
\alpha _{m}\beta _{1}\cdots \beta _{m^{\prime }}}^{\mu _{1}\cdots \mu _{\ell
}}$ must be constructed from tensor structures made of $u^{\mu }$ and the
metric tensor $g^{\mu \nu }$, or, equivalently, $u^{\mu }$ and $\Delta ^{\mu
\nu }$. Furthermore, $\left( \mathcal{N}_{rnn^{\prime }}\right) _{\alpha
_{1}\cdots \alpha _{m}\beta _{1}\cdots \beta _{m^{\prime }}}^{\mu _{1}\cdots
\mu _{\ell }}$ must be orthogonal to $u^{\mu }$ which implies that it can
only be constructed from combinations of elementary projection operators, $%
\Delta ^{\mu \nu}$. This already constrains the rank of the tensor, $%
\ell + m+m^{\prime } $, to be an even number. Finally, it must satisfy the
following property: 
\begin{equation}
\Delta _{\mu _{1}\cdots \mu _{\ell }}^{\mu _{1}^{\prime }\cdots \mu _{\ell
}^{\prime }}\Delta _{\alpha _{1}^{\prime }\cdots \alpha _{m}^{\prime
}}^{\alpha _{1}\cdots \alpha _{m}}\Delta _{\beta _{1}^{\prime }\cdots \beta
_{m^{\prime }}^{\prime }}^{\beta _{1}\cdots \beta _{m^{\prime }}}\left( 
\mathcal{N}_{rnn^{\prime }}\right) _{\alpha _{1}\cdots \alpha _{m}\beta
_{1}\cdots \beta _{m^{\prime }}}^{\mu _{1}\cdots \mu _{\ell }}=\left( 
\mathcal{N}_{rnn^{\prime }}\right) _{\alpha _{1}^{\prime }\cdots \alpha
_{m}^{\prime }\beta _{1}^{\prime }\cdots \beta _{m^{\prime }}^{\prime
}}^{\mu _{1}^{\prime }\cdots \mu _{\ell }^{\prime }}.  \label{great property}
\end{equation}

For our purposes it is sufficient to calculate terms that are of second
order in inverse Reynolds number, i.e., the terms $\mathcal{R}$, $\mathcal{R}%
^{\mu }$, and $\mathcal{R}^{\mu \nu }$. Therefore, we only need to consider
the cases $\ell =0$, $\ell =1$, and $\ell =2$. Since the actual deduction of
the nonlinear collision integrals is complicated, this task is relegated to
Appendix \ref{collision_tensors} and here we shall only give the final
results.

The scalar nonlinear collision integral from Eq.\ (\ref{NonLin_collint}) is
given by 
\begin{align}
N_{r-1}&\equiv \sum_{m^{\prime }=0}^{\infty }\sum_{m=0}^{m^{\prime
}}\sum_{n=0}^{N_{m}}\sum_{n^{\prime }=0}^{N_{m^{\prime }}}\left( \mathcal{N}%
_{rnn^{\prime }}\right) _{\alpha _{1}\cdots \alpha _{m}\beta _{1}\cdots
\beta _{m^{\prime }}}\rho _{n}^{\alpha _{1}\cdots \alpha _{m}}\rho
_{n^{\prime }}^{\beta _{1}\cdots \beta _{m^{\prime }}}  \notag \\
&=\sum_{n=0}^{N_{0}}\sum_{n^{\prime }=0}^{N_{0}}\mathcal{C}_{rnn^{\prime
}}^{0\left( 0,0\right) }\rho _{n}\rho_{n^{\prime }} +
\sum_{m=1}^{\infty}\sum_{n=0}^{N_{m}}\sum_{n^{\prime }=0}^{N_{m}}\mathcal{C}%
_{rnn^{\prime }}^{0\left( m,m\right) }\rho _{n}^{\alpha _{1}\cdots \alpha
_{m}}\rho _{n^{\prime },\alpha _{1}\cdots \alpha _{m}},  \label{Nr_scalar}
\end{align}%
where $\mathcal{C}_{rnn^{\prime }}^{0\left( m,m\right) }$ is the special
case $\ell =0$ of a more general coefficient 
\begin{align}
\mathcal{C}_{rnn^{\prime }}^{\ell \left( m,m+\ell \right) }& =\frac{1}{%
\left( 2m+2\ell +1\right) \nu }\int_{f}E_{\mathbf{k}}^{r-1}k^{\left\langle
\mu _{1}\right. }\cdots k^{\left. \mu _{\ell }\right\rangle }  \notag \\
& \times \left[ \mathcal{H}_{\mathbf{p}n}^{\left( m\right) }\,\mathcal{H}_{%
\mathbf{p}^{\prime }n^{\prime }}^{\left( m+\ell \right) }\,p^{\left\langle
\nu _{1}\right. }\cdots p^{\left. \nu _{m}\right\rangle }\,p_{\left\langle
\mu _{1}\right. }^{\prime }\cdots p_{\mu _{\ell }}^{\prime }\,p_{\nu
_{1}}^{\prime }\cdots p_{\left. \nu _{m}\right\rangle }^{\prime }\right. 
\notag \\
& \left. +\left( 1-\delta _{m,m+\ell}\right) \mathcal{H}_{\mathbf{p}^{\prime
}n}^{\left( m\right) }\,\mathcal{H}_{\mathbf{p}n^{\prime }}^{\left( m+\ell
\right) }\,p^{\prime \left\langle \nu _{1}\right. }\cdots p^{\prime \left.
\nu _{m}\right\rangle }\,p_{\left\langle \mu _{1}\right. }\cdots p_{\mu
_{\ell }}\,p_{\nu _{1}}\cdots p_{\left. \nu _{m}\right\rangle }\right. 
\notag \\
& \left. -\mathcal{H}_{\mathbf{k}n}^{\left( m\right) }\,\mathcal{H}_{\mathbf{%
k}^{\prime }n^{\prime }}^{\left( m+\ell \right) }\,k^{\left\langle \nu
_{1}\right. }\cdots k^{\left. \nu _{m}\right\rangle }\,k_{\left\langle \mu
_{1}\right. }^{\prime }\cdots k_{\mu _{\ell }}^{\prime }\,k_{\nu
_{1}}^{\prime }\cdots k_{\left. \nu _{m}\right\rangle }^{\prime }\right. 
\notag \\
& \left. -\left( 1-\delta _{m,m+\ell}\right) \mathcal{H}_{\mathbf{k}^{\prime
}n}^{\left( m\right) }\,\mathcal{H}_{\mathbf{k}n^{\prime }}^{\left( m+\ell
\right) }\,k^{\prime \left\langle \nu _{1}\right. }\cdots k^{\prime \left.
\nu _{m}\right\rangle }\,k_{\left\langle \mu _{1}\right. }\cdots k_{\mu
_{\ell }}k_{\nu _{1}}\cdots k_{\left. \nu _{m}\right\rangle }\right] .
\label{C_lm}
\end{align}

Similarly, the nonlinear collision term for $\ell =1$ becomes,%
\begin{align}
N_{r-1}^{\mu }&\equiv \sum_{m^{\prime }=0}^{\infty }\sum_{m=0}^{m^{\prime
}}\sum_{n=0}^{N_{m}}\sum_{n^{\prime }=0}^{N_{m^{\prime }}}\left( \mathcal{N}%
_{rnn^{\prime }}\right) _{\alpha _{1}\cdots \alpha _{m}\beta _{1}\cdots
\beta _{m^{\prime }}}^{\mu }\rho _{n}^{\alpha _{1}\cdots \alpha _{m}}\rho
_{n^{\prime }}^{\beta _{1}\cdots \beta _{m^{\prime }}}  \notag \\
&=\sum_{n=0}^{N_{0}}\sum_{n^{\prime }=0}^{N_{1}}\mathcal{C}_{rnn^{\prime
}}^{1\left( 0,1\right) }\rho _{n}\rho_{n^{\prime }}^{\mu } +
\sum_{m=1}^{\infty }\sum_{n=0}^{N_{m}}\sum_{n^{\prime }=0}^{N_{m+1}}\mathcal{%
C}_{rnn^{\prime }}^{1\left( m,m+1\right) }\rho _{n}^{\alpha _{1}\cdots
\alpha _{m}}\rho _{n^{\prime },\alpha _{1}\cdots \alpha _{m}}^{\mu }.
\label{Nr_vector}
\end{align}%
where the coefficient $\mathcal{C}_{rnn^{\prime }}^{1\left( m,m+1\right) }$
is the $\ell =1$ case of the general coefficient\ $\mathcal{C}_{rnn^{\prime
}}^{\ell \left( m,m+\ell \right) }$ \ introduced in Eq.\ (\ref{C_lm}).

Finally, the rank-2 tensor terms are obtained taking $\ell =2$, 
\begin{align}
N_{r-1}^{\mu \nu }& \equiv \sum_{m^{\prime }=0}^{\infty
}\sum_{m=0}^{m^{\prime }}\sum_{n=0}^{N_{m}}\sum_{n^{\prime
}=0}^{N_{m^{\prime }}}\left( \mathcal{N}_{rnn^{\prime }}\right) _{\alpha
_{1}\cdots \alpha _{m}\beta _{1}\cdots \beta _{m^{\prime }}}^{\mu \nu }\rho
_{n}^{\alpha _{1}\cdots \alpha _{m}}\rho _{n^{\prime }}^{\beta _{1}\cdots
\beta _{m^{\prime }}}  \notag \\
& =\sum_{m=0}^{\infty }\sum_{n=0}^{N_{m+2}}\sum_{n^{\prime }=0}^{N_{m}}%
\mathcal{C}_{rnn^{\prime }}^{2\left( m,m+2\right) }\rho _{n}^{\alpha
_{1}\cdots \alpha _{m}}\rho _{n^{\prime },\alpha _{1}\cdots \alpha
_{m}}^{\mu \nu }  \notag \\
& + \sum_{n=0}^{N_{1}}\sum_{n^{\prime}=0}^{N_{1}} \mathcal{D}_{rnn^{\prime
}}^{2\left( 11\right) }\rho_{n}^{\left\langle \mu \right. }\rho _{n^{\prime
}}^{\left. \nu \right\rangle } +\sum_{m=2}^{\infty
}\sum_{n=0}^{N_{m}}\sum_{n^{\prime }=0}^{N_{m}}\mathcal{D}_{rnn^{\prime
}}^{2\left( mm\right) }\rho _{n}^{\alpha _{2}\cdots \alpha _{m}\left\langle
\mu \right. }\rho _{n^{\prime },\alpha _{2}\cdots \alpha _{m}}^{\left. \nu
\right\rangle },  \label{Nr_tensor}
\end{align}%
where $\mathcal{C}_{rnn^{\prime }}^{2\left( m,m+2\right) }$ can be
calculated from Eq.\ (\ref{C_lm}) and we introduced another coefficient, 
\begin{eqnarray}
\mathcal{D}_{rnn^{\prime }}^{2\left( mm\right) } &=&\frac{1}{d^{\left(
m\right) }\nu }\int_{f}E_{\mathbf{k}}^{r-1}k^{\left\langle \mu \right.
}k^{\left. \nu \right\rangle }  \notag \\
&\times &\left( \mathcal{H}_{\mathbf{p}n}^{\left( m\right) }\,\mathcal{H}_{%
\mathbf{p}^{\prime }n^{\prime }}^{\left( m\right) }\,p_{\left\langle \mu
\right. }p^{\beta _{q+1}}\cdots p^{\left. \beta _{m}\right\rangle
}\,p_{\left\langle \nu \right. }^{\prime }p_{\beta _{q+1}}^{\prime }\cdots
p_{\left. \beta _{m}\right\rangle }^{\prime }-\mathcal{H}_{\mathbf{k}%
n}^{\left( m\right) }\,\mathcal{H}_{\mathbf{k}^{\prime }n^{\prime }}^{\left(
m\right) }\,k_{\left\langle \mu \right. }k^{\beta _{q+1}}\cdots k^{\left.
\beta _{m}\right\rangle }\,k_{\left\langle \nu \right. }^{\prime }k_{\beta
_{q+1}}\cdots k_{\left. \beta _{m}\right\rangle }^{\prime }\right) .
\label{DDDDD}
\end{eqnarray}

The normalization $d^{\left( m\right) }$ is complicated and is discussed 
in Appendix \ref{collision_tensors} together with other details of
the derivation of the nonlinear collision term.

\section{Transport coefficients in the 14--moment approximation}

\label{BTE_14moment}

In this section we calculate the previously introduced coefficients $%
\mathcal{A}_{rn}^{\left( \ell \right) }$, $\mathcal{C}_{rnn^{\prime
}}^{0\left( mm\right) }$, $\mathcal{C}_{rnn^{\prime }}^{1\left( m,m+1\right)
}$,$\ \mathcal{C}_{rnn^{\prime }}^{2\left( m,m+2\right) }$, and $\mathcal{D}%
_{rnn^{\prime }}^{2\left( m,m\right) }$ in the 14--moment approximation. As
shown in Refs.\ \cite{Denicol:2012cn,Denicol:2012es}, this corresponds to
the truncation $N_{0}=2,\,N_{1}=1,\,N_{2}=0$. This implies that the
following irreducible moments appear: $\rho _{0}=-3\Pi /m^{2}$, $\rho _{1}=0$%
, $\rho _{2}=0$, $\rho _{0}^{\mu }=n^{\mu }$, $\rho _{1}^{\mu }=0$, and $%
\rho _{0}^{\mu \nu }=\pi ^{\mu \nu }$. As one can see, they are uniquely
related to the dissipative quantities.

Before proceeding and for the sake of later convenience, we
re-express the coefficients $\mathcal{H}_{\mathbf{k}n}^{\left( \ell \right) }
$ using Eqs.\ (\ref{H_kn}) and (\ref{P_kn}) as 
\begin{equation}
\mathcal{H}_{\mathbf{k}n}^{\left( \ell \right) }\equiv \frac{W^{\left( \ell
\right) }}{\ell !}\sum_{k=n}^{N_{\ell }}\sum_{r=0}^{k}a_{kr}^{(\ell
)}a_{kn}^{(\ell )}E_{\mathbf{k}}^{r}=\sum_{r=n}^{N_{\ell }}A_{rn}^{\left(
\ell \right) }E_{\mathbf{k}}^{r}+\sum_{r=0}^{n-1}A_{nr}^{\left( \ell \right)
}E_{\mathbf{k}}^{r},  \label{H_kn_expanded}
\end{equation}%
where 
\begin{eqnarray}
A_{rn}^{(\ell )} &=&\frac{W^{\left( \ell \right) }}{\ell !}%
\sum_{k=r}^{N_{\ell }}a_{kr}^{(\ell )}a_{kn}^{(\ell )}.  \label{A_l_rn}
\end{eqnarray}%
Note that, for $n=0$,  the second sum in Eq.\ (\ref%
{H_kn_expanded}) identically vanishes, which greatly simplifies the calculation of the collision
integral.

Furthermore, from the definition of the irreducible moments and using Eqs.\ (%
\ref{expansion1}) -- (\ref{P_kn}) together with the orthogonality condition (%
\ref{orthogonality1}) we obtain the following general result, 
\begin{eqnarray}
\rho _{r}^{\mu _{1}\cdots \mu _{\ell }} &\equiv &\frac{\ell !}{\left( 2\ell
+1\right) !!}\sum_{n=0}^{N_{\ell }}\rho _{n}^{\mu _{1}\cdots \mu _{\ell
}}\int dKE_{\mathbf{k}}^{r}\left( \Delta ^{\alpha \beta }k_{\alpha }k_{\beta
}\right) ^{\ell }\mathcal{H}_{\mathbf{k}n}^{\left( \ell \right) }f_{0\mathbf{%
k}}\tilde{f}_{0\mathbf{k}}  \notag \\
&=&\left( -1\right) ^{\ell }\ell !\sum_{n=0}^{N_{\ell }}\rho _{n}^{\mu
_{1}\cdots \mu _{\ell }}\left( \sum_{r^{\prime }=n}^{N_{\ell }}A_{r^{\prime
}n}^{\left( \ell \right) }J_{r+r^{\prime }+2\ell ,\ell }+\sum_{r^{\prime
}=0}^{n-1}A_{n r^{\prime }}^{\left( \ell \right) }J_{r+r^{\prime }+2\ell
,\ell }\right) .
\end{eqnarray}%
where we used Eq.\ (\ref{H_kn_expanded}) in the last step.\textbf{\ }%
Therefore, truncating the above general result in the 14--moment
approximation we obtain 
\begin{eqnarray}
\rho _{r} &\equiv &\gamma _{r}^{\Pi }\rho _{0}=-\frac{3}{m^{2}}\left(
A_{00}^{\left( 0\right) }J_{r,0}+A_{10}^{\left( 0\right)
}J_{r+1,0}+A_{20}^{\left( 0\right) }J_{r+2,0}\right) \Pi ,
\label{rho_r_recursive} \\
\rho _{r}^{\mu } &\equiv &\gamma _{r}^{n}\rho _{0}^{\mu }=-\left(
A_{00}^{\left( 1\right) }J_{r+2,1}+A_{10}^{\left( 1\right) }J_{r+3,1}\right)
n^{\mu },  \label{rho_mu_r_recursive} \\
\rho _{r}^{\mu \nu } &\equiv &\gamma _{r}^{\pi }\rho _{0}^{\mu \nu }=\left(
2A_{00}^{\left( 2\right) }J_{r+4,2}\right) \pi ^{\mu \nu },
\label{rho_mu_nu_r_recursive}
\end{eqnarray}%
where for $r=0$ we obviously have $\gamma _{0}^{\Pi }=\gamma _{0}^{n}=\gamma
_{0}^{\pi }=1$. The coefficients $A_{20}^{\left( 0\right) }$, $%
A_{10}^{\left( 1\right) }$, $A_{00}^{\left( 2\right) }$, as well as $%
A_{00}^{\left( 0\right) }$, $A_{10}^{\left( 0\right) }$, $A_{20}^{\left(
0\right) }$ are calculated from Eq.\ (\ref{A_l_rn}) and listed in Appendix %
\ref{exp_coefficients}. These linear relations between the moments are the
main result of the 14--moment approximation, which was also obtained in
Ref.\ \cite{Denicol:2012es}.

It is straightforward to show using Eqs.\ (\ref{Arn_tensor}), (\ref{A_rn}),
and (\ref{H_kn_expanded}) that the $\mathcal{A}_{rn}^{\left( \ell \right) }$
coefficients of the linear collision term can be expressed in terms of $%
A_{rn}^{(\ell)}$. For $\ell=0$ where, in the 14--moment approximation, $%
N_{0}=2$, the coefficient is 
\begin{align}
\mathcal{A}_{r0}^{\left( 0\right) }& \equiv A_{20}^{\left( 0\right) }\frac{1%
}{\nu }\int_{f}E_{\mathbf{k}}^{r-1}\left( E_{\mathbf{p}}^{2} +E_{\mathbf{p}%
^{\prime }}^{2} - E_{\mathbf{k}}^{2}-E_{\mathbf{k}^{\prime }}^{2}\right)
=A_{20}^{\left( 0\right) }X_{\left( r-3\right) }^{\mu \nu \alpha \beta
}u_{\mu }u_{\nu }u_{\alpha }u_{\beta },  \label{A0_r0}
\end{align}%
where the integrals proportional to $A_{00}^{\left( 0\right) }\int_{f}
\left( 1+1 - 1-1\right) =0$ and $A_{10}^{\left( 0\right) }\int_{f}\left( E_{%
\mathbf{p}}+E_{\mathbf{p}^{\prime }} - E_{\mathbf{k}}-E_{\mathbf{k}^{\prime
}}\right) =0$ vanish due to particle number and energy conservation in
binary collisions. Here, we introduced the following rank-4 tensor 
\begin{equation}
X_{\left( r\right) }^{\mu \nu \alpha \beta }=\frac{1}{\nu }\int_{f}E_{%
\mathbf{k}}^{r}k^{\mu }k^{\nu }\left( p^{\alpha }p^{\beta }+p^{\prime \alpha
}p^{\prime \beta } - k^{\alpha }k^{\beta } - k^{\prime \alpha }k^{\prime
\beta}\right) ,  \label{X_4rank_tensor}
\end{equation}%
which is symmetric upon the interchange of indices $\left( \mu ,\nu \right) $
and $\left( \alpha ,\beta \right) $, i.e., $X_{\left( r\right) }^{\mu \nu
\alpha \beta }=X_{\left( r\right) }^{\left( \mu \nu \right) \left( \alpha
\beta \right) }$, and it is also traceless in the latter indices, $X_{\left(
r\right) }^{\mu \nu \alpha \beta }g_{\alpha \beta }=0$.

Similarly, for $\ell =1$ we have 
\begin{align}
\mathcal{A}_{r0}^{\left( 1\right) }& \equiv A_{10}^{\left( 1\right) }\frac{1%
}{3\nu }\int_{f}E_{\mathbf{k}}^{r-1}k^{\left\langle \mu \right\rangle
}\left( E_{\mathbf{p}}p_{\left\langle \mu \right\rangle }+E_{\mathbf{p}%
^{\prime }}\,p_{\left\langle \mu \right\rangle }^{\prime }-E_{\mathbf{k}%
}\,k_{\left\langle \mu \right\rangle }-E_{\mathbf{k}^{\prime
}}\,k_{\left\langle \mu \right\rangle }^{\prime }\right) =A_{10}^{\left(
1\right) }\frac{1}{3}X_{\left( r-2\right) }^{\mu \nu \alpha \beta }u_{\left(
\mu \right. }\Delta _{\left. \nu \right) \left( \alpha \right. }u_{\left.
\beta \right) },  \label{A1_r0}
\end{align}%
where $A_{00}^{\left( 1\right) }\int_{f}\left( p_{\left\langle \mu
\right\rangle }+p_{\left\langle \mu \right\rangle }^{\prime
}-k_{\left\langle \mu \right\rangle }-k_{\left\langle \mu \right\rangle
}^{\prime }\right) =0$ vanishes due to 3-momentum conservation.

Finally, for $\ell =2$ we obtain 
\begin{align}
\mathcal{A}_{r0}^{\left( 2\right) }& \equiv A_{00}^{\left( 2\right) }\frac{1%
}{5\nu }\int_{f}E_{\mathbf{k}}^{r-1}k^{\left\langle \mu \right. }k^{\left.
\nu \right\rangle }\left( p_{\left\langle \mu \right. }p_{\left. \nu
\right\rangle }+p_{\left\langle \mu \right. }^{\prime }p_{\left. \nu
\right\rangle }^{\prime }-k_{\left\langle \mu \right. }k_{\left. \nu
\right\rangle }-k_{\left\langle \mu \right. }^{\prime }k_{\left. \nu
\right\rangle }^{\prime }\right) =A_{00}^{\left( 2\right) }\frac{1}{5}%
X_{\left( r-1\right) }^{\mu \nu \alpha \beta }\Delta _{\mu \nu \alpha \beta
}.  \label{A2_r0}
\end{align}%
Recalling that $L_{r-1}^{\left\langle \mu _{1}\cdots \mu _{\ell
}\right\rangle }=\sum_{n=0}^{N_{\ell }}\mathcal{A}_{rn}^{\left( \ell \right)
}\rho _{n}^{\mu _{1}\cdots \mu _{\ell }}$, these results lead to the linear
collision terms in the 14--moment approximation, 
\begin{align}
L_{r-1}& =\sum_{n=0}^{2}\mathcal{A}_{rn}^{\left( 0\right) }\rho _{n}\equiv 
\mathcal{A}_{r0}^{\left( 0\right) }\rho _{0}=-A_{20}^{\left( 0\right)
}X_{\left( r-3\right) ,1}\;\frac{3}{m^{2}}\Pi , \\
L_{r-1}^{\left\langle \mu \right\rangle }& =\sum_{n=0}^{1}\mathcal{A}%
_{rn}^{\left( 1\right) }\rho _{n}^{\mu }\equiv \mathcal{A}_{r0}^{\left(
1\right) }\rho _{0}^{\mu }=A_{10}^{\left( 1\right) }X_{\left( r-2\right)
,3}\;n^{\mu },  \label{L_mu} \\
L_{r-1}^{\left\langle \mu \nu \right\rangle }& =\mathcal{A}_{r0}^{\left(
2\right) }\rho _{0}^{\mu \nu }=A_{00}^{\left( 2\right) }X_{\left( r-1\right)
,4}\;\pi ^{\mu \nu }.  \label{L_mu_nu}
\end{align}%
Here, we denoted the different tensor projections as $X_{\left( r\right)
,1}=X_{\left( r\right) }^{\mu \nu \alpha \beta }u_{\mu }u_{\nu }u_{\alpha
}u_{\beta }$, $X_{\left( r\right) ,3}=\frac{1}{3}X_{\left( r\right) }^{\mu
\nu \alpha \beta }u_{\left( \mu \right. }\Delta _{\left. \nu \right) \left(
\alpha \right. }u_{\left. \beta \right) }$, and $X_{\left( r\right) ,4}=%
\frac{1}{5}X_{\left( r\right) }^{\mu \nu \alpha \beta }\Delta _{\mu \nu
\alpha \beta }$.

Now, with the help of these formulas and using Eqs.\ (\ref{rho_r_recursive}%
) -- (\ref{rho_mu_nu_r_recursive}) the coefficients of bulk viscosity,
particle diffusion coefficient, and shear viscosity, as well as the
corresponding relaxation times can be calculated,%
\begin{eqnarray}
\zeta ^{r} &=&\frac{\alpha _{r}^{\left( 0\right) }}{A_{20}^{\left( 0\right)
}X_{\left( r-3\right) ,1}},\ \tau _{\Pi }^{r}=-\frac{\gamma _{r}^{\Pi }}{%
A_{20}^{\left( 0\right) }X_{\left( r-3\right) ,1}}, \\
\kappa ^{r} &=&-\frac{\alpha _{r}^{\left( 1\right) }}{A_{10}^{\left(
1\right) }X_{\left( r-2\right) ,3}},\ \tau _{n}^{r}=-\frac{\gamma _{r}^{n}}{%
A_{10}^{\left( 1\right) }X_{\left( r-2\right) ,3}}, \\
\eta ^{r} &=&-\frac{\alpha _{r}^{\left( 2\right) }}{A_{00}^{\left( 2\right)
}X_{\left( r-1\right) ,4}},\ \tau _{\pi }^{r}=-\frac{\gamma _{r}^{\pi }}{%
A_{00}^{\left( 2\right) }X_{\left( r-1\right) ,4}},
\end{eqnarray}%
where $\alpha _{r}^{\left( 0\right) }$, $\alpha _{r}^{\left( 1\right) }$,
and $\alpha _{r}^{\left( 2\right) }$ were defined in Eqs.\ (\ref{alpha_r_0}%
) -- (\ref{alpha_r_2}) while $\gamma _{r}^{\Pi }$, $\gamma _{r}^{n}$, and $%
\gamma _{r}^{\pi }$ are listed in Eqs.\ (\ref{rho_r_recursive}) -- (\ref%
{rho_mu_nu_r_recursive}).

We now compute the nonlinear collision terms. With Eq.\ (\ref{H_kn_expanded}%
), the scalar contribution (\ref{C_lm}) is 
\begin{align}
\mathcal{C}_{rnn^{\prime }}^{0\left( mm\right) } & =\frac{1}{\left(
2m+1\right) \nu }\int_{f}E_{\mathbf{k}}^{r-1}  \notag \\
& \times \left[ \left( \sum_{i=n}^{N_{m}}A_{in}^{\left( m\right) }E_{\mathbf{%
p}}^{i}+\sum_{i=0}^{n-1}A_{ni}^{\left( m\right) }E_{\mathbf{p}}^{i}\right)
\left( \sum_{i^{\prime }=n^{\prime }}^{N_{m}}A_{i^{\prime }n^{\prime
}}^{\left( m\right) }E_{\mathbf{p}^{\prime }}^{i^{\prime }}+\sum_{i^{\prime
}=0}^{n^{\prime }-1}A_{n^{\prime }i^{\prime }}^{\left( m\right) }E_{\mathbf{p%
}^{\prime }}^{i^{\prime }}\right) p^{\left\langle \mu _{1}\right. }\cdots
p^{\left. \mu _{m}\right\rangle }p_{\left\langle \mu _{1}\right. }^{\prime
}\cdots p_{\left. \mu _{m}\right\rangle }^{\prime }\right.  \notag \\
& \left. -\left( \sum_{i=n}^{N_{m}}A_{in}^{\left( m\right) }E_{\mathbf{k}%
}^{i}+\sum_{i=0}^{n-1}A_{ni}^{\left( m\right) }E_{\mathbf{k}}^{i}\right)
\left( \sum_{i^{\prime }=n^{\prime }}^{N_{m}}A_{i^{\prime }n^{\prime
}}^{\left( m\right) }E_{\mathbf{k}^{\prime }}^{i^{\prime }}+\sum_{i^{\prime
}=0}^{n^{\prime }-1}A_{n^{\prime }i^{\prime }}^{\left( m\right) }E_{\mathbf{k%
}^{\prime }}^{i^{\prime }}\right) k^{\left\langle \mu _{1}\right. }\cdots
k^{\left. \mu _{m}\right\rangle }k_{\left\langle \mu _{1}\right. }^{\prime
}\cdots k_{\left. \mu _{m}\right\rangle }^{\prime }\right] .
\end{align}%
As noted before, in the 14--moment approximation terms proportional to $%
\sum_{i=0}^{n-1}A_{ni}^{\left( m\right) }$ vanish, hence Eq.\ (\ref%
{Nr_scalar}) leads to%
\begin{eqnarray}
N_{r-1} &=& \frac{9}{m^{4}}\mathcal{C}_{r00}^{0\left( 00\right) }\Pi ^{2}+%
\mathcal{C}_{r00}^{0\left( 11\right) }n^{\mu }n_{\mu }+\mathcal{C}%
_{r00}^{0\left( 22\right) }\pi ^{\mu \nu }\pi _{\mu \nu },
\end{eqnarray}%
where we used Eqs.\ (\ref{rho_r_recursive}) -- (\ref{rho_mu_nu_r_recursive})
for the $\rho _{n}^{\alpha _{1}\cdots \alpha _{m}}$'s. The coefficients in
the above equation are 
\begin{align}
\mathcal{C}_{r00}^{0\left( 00\right) }& =\left( A_{10}^{\left( 0\right)
}\right) ^{2}Y_{2\left( r-3\right) ,1}+\left( A_{20}^{\left( 0\right)
}\right) ^{2}Y_{5\left( r-3\right) ,1}+\left( A_{00}^{\left( 0\right)
}A_{20}^{\left( 0\right) }\right) X_{\left( r-3\right) ,1}+\left(
A_{10}^{\left( 0\right) }A_{20}^{\left( 0\right) }\right) Y_{3\left(
r-3\right) ,1},  \label{B00} \\
\mathcal{C}_{r00}^{0\left( 11\right) }& =-\frac{1}{3}\left( A_{00}^{\left(
1\right) }\right) ^{2}Y_{2\left( r-3\right) ,1}+\left( A_{00}^{\left(
1\right) }A_{10}^{\left( 1\right) }\right) Y_{3\left( r-3\right) ,5}+\left(
A_{10}^{\left( 1\right) }\right) ^{2}Y_{5\left( r-3\right) ,5},  \label{B11}
\\
\mathcal{C}_{r00}^{0\left( 22\right) }& =2\left( A_{00}^{\left( 2\right)
}\right) ^{2}Y_{5\left( r-3\right) ,9},  \label{B22}
\end{align}%
where the detailed derivation is given in Appendix \ref{2nd_order_coll_int}.
Comparing the above result to Eq.\ (\ref{R_scalar}) and taking
into account the corresponding relaxation time from Eq.\ (\ref{Final1}), we
obtain 
\begin{align}
\varphi_{1}& =\frac{9}{m^{4}}\,\frac{\tau _{\Pi }^{r}}{\gamma _{r}^{\Pi }}\,%
\mathcal{C}_{r00}^{0\left( 00\right) }, \\
\varphi _{2}& =\frac{\tau _{\Pi }^{r}}{\gamma _{r}^{\Pi }}\,\mathcal{C}%
_{r00}^{0\left( 11\right) }, \\
\varphi _{3}& =\frac{\tau _{\Pi }^{r}}{\gamma _{r}^{\Pi }}\,\mathcal{C}%
_{r00}^{0\left( 22\right) }.
\end{align}

Similarly, the vector term (\ref{Nr_vector}) in the 14--moment approximation
leads to the following formula%
\begin{eqnarray}
N_{r-1}^{\mu } &=& \mathcal{C}_{r00}^{1\left( 12\right) }n^{\nu }\pi _{\nu
}^{\mu }-\frac{3}{m^{2}}\mathcal{C}_{r00}^{1\left( 01\right) }\Pi n^{\mu },
\end{eqnarray}%
where%
\begin{align}
\mathcal{C}_{r00}^{1\left( 12\right) }& =2\left[ \left( A_{00}^{\left(
1\right) }A_{00}^{\left( 2\right) }\right) Y_{3\left( r-2\right) ,8}+\left(
A_{10}^{\left( 1\right) }A_{00}^{\left( 2\right) }\right) Y_{4\left(
r-2\right) ,8}\right] , \\
\mathcal{C}_{r00}^{1\left( 01\right) }& =\left( A_{00}^{\left( 0\right)
}A_{10}^{\left( 1\right) }\right) X_{\left( r-2\right) ,3}+\left(
A_{10}^{\left( 0\right) }A_{00}^{\left( 1\right) }\right) Y_{1\left(
r-2\right) ,3}
+ \left( A_{10}^{\left( 0\right) }A_{10}^{\left( 1\right) }\right)Y_{3\left( r-2\right) ,3}  
\notag \\
& - \left( A_{20}^{\left( 0\right) }A_{00}^{\left( 1\right)
}\right) \left(
3Y_{3\left( r-2\right) ,6}+2Y_{3\left( r-2\right) ,8}\right) +\left(
A_{20}^{\left( 0\right) }A_{10}^{\left( 1\right) }\right) Y_{4\left(
r-2\right) ,3}.  \label{C101}
\end{align}%
Now, recalling Eq.\ (\ref{R_vector}) we obtain 
\begin{align}
\varphi _{4}& =\frac{\tau _{n}^{r}}{\gamma _{r}^{n}}\,\mathcal{C}%
_{r00}^{1\left( 12\right) }, \\
\varphi _{5}& =-\frac{3}{m^{2}}\frac{\tau _{n}^{r}}{\gamma _{r}^{n}}\,%
\mathcal{C}_{r00}^{1\left( 01\right) }.
\end{align}

Finally, the tensor term (\ref{Nr_tensor}) is 
\begin{eqnarray}
N_{r-1}^{\mu \nu } &=&-\frac{3}{m^{2}}\mathcal{C}_{r00}^{2\left( 0,2\right)
}\Pi \pi ^{\mu \nu }+\mathcal{D}_{r00}^{2\left( 2,2\right) }\pi ^{\lambda
\left\langle \mu \right. }\pi _{\lambda }^{\left. \nu \right\rangle }+%
\mathcal{D}_{r00}^{2\left( 1,1\right) }n^{\left\langle \mu \right.
}n^{\left. \nu \right\rangle },
\end{eqnarray}%
where 
\begin{align}
\mathcal{C}_{r00}^{2\left( 02\right) }& =A_{00}^{\left( 2\right) }\left[
A_{00}^{\left( 0\right) }X_{\left( r-1\right) ,4}+A_{10}^{\left( 0\right)
}Y_{3\left( r-1\right) ,4}+A_{20}^{\left( 0\right) }Y_{4\left( r-1\right) ,4}%
\right] , \\
\mathcal{D}_{r00}^{2\left( 22\right) }& =8\left( A_{00}^{\left( 2\right)
}\right) ^{2}Y_{5\left( r-1\right) ,11}, \\
\mathcal{D}_{r00}^{2\left( 11\right) }& =\left( A_{00}^{\left( 1\right)
}\right) ^{2}Y_{2\left( r-1\right) ,4}+2\left( A_{00}^{\left( 1\right)
}A_{10}^{\left( 1\right) }\right) Y_{3\left( r-1\right) ,7}+2\left(
A_{10}^{\left( 1\right) }\right) ^{2}Y_{5\left( r-1\right) ,7},
\end{align}%
and a comparison with Eq.\ (\ref{R_tensor}) yields 
\begin{align}
\varphi _{6}& =-\frac{3}{m^{2}}\frac{\tau _{\pi }^{r}}{\gamma _{r}^{\pi }}%
\mathcal{C}_{r00}^{2\left( 0,2\right) }, \\
\varphi _{7}& =\frac{\tau _{\pi }^{r}}{\gamma _{r}^{\pi }}\mathcal{D}%
_{r00}^{2\left( 2,2\right) }, \\
\varphi _{8}& =\frac{\tau _{\pi }^{r}}{\gamma _{r}^{\pi }}\mathcal{D}%
_{r00}^{2\left( 1,1\right) }.
\end{align}

In order to calculate these coefficients we have introduced five new
tensors, which are similar to $X_{\left( r\right) }^{\mu \nu \alpha \beta }$
and are given as follows, 
\begin{align}
Y_{1\left( r\right) }^{\mu \nu \alpha \beta }& =\frac{1}{\nu }\int_{f}E_{%
\mathbf{k}}^{r}k^{\mu }k^{\nu }\left( p^{\alpha }p^{\prime \beta }+p^{\prime
\alpha }p^{\beta }-k^{\alpha }k^{\prime \beta }-k^{\prime \alpha }k^{\beta
}\right) ,  \label{Y1_4rank_tens} \\
Y_{2\left( r\right) }^{\mu \nu \alpha \beta }& =\frac{1}{\nu }\int_{f}E_{%
\mathbf{k}}^{r}k^{\mu }k^{\nu }\left[ p^{\alpha }p^{\prime \beta }-k^{\alpha
}k^{\prime \beta }\right] ,  \label{Y2_4rank_tens} \\
Y_{3\left( r\right) }^{\mu \nu \alpha \beta \kappa }& =\frac{1}{\nu }%
\int_{f}E_{\mathbf{k}}^{r}k^{\mu }k^{\nu }\left( p^{\alpha }p^{\prime \beta
}p^{\prime \kappa }+p^{\prime \alpha }p^{\beta }p^{\kappa }-k^{\alpha
}k^{\prime \beta }k^{\prime \kappa }-k^{\prime \alpha }k^{\beta }k^{\kappa
}\right) ,  \label{Y3_5rank_tens} \\
Y_{4\left( r\right) }^{\mu \nu \alpha \beta \kappa \lambda }& =\frac{1}{\nu }%
\int_{f}E_{\mathbf{k}}^{r}k^{\mu }k^{\nu }\left( p^{\alpha }p^{\beta
}p^{\prime \kappa }p^{\prime \lambda }+p^{\prime \alpha }p^{\prime \beta
}p^{\kappa }p^{\lambda }-k^{\alpha }k^{\beta }k^{\prime \kappa }k^{\prime
\lambda }-k^{\prime \alpha }k^{\prime \beta }k^{\kappa }k^{\lambda }\right) ,
\label{Y4_6rank_tens} \\
Y_{5\left( r\right) }^{\mu \nu \alpha \beta \kappa \lambda }& =\frac{1}{\nu }%
\int_{f}E_{\mathbf{k}}^{r}k^{\mu }k^{\nu }\left( p^{\alpha }p^{\beta
}p^{\prime \kappa }p^{\prime \lambda }-k^{\alpha }k^{\beta }k^{\prime \kappa
}k^{\prime \lambda }\right) .  \label{Y5_6rank_tens}
\end{align}%
Note that the $Y_{i\left( r\right) ,j}$ terms in the previous equations are
different contractions of these five tensors. Our notation is such that the $%
i$-index specifies the tensor while the $j$-index labels a particular
contraction. More details are given in Appendix \ref{tens_decompositions}.

We have shown earlier that the coefficients in the equations of motion
depend explicitly on the choice of the moment, i.e., the index $r$.
Therefore, once the 14--moment approximation is enforced any moment of the
Boltzmann equation leads to a closed set of equations, but when calculating
the coefficients of the nonlinear collision integrals one has to account for
the exact form of the relaxation equations which follow from Eqs.\ (\ref%
{rho_dot}) -- (\ref{rho_mu_nu_dot}) using Eqs.\ (\ref{rho_r_recursive}) -- (\ref%
{rho_mu_nu_r_recursive}). As an example we quote the equations for the particle
diffusion current and shear-stress tensor for arbitrary $r$,%
\begin{eqnarray}
\tau _{n}^{r}\dot{n}^{\left\langle \mu \right\rangle }+n^{\mu }+\frac{3}{%
m^{2}}\frac{\tau _{n}^{r}\mathcal{C}_{r00}^{1\left( 01\right) }}{\gamma
_{r}^{n}}\Pi n^{\mu }-\frac{\tau _{n}^{r}\mathcal{C}_{r00}^{1\left(
12\right) }}{\gamma _{r}^{n}}n^{\nu }\pi _{\nu }^{\mu } &=&\kappa ^{r}\nabla
^{\mu }\alpha _{0}+\ldots . \\
\tau _{\pi }^{r}\pi ^{\left\langle \mu \nu \right\rangle }+\pi ^{\mu \nu }+%
\frac{3}{m^{2}}\frac{\tau _{\pi }^{r}\mathcal{C}_{r00}^{2\left( 0,2\right) }%
}{\gamma _{r}^{\pi }}\Pi \pi ^{\mu \nu }-\frac{\tau _{\pi }^{r}\mathcal{D}%
_{r00}^{2\left( 2,2\right) }}{\gamma _{r}^{\pi }}\pi ^{\lambda \left\langle
\mu \right. }\pi _{\lambda }^{\left. \nu \right\rangle }-\frac{\tau _{\pi
}^{r}\mathcal{D}_{r00}^{2\left( 1,1\right) }}{\gamma _{r}^{\pi }}%
n^{\left\langle \mu \right. }n^{\left. \nu \right\rangle } &=&2\eta
^{r}\sigma ^{\mu \nu }+\cdots
\end{eqnarray}%
Note that in order to recover Eqs.\ (\ref{Final1}) -- (\ref{Final3}) one must
take $r=0$, i.e., $\zeta =\zeta ^{0}$, $\kappa =\kappa ^{0}$, $\eta =\eta
^{0}$, $\tau _{\Pi }=\tau _{\Pi }^{0}$, $\tau _{n}=\tau _{n}^{0}$ and $\tau
_{\pi }=\tau _{\pi }^{0}$.

Using the above relations it was already shown in Ref.\ \cite{Denicol:2012es}
how to derive the equations of motion and calculate the transport
coefficients for different choices of the moments corresponding to the
traditional method by Israel and Stewart \cite{Israel:1979wp} and to the one
proposed by Denicol, Koide, and Rischke (DKR) \cite{Denicol:2010xn}. Here,
we shall follow this recipe and calculate the coefficients of the nonlinear
collision integral in both cases.

The equations of motion derived by Israel and Stewart \cite{Israel:1979wp}
can be obtained by the choice $\rho _{3}=-\frac{3}{m^{2}}\gamma _{3}^{\Pi
}\Pi $, $\rho _{2}^{\mu }=\gamma _{2}^{n}n^{\mu }$, $\rho _{1}=\gamma
_{1}^{\pi }\pi ^{\mu \nu }$ in Eqs.\ (\ref{rho_r_recursive}) -- (\ref%
{rho_mu_nu_r_recursive}) and substituting these values into the equations of
motion (\ref{rho_dot}) -- (\ref{rho_mu_nu_dot}). Therefore, in the IS theory
all coefficients need to be calculated with $r=3$ for the scalar moments, $%
r=2$ for vector moments, and $r=1$ for second-rank tensor moments. In
contrast, the choice of DKR is to use $r=0$ for all equations and
coefficients.

We explicitly compute some of these coefficients in the ultrarelativistic
limit, $m\beta _{0}\rightarrow 0$, for a classical gas ($a=0$) with fixed
cross section. Since in this limit $\Pi =0$, we do not need to compute the
coefficients $\varphi _{1},\varphi _{2}$, and $\varphi _{3}$ appearing in
the term $\mathcal{R}$, Eq.\ (\ref{R_scalar}), which enters the equation of
motion (\ref{Final1}) for the bulk viscous pressure. Furthermore, $\varphi
_{5},\varphi _{6}$ are coefficients in terms which are proportional to $\Pi $%
, and thus also need not be computed. The remaining, non-vanishing
coefficients corresponding to the DKR choice $r=0$ are simply denoted as $%
\varphi _{4}$, $\varphi _{7}$, and $\varphi _{8}$, while the ones
corresponding to the IS choice are denoted by $\varphi _{4}^{IS}$ for $r=2$,
while $\varphi _{7}^{IS}$ and $\varphi _{8}^{IS}$ for $r=1$. They read 
\begin{align}
\varphi _{4}& =\frac{1}{25}P_{0}^{-1},\  & \varphi _{4}^{IS}& =\frac{1}{4}%
P_{0}^{-1}, \\
\varphi _{7}& =\frac{9}{70}P_{0}^{-1},\  & \varphi _{7}^{IS}& =\frac{1}{5}%
P_{0}^{-1}, \\
\varphi _{8}& =\frac{8}{5\beta _{0}^{2}}P_{0}^{-1},\ & \varphi _{8}^{IS}& =-%
\frac{4}{5\beta _{0}^{2}}P_{0}^{-1}.
\end{align}%
We observe that only $\varphi _{8}$ (multiplying $n^{\langle
\mu }n^{\nu \rangle }$) differs in sign between the DKR and IS choices, with
the absolute magnitude of the latter being half as large. The coefficients
$\varphi _{7}$ (multiplying $\pi ^{\lambda \left\langle \mu
\right. }\pi _{\lambda }^{\left. \nu \right\rangle }$) are
approximately of the same magnitude for both choices, while $\varphi _{4}$ (multiplying $%
n_{\nu }\pi ^{\mu \nu }$) is more than six times smaller in DKR than in
IS theory. The implications of these results have already been discussed in
the Introduction and Conclusions for the DKR choice.

Finally, for further reference, we also quote the coefficients of particle
diffusion and shear viscosity, 
\begin{align}
\kappa & =\frac{3}{16}n_{0}\lambda _{\mathrm{mfp}},\  & \kappa ^{IS}& =\frac{%
1}{8}n_{0}\lambda _{\mathrm{mfp}}, \\
\eta & =\frac{4}{3}P_{0}\lambda _{\mathrm{mfp}},\  & \eta ^{IS}& =\frac{6}{5}%
P_{0}\lambda _{\mathrm{mfp}},
\end{align}%
where $\lambda _{\mathrm{mfp}}=1/(n_{0}\sigma _{T})$ is the mean-free path
and $\sigma _{T}$ is the total cross section. Note that all remaining
transport coefficients from Eq.\ (\ref{14_moment_terms}) were already
computed in Ref.\ \cite{Denicol:2012es} for both the DKR and IS choices and it
was shown that the differences are of the order of $%
10-30\%$. Therefore, the only coefficients that change considerably from one
formalism to the other are $\varphi _{4}$ and $\varphi _{8}$.

In closing we remark that the numerical solutions of both the IS and DKR
theories were compared to the numerical solutions of the Boltzmann equation
in various cases \cite%
{Huovinen:2008te,Denicol:2010xn,Denicol:2012vq,Bouras:2009vs,Bouras:2010hm,Florkowski:2013lya}. 
These investigations showed the advantages of the DKR choice for the
corresponding coefficients, which leads to a far better agreement with
numerical solutions of the Boltzmann equation than the IS theory.


\begin{acknowledgments}
This work was supported by the Helmholtz International Center for FAIR
within the framework of the LOEWE program launched by the State of Hesse. 
G.S.\ Denicol is supported by a Banting Fellowship of the Natural Sciences and Engineering Research
Council of Canada. The work of H.\ Niemi was supported by Academy of Finland,
Project No. 133005.
E.\ Moln\'ar was partially supported by the European Union and the European Social Fund 
through project Supercomputer, the national virtual lab (grant no.: TAMOP-4.2.2.C-11/1/KONV-2012-0010),
as well as by TAMOP 4.2.4. A/2-11-1-2012-0001 National Excellence Program (A2-MZPD\"O-13-0042).

The authors thank R.\ Paatelainen for his help with FeynCalc.
\end{acknowledgments}

\appendix

\section{Thermodynamic integrals}

\label{thermo_integrals}

Following Refs.\ \cite{Israel:1979wp,Muronga:2006zw} we introduced the
following equilibrium moments of tensor rank $n$,%
\begin{equation}
I_{n}^{\mu _{1}\cdots \mu _{n}}\equiv \left\langle k^{\mu _{1}}\cdots k^{\mu
_{n}}\right\rangle _{0}=\sum_{q=0}^{\left[ n/2\right] }\left( -1\right)
^{q}b_{nq}I_{nq}\Delta ^{\left( \mu _{1}\mu _{2}\right. }\cdots \Delta ^{\mu
_{2q-1}\mu _{2q}}u^{\mu _{2q+1}}\cdots u^{\left. \mu _{n}\right) },
\label{I_n_moment}
\end{equation}%
where $n$, $q$ are natural numbers and $[n/2]$ is the largest integer not
exceeding $n/2$, cf.\ Eq.\ (A8) in Ref.\ \cite{Israel:1979wp}. The
parentheses $(\ldots )$ around indices denote symmetrization. For an
arbitrary tensor of rank $n$, this operation is defined by $A^{\left( \mu
_{1}\cdots \mu _{n}\right) }=\frac{1}{n!}\sum_{\wp _{\mu }}A^{\mu _{1}\mu
_{2}\cdots \mu _{n}}$, where $\wp _{\mu }$ denotes all possible permutations
of the $\mu $-indices.

The $b_{nq}$ coefficient is equal to the number of permutations in the set $%
\wp_\mu$, which lead to identical tensor products of the $u^{\mu }$ and $%
\Delta ^{\mu \nu }$ projectors \cite{Israel:1979wp}, 
\begin{equation}
b_{nq}\equiv \frac{n!}{2^{q}q!\left( n-2q\right) !}=\frac{n!\left(
2q-1\right) !!}{\left( 2q\right) !\left( n-2q\right) !},  \label{a_nq}
\end{equation}%
see Eq.\ (A2) of Ref.\ \cite{Israel:1979wp}.

The thermodynamic integrals $I_{nq}$ and $J_{nq}$ were defined in Eqs.\ (\ref%
{I_nq}), (\ref{J_nq}):%
\begin{eqnarray}
I_{nq} &=&\frac{\left( -1\right) ^{q}}{\left( 2q+1\right) !!}\int dK\left(
E_{\mathbf{k}}\right) ^{n-2q}\left( \Delta ^{\alpha \beta }k_{\alpha
}k_{\beta }\right) ^{q}f_{0\mathbf{k}},  \label{I_nq_appendix} \\
J_{nq} &=&\frac{\left( -1\right) ^{q}}{\left( 2q+1\right) !!}\int dK\left(
E_{\mathbf{k}}\right) ^{n-2q}\left( \Delta ^{\alpha \beta }k_{\alpha
}k_{\beta }\right) ^{q}f_{0\mathbf{k}}\tilde{f}_{0\mathbf{k}}.
\end{eqnarray}%
Replacing $\left( \Delta ^{\alpha \beta }k_{\alpha }k_{\beta }\right)
^{q}=\left( m^{2}-E_{\mathbf{k}}^{2}\right) ^{q}$ we get the following
recursion relations for $0\leq q\leq n/2$,%
\begin{eqnarray}
I_{n+2,q} &=&m^{2}I_{n,q}+\left( 2q+3\right) I_{n+2,q+1},
\label{I_nq_recursive} \\
J_{n+2,q} &=&m^{2}J_{n,q}+\left( 2q+3\right) J_{n+2,q+1},
\end{eqnarray}%
while an integration by parts of Eq.\ (\ref{I_nq_appendix}) leads to the
following relation, 
\begin{equation}
\beta _{0}J_{nq}=I_{n-1,q-1}+\left( n-2q\right) I_{n-1,q}.
\end{equation}%
Furthermore, 
\begin{eqnarray}
dI_{nq}(\alpha _{0},\beta _{0}) &\equiv &\frac{\partial I_{nq}}{\partial
\alpha _{0}}\,d\alpha _{0}+\frac{\partial I_{nq}}{\partial \beta _{0}}%
\,d\beta _{0},  \notag \\
&=&J_{nq}\,d\alpha _{0}-J_{n+1,q}\,d\beta _{0},
\end{eqnarray}%
with a similar relation for $dJ_{nq}(\alpha _{0},\beta _{0})$.

\section{Irreducible tensors}

\label{irreducible_tensors}

We define the following projection operator \cite{deGroot} 
\begin{equation}
\Delta ^{\mu _{1}\cdots \mu _{n}\,\nu _{1}\cdots \nu _{n}}=\sum_{q=0}^{\left[
n/2\right] }c_{nq}\Phi _{(nq)}^{\mu _{1}\cdots \mu _{n}\nu _{1}\cdots \nu
_{n}},  \label{general_projection_tensor}
\end{equation}%
where the coefficients are given by 
\begin{equation}
c_{nq}=\left( -1\right) ^{q}\frac{\left( n!\right) ^{2}}{\left( 2n\right) !}%
\frac{\left( 2n-2q\right) !}{q!\left( n-q\right) !\left( n-2q\right) !},
\end{equation}%
and 
\begin{equation}
\Phi _{(nq)}^{\mu _{1}\cdots \mu _{n}\,\nu _{1}\cdots \nu _{n}}=\left(
n-2q\right) !\left( \frac{2^{q}q!}{n!}\right) ^{2}\sum_{\wp _{\mu }\wp _{\nu
}}\Delta ^{\mu _{1}\mu _{2}}\cdots \Delta ^{\mu _{2q-1}\mu _{2q}}\Delta
^{\nu _{1}\nu _{2}}\cdots \Delta ^{\nu _{2q-1}\nu _{2q}}\Delta ^{\mu
_{2q+1}\nu _{2q+1}}\cdots \Delta ^{\mu _{n}\nu _{n}}.
\end{equation}%
The summation is taken over all \emph{distinct\/} permutations of $\mu $-
and $\nu $-type indices (without mutual exchange of these types of indices).
The prefactor is the inverse of the number of distinct permutations. This
can be seen as follows: the \emph{total\/} number of permutations of $\mu $-
and $\nu $-type indices is $\left( n!\right) ^{2}$. In order to obtain the
number of \emph{distinct\/} permutations, we have to divide this by the
following three numbers: $(2^{q})^{2}$ permutations lead to terms which only
differ by a trivial permutation of indices on the same projector, e.g.\ $%
\Delta ^{\mu _{1}\mu _{2}}=\Delta ^{\mu _{2}\mu _{1}}$; $(q!)^{2}$ terms
just correspond to a pairwise exchange of indices between projectors, e.g.\ $%
\Delta ^{\mu _{1}\mu _{2}}\Delta ^{\mu _{3}\mu _{4}}=\Delta ^{\mu _{3}\mu
_{4}}\Delta ^{\mu _{1}\mu _{2}}$; for any given distribution of $\mu $-type
indices in the product $\Delta ^{\mu _{2q+1}\nu _{2q+1}}\cdots \Delta ^{\mu
_{n}\nu _{n}}$, there are $\left( n-2q\right) !$ possible ways to distribute
the $\nu $-type indices, which lead to the same product of projectors.

The projection operator has the following properties [for details, see Ref.\ 
\cite{deGroot}]:

\begin{enumerate}
\item[(i)] It is separately symmetric upon interchange of $\mu$- or $\nu$%
-type indices, 
\begin{equation}
\Delta ^{\mu _{1}\cdots \mu _{n}\,\nu _{1}\cdots \nu _{n}}=\Delta ^{\left(
\mu _{1}\cdots \mu _{n}\right) \left( \nu _{1}\cdots \nu _{n}\right) }.
\end{equation}

\item[(ii)] It is traceless upon contraction of $\mu$- or $\nu$-type
indices, 
\begin{equation}
\Delta ^{\mu _{1}\cdots \mu _{n}\nu _{1}\cdots \nu _{n}}g_{\mu _{i}\mu _{j}}
= \Delta ^{\mu _{1}\cdots \mu _{n}\nu _{1}\cdots \nu _{n}}g_{\nu _{i}\nu
_{j}}=0,  \label{tracelessness}
\end{equation}
for any pair of indices $\mu_i,\mu_j $ or $\nu_i,\nu_j $, where $1\leq
i,j\leq n$.

\item[(iii)] The complete contraction is 
\begin{equation}
\Delta _{\mu _{1}\cdots \mu _{\ell }}^{\mu _{1}\cdots \mu _{\ell }}= 2\ell
+1\; .  \label{total_contraction}
\end{equation}
\end{enumerate}

The irreducible tensors $k^{\left\langle \mu _{1}\right. }\cdots k^{\left.
\mu _{\ell }\right\rangle }$ defined in Eq.\ (\ref{k_irreducible}) are 
\begin{equation}
k^{\left\langle \mu _{1}\right. }\cdots k^{\left. \mu _{\ell }\right\rangle
}=\Delta _{\nu _{1}\cdots \nu _{\ell }}^{\mu _{1}\cdots \mu _{\ell }}k^{\nu
_{1}}\cdots k^{\nu _{\ell }},
\end{equation}%
where $\Delta _{\nu _{1}\cdots \nu _{\ell }}^{\mu
_{1}\cdots \mu _{\ell }}\equiv \Delta ^{\mu _{1}\cdots \mu _{\ell }\alpha
_{1}\cdots \alpha _{\ell }}g_{\alpha _{1}\nu _{1}}\cdots g_{\alpha _{\ell
}\nu _{\ell }}$. Furthermore, the tensors $k^{\left\langle \mu _{1}\right.
}\cdots k^{\left. \mu _{m}\right\rangle }$ satisfy the following
orthogonality condition, 
\begin{equation}
\int dKF_{\mathbf{k}}k^{\left\langle \mu _{1}\right. }\cdots k^{\left. \mu
_{m}\right\rangle }k^{\left\langle \nu _{1}\right. }\cdots k^{\left. \nu
_{n}\right\rangle } =\frac{m!\, \delta _{mn}}{\left( 2m+1\right) !!}\Delta
^{\mu _{1}\cdots \mu _{m}\nu _{1}\cdots \nu _{m}}\int dKF_{\mathbf{k}}\left(
\Delta ^{\alpha \beta }k_{\alpha }k_{\beta }\right) ^{m},
\label{orthogonality1}
\end{equation}%
where $F_{\mathbf{k}}$ is an arbitrary scalar function of $E_{\mathbf{k}}$.

Let us explicitly write down the projection operators (\ref%
{general_projection_tensor}) which are needed for our calculations. The
first one follows from Eq.\ (\ref{general_projection_tensor}) for $n=1$,
which defines the elementary projection operator, $\Delta ^{\mu _{1}\nu
_{1}} $, and hence for any 4-vector we have 
\begin{equation}
A^{\left\langle \mu _{1}\right\rangle }=\Delta ^{\mu _{1}\nu _{1}}A_{\nu
_{1}}.
\end{equation}%
The next one is given for $n=2$, which defines the symmetric, traceless, and
orthogonal projection in case of arbitrary second-rank tensors, 
\begin{equation}
\Delta ^{\mu _{1}\mu _{2}\nu _{1}\nu _{2}}=\Delta ^{\mu _{1}\left( \nu
_{1}\right. }\Delta ^{\left. \nu _{2}\right) \mu _{2}}-\frac{1}{3}\Delta
^{\mu _{1}\mu _{2}}\Delta ^{\nu _{1}\nu _{2}},
\end{equation}%
hence for any second-rank tensor formed from the dyadic product of two
4-vectors, $A^{\mu _{1}}$ and $A^{\mu _{2}}$, we obtain 
\begin{equation}
A^{\left\langle \mu _{1}\right. }A^{\left. \mu _{2}\right\rangle
}=A^{\left\langle \mu _{1}\right\rangle }A^{\left\langle \mu
_{2}\right\rangle }-\frac{1}{3}\Delta ^{\mu _{1}\mu _{2}}\left( \Delta
^{\alpha \beta }A_{\alpha }A_{\beta }\right) .
\end{equation}%
The case $n=3$ leads to%
\begin{align}
\Delta ^{\mu _{1}\mu _{2}\mu _{3}\nu _{1}\nu _{2}\nu _{3}}& =\frac{1}{3}%
\left( \Delta ^{\mu _{1}\nu _{1}}\Delta ^{\mu _{2}\left( \nu _{2}\right.
}\Delta ^{\left. \nu _{3}\right) \mu _{3}}+\Delta ^{\mu _{1}\nu _{2}}\Delta
^{\mu _{2}\left( \nu _{1}\right. }\Delta ^{\left. \nu _{3}\right) \mu
_{3}}+\Delta ^{\mu _{1}\nu _{3}}\Delta ^{\mu _{2}\left( \nu _{2}\right.
}\Delta ^{\left. \nu _{1}\right) \mu _{3}}\right)  \notag \\
& -\frac{3}{5}\Delta ^{\left( \mu _{1}\mu _{2}\right. }\Delta ^{\left. \mu
_{3}\right) \left( \nu _{3}\right. }\Delta ^{\left. \nu _{1}\nu _{2}\right)
},
\end{align}%
and so for any rank-3 tensor formed from the dyadic product of the 4-vectors 
$A^{\mu _{1}}$, $A^{\mu _{2}}$, and $A^{\mu _{3}}$ we obtain 
\begin{equation}
A^{\left\langle \mu _{1}\right. }A^{\mu _{2}}A^{\left. \mu _{3}\right\rangle
}=A^{\left\langle \mu _{1}\right\rangle }A^{\left\langle \mu
_{2}\right\rangle }A^{\left\langle \mu _{3}\right\rangle }-\frac{1}{5}\left(
\Delta ^{\mu _{1}\mu _{2}}A^{\left\langle \mu _{3}\right\rangle }+\Delta
^{\mu _{1}\mu _{3}}A^{\left\langle \mu _{2}\right\rangle }+\Delta ^{\mu
_{2}\mu _{3}}A^{\left\langle \mu _{1}\right\rangle }\right) \left( \Delta
^{\alpha \beta }A_{\alpha }A_{\beta }\right) .
\end{equation}%
Finally, for $n=4$ Eq.\ (\ref{general_projection_tensor}) leads to%
\begin{align}
\Delta ^{\mu _{1}\mu _{2}\mu _{3}\mu _{4}\nu _{1}\nu _{2}\nu _{3}\nu _{4}}& =%
\frac{1}{4!}\sum_{\wp _{\mu }\wp _{\nu }}\Delta ^{\mu _{1}\nu _{1}}\Delta
^{\mu _{2}\nu _{2}}\Delta ^{\mu _{3}\nu _{3}}\Delta ^{\mu _{4}\nu _{4}}-%
\frac{3}{14}\Delta ^{\left( \mu _{1}\mu _{2}\right. }\Delta ^{\left. \mu
_{3}\right) \left( \nu _{3}\right. }\Delta ^{\nu _{1}\nu _{2}}\Delta
^{\left. \nu _{4}\right) \mu _{4}}  \notag \\
& -\frac{3}{14}\Delta ^{\left( \mu _{1}\mu _{2}\right. }\Delta ^{\left. \mu
_{4}\right) \left( \nu _{3}\right. }\Delta ^{\nu _{1}\nu _{2}}\Delta
^{\left. \nu _{4}\right) \mu _{3}}-\frac{3}{14}\Delta ^{\left( \mu _{1}\mu
_{3}\right. }\Delta ^{\left. \mu _{4}\right) \left( \nu _{3}\right. }\Delta
^{\nu _{1}\nu _{2}}\Delta ^{\left. \nu _{4}\right) \mu _{2}}  \notag \\
& -\frac{3}{14}\Delta ^{\left( \mu _{2}\mu _{3}\right. }\Delta ^{\left. \mu
_{4}\right) \left( \nu _{3}\right. }\Delta ^{\nu _{1}\nu _{2}}\Delta
^{\left. \nu _{4}\right) \mu _{1}}+\frac{3}{35}\Delta ^{\left( \mu _{1}\mu
_{2}\right. }\Delta ^{\left. \mu _{3}\mu _{4}\right) }\Delta ^{\left( \nu
_{1}\nu _{2}\right. }\Delta ^{\left. \nu _{3}\nu _{4}\right) },
\end{align}%
and 
\begin{align}
A^{\left\langle \mu _{1}\right. }A^{\mu _{2}}A^{\mu _{3}}A^{\left. \mu
_{4}\right\rangle }& =A^{\left\langle \mu _{1}\right\rangle }A^{\left\langle
\mu _{2}\right\rangle }A^{\left\langle \mu _{3}\right\rangle
}A^{\left\langle \mu _{4}\right\rangle }-\frac{3}{14}\Delta ^{\left( \mu
_{1}\mu _{2}\right. }A^{\left. \left\langle \mu _{3}\right\rangle \right)
}A^{\left\langle \mu _{4}\right\rangle }\left( \Delta ^{\alpha \beta
}A_{\alpha }A_{\beta }\right)  \notag \\
& -\frac{3}{14}\Delta ^{\left( \mu _{1}\mu _{2}\right. }A^{\left.
\left\langle \mu _{4}\right\rangle \right) }A^{\left\langle \mu
_{3}\right\rangle }\left( \Delta ^{\alpha \beta }A_{\alpha }A_{\beta
}\right) -\frac{3}{14}\Delta ^{\left( \mu _{1}\mu _{4}\right. }A^{\left.
\left\langle \mu _{3}\right\rangle \right) }A^{\left\langle \mu
_{2}\right\rangle }\left( \Delta ^{\alpha \beta }A_{\alpha }A_{\beta }\right)
\notag \\
& -\frac{3}{14}\Delta ^{\left( \mu _{4}\mu _{2}\right. }A^{\left.
\left\langle \mu _{3}\right\rangle \right) }A^{\left\langle \mu
_{1}\right\rangle }\left( \Delta ^{\alpha \beta }A_{\alpha }A_{\beta
}\right) +\frac{3}{35}\Delta ^{\left( \mu _{1}\mu _{2}\right. }\Delta
^{\left. \mu _{3}\mu _{4}\right) }\left( \Delta ^{\alpha \beta }A_{\alpha
}A_{\beta }\right) ^{2}.
\end{align}

Note that we also use the notation with mixed indices such as,%
\begin{eqnarray}
A^{\left\langle \mu _{1}\right. }A^{\left. \mu _{2}\right\rangle }\Delta
_{\mu _{2}\nu _{1}} &=&A^{\left\langle \mu _{1}\right. }A_{\left. \nu
_{1}\right\rangle }, \\
A^{\left\langle \mu _{1}\right. }A^{\mu _{2}}A^{\mu _{3}}A^{\left. \mu
_{4}\right\rangle }\Delta _{\mu _{3}\mu _{4}\nu _{1}\nu _{2}}
&=&A^{\left\langle \mu _{1}\right. }A^{\mu _{2}}A_{\nu _{1}}A_{\left. \nu
_{2}\right\rangle }.
\end{eqnarray}

\section{Reduction of collision tensors}

\label{collision_tensors}

In this Appendix, we show how to derive the general structure of the
collision integrals introduced in the main text. As already discussed, the
tensor structure of $\left( \mathcal{N}_{rnn^{\prime }}\right) _{\alpha
_{1}\cdots \alpha _{m}\beta _{1}\cdots \beta _{m^{\prime }}}^{\mu _{1}\cdots
\mu _{\ell }}$ can only be constructed from tensors formed using projection
operators $\Delta ^{\mu \nu }$. We start by collecting all possible
combinations of projection operators that can appear in $\left( \mathcal{N}%
_{rnn^{\prime }}\right) _{\alpha _{1}\cdots \alpha _{m}\beta _{1}\cdots
\beta _{m^{\prime }}}^{\mu _{1}\cdots \mu _{\ell }}$:

\begin{enumerate}
\item[(i)] Terms where all $\mu $--type indices pair up on projectors, all $%
\alpha $--type indices pair up on projectors, and all $\beta $--type
indices pair up on projectors, e.g. 
\begin{equation}
\Delta ^{\mu _{1}\mu _{2}}\cdots \Delta ^{\mu _{\ell -1}\mu _{\ell }}\Delta
_{\alpha _{1}\alpha _{2}}\cdots \Delta _{\alpha _{m-1}\alpha _{m}}\Delta
_{\beta _{1}\beta _{2}}\cdots \Delta _{\beta _{m^{\prime }-1}\beta
_{m^{\prime }}}\;.
\end{equation}%
All possible permutations of the $\mu ,\alpha ,\beta $--type indices among
themselves are allowed. In this case, $\ell $, $m$, and $m^{\prime }$ must
all be even.

\item[(ii)] Terms where at least one $\mu $--type index pairs with an $%
\alpha $--type index on a projector, or one $\mu $--type index pairs with a $%
\beta $--type index on a projector, or one $\alpha $--type index pairs with
a $\beta $--type index on a projector, e.g. 
\begin{eqnarray}
&&\Delta _{\alpha _{1}}^{\mu _{1}}\Delta ^{\mu _{2}\mu _{3}}\cdots \Delta
^{\mu _{\ell -1}\mu _{\ell }}\Delta _{\alpha _{2}\alpha _{3}}\cdots \Delta
_{\alpha _{m-1}\alpha _{m}}\Delta _{\beta _{1}\beta _{2}}\cdots \Delta
_{\beta _{m^{\prime }-1}\beta _{m^{\prime }}}\;, \\
&&\Delta _{\beta _{1}}^{\mu _{1}}\Delta ^{\mu _{2}\mu _{3}}\cdots \Delta
^{\mu _{\ell -1}\mu _{\ell }}\Delta _{\alpha _{1}\alpha _{2}}\cdots \Delta
_{\alpha _{m-1}\alpha _{m}}\Delta _{\beta _{2}\beta _{3}}\cdots \Delta
_{\beta _{m^{\prime }-1}\beta _{m^{\prime }}}\;, \\
&&\Delta _{\alpha _{1}\beta _{1}}\Delta ^{\mu _{1}\mu _{2}}\cdots \Delta
^{\mu _{\ell -1}\mu _{\ell }}\Delta _{\alpha _{2}\alpha _{3}}\cdots \Delta
_{\alpha _{m-1}\alpha _{m}}\Delta _{\beta _{2}\beta _{3}}\cdots \Delta
_{\beta _{m^{\prime }-1}\beta _{m^{\prime }}}\;.
\end{eqnarray}%
Again, all possible permutations of the $\mu $--type, $\alpha $--type, and $%
\beta $--type indices are allowed.

\item[(iii)] Terms where \emph{each} $\mu $--type, $\alpha $--type, and $%
\beta $--type index pairs up with an index of another type. To guarantee
that the $\mu $--type indices have sufficiently many partners among the
other two types of indices, one must have $\ell \leq m+m^{\prime }$.
Similarly, in order for the $\alpha $--type indices to pair up in this way,
we have to require $m\leq \ell +m^{\prime }$. Finally, for the $\beta $%
--type indices we need the condition $m^{\prime }\leq \ell +m$. In this
case, only projectors of the type $\Delta _{\alpha _{j}}^{\mu _{i}}$, $%
\Delta _{\beta _{j}}^{\mu _{i}}$ or $\Delta _{\alpha _{i}\beta _{j}}$ exist,
with no left-over projectors containing indices of the same type. Such terms
have the form 
\begin{equation}
\Delta _{\alpha _{p}}^{\mu _{i}}\Delta _{\beta _{q}}^{\mu _{j}}\Delta
_{\alpha _{r}\beta _{s}}\cdots \;.
\end{equation}%
Again, all permutations of the $\mu ,\alpha ,\beta $--type indices among
themselves are allowed.
\end{enumerate}

It is important to emphasize that terms of the type (i) and (ii) by
themselves do not satisfy the property (\ref{great property}), since they
are not traceless. This can also be seen from the fact that any term which
contains at least one projector of the type $\Delta ^{\mu _{i}\mu _{j}}$, $%
\Delta _{\alpha _{p}\alpha _{q}}$, or $\Delta _{\beta _{r}\beta _{s}}$
vanishes when contracted with $\Delta _{\mu _{1}\cdots \mu _{\ell }}^{\mu
_{1}^{\prime }\cdots \mu _{\ell }^{\prime }}\Delta _{\alpha _{1}^{\prime
}\cdots \alpha _{m}^{\prime }}^{\alpha _{1}\cdots \alpha _{m}}\Delta _{\beta
_{1}^{\prime }\cdots \beta _{m^{\prime }}^{\prime }}^{\beta _{1}\cdots \beta
_{m^{\prime }}}$ . Thus, $\left( \mathcal{N}_{rnn^{\prime }}\right) _{\alpha
_{1}\cdots \alpha _{m}\beta _{1}\cdots \beta _{m^{\prime }}}^{\mu _{1}\cdots
\mu _{\ell }}$ cannot be solely constructed from terms of type (i) and
(ii) and there must be at least one term of type (iii).

Therefore, terms of type (iii) are of special importance in this derivation
and it is convenient to further discuss some of their properties. The
inequalities that constrain terms of type (iii), i.e., $\ell \leq
m+m^{\prime }$, $m\leq \ell +m^{\prime }$, $m^{\prime }\leq \ell +m$, can be
solved and lead to%
\begin{equation}
\ell =q+r,\text{ \ }m=p+r,\text{ \ }m^{\prime }=p+q,
\end{equation}%
with $p,q,r=0,1,2,\ldots $ . Since the $\ell $ index is always fixed in the
summations appearing in Eq.\ (\ref{NonLin_collint}), one can re-express the
above equations as%
\begin{equation}
m=p-q+\ell ,\text{ \ }m^{\prime }=p+q,\text{ \ }q\leq \ell \text{ }.
\label{equalities}
\end{equation}%
For our purposes it is sufficient to calculate terms of second order in
inverse Reynolds number in the terms $\mathcal{R}$, $\mathcal{R}^{\mu }$,
and $\mathcal{R}^{\mu \nu }$. Therefore, we only need to consider the cases $%
\ell =0$, $\ell =1$, and $\ell =2$:

\subsection{$\ell = 0$}

If $\ell =0$, the equalities (\ref{equalities}) imply that $m^{\prime
}=m=0,1,\ldots $ and, consequently, one must have%
\begin{equation}
\left( \mathcal{N}_{rnn^{\prime }}\right) _{\alpha _{1}\cdots \alpha
_{m}\beta _{1}\cdots \beta _{m^{\prime }}}=\delta _{mm^{\prime }}\mathcal{C}%
_{\left( 0\right) }\Delta _{(\alpha _{1}\beta _{1}}\cdots \Delta _{\alpha
_{m}\beta _{m})}+[\mbox{terms of type (i)
and (ii)}]\;.  \label{ScalarDerivation}
\end{equation}%
Contracting Eq.\ (\ref{ScalarDerivation}) with $\Delta _{\alpha _{1}^{\prime
}\cdots \alpha _{m}^{\prime }}^{\alpha _{1}\cdots \alpha _{m}}\Delta _{\beta
_{1}^{\prime }\cdots \beta _{m^{\prime }}^{\prime }}^{\beta _{1}\cdots \beta
_{m^{\prime }}}$ and using Eq.\ (\ref{great property}), we prove that%
\begin{equation}
\left( \mathcal{N}_{rnn^{\prime }}\right) _{\alpha _{1}\cdots \alpha
_{m}\beta _{1}\cdots \beta _{m^{\prime }}}=\delta _{mm^{\prime }}\mathcal{C}%
_{\left( 0\right) }\Delta _{\alpha _{1}\cdots \alpha _{m}\beta _{1}\cdots
\beta _{m}}\;,
\end{equation}%
where $\mathcal{C}_{\left( 0\right) }$ is the trace of $\left( \mathcal{N}%
_{rnn^{\prime }}\right) _{\alpha _{1}\cdots \alpha _{m}\beta _{1}\cdots
\beta _{m^{\prime }}}$, 
\begin{eqnarray}
\mathcal{C}_{\left( 0\right) } &\equiv& \left[ \Delta _{\beta _{1}\cdots
\beta _{m}}^{\alpha _{1}\cdots \alpha _{m}}\Delta _{\alpha _{1}\cdots \alpha
_{m}}^{\beta _{1}\cdots \beta _{m}}\right] ^{-1}\Delta ^{\alpha _{1}\cdots
\alpha _{m}\beta _{1}\cdots \beta _{m}}\left( \mathcal{N}_{rnn^{\prime
}}\right) _{\alpha _{1}\cdots \alpha _{m}\beta _{1}\cdots \beta_{m}}  \notag\\
&=&\frac{1}{\left( 2m+1\right) \nu }\int_{f}E_{\mathbf{k}}^{r-1}\left( 
\mathcal{H}_{\mathbf{p}n}^{\left( m\right) }\,\mathcal{H}_{\mathbf{p}%
^{\prime }n^{\prime }}^{\left( m\right) }\,p^{\left\langle \mu _{1}\right.
}\cdots p^{\left. \mu _{m}\right\rangle }\,p_{\left\langle \mu _{1}\right.
}^{\prime }\cdots p_{\left. \mu _{m}\right\rangle }^{\prime }-\mathcal{H}_{%
\mathbf{k}n}^{\left( m\right) }\,\mathcal{H}_{\mathbf{k}^{\prime }n^{\prime
}}^{\left( m\right) }\,k^{\left\langle \mu _{1}\right. }\cdots k^{\left. \mu
_{m}\right\rangle }\,k_{\left\langle \mu _{1}\right. }^{\prime }\cdots
k_{\left. \mu _{m}\right\rangle }^{\prime }\right) . \quad
\end{eqnarray}

The coefficient $\mathcal{C}_{\left( 0\right) }=\mathcal{C}_{rnn^{\prime
}}^{0\left( mm\right) }$, the $\ell =0$ case of Eq.\ (\ref{C_lm}). Thus, we
obtain Eq.\ (\ref{Nr_scalar}) for the scalar nonlinear collision integral.

\subsection{$\ell=1$}

For\ $\ell =1$, Eqs.\ (\ref{equalities}) imply that $m^{\prime }=m+1$, and,
consequently, 
\begin{equation}
\left( \mathcal{N}_{rnn^{\prime }}\right) _{\alpha _{1}\cdots \alpha
_{m}\beta _{1}\cdots \beta _{m^{\prime }}}^{\mu }=\delta _{m+1,m^{\prime }}%
\mathcal{C}_{\left( 1\right) }\Delta _{(\beta _{1}}^{\mu }\Delta _{\beta
_{2}\alpha _{1}}\cdots \Delta _{\beta _{m+1}\alpha _{m})}+[%
\mbox{terms of type (i)
and (ii)}]\;.  \label{VectorDerivation}
\end{equation}

All permutations of the $\alpha $--indices and $\beta $--indices among
themselves are allowed, while permutations of the $\alpha $--indices with
the $\beta $--indices are forbidden. Contracting Eq.\ (\ref{VectorDerivation}%
) with $\Delta _{\mu }^{\mu ^{\prime }}\Delta _{\alpha _{1}^{\prime }\cdots
\alpha _{m}^{\prime }}^{\alpha _{1}\cdots \alpha _{m}}\Delta _{\beta
_{1}^{\prime }\cdots \beta _{m+1}^{\prime }}^{\beta _{1}\cdots \beta _{m+1}}$
and using Eq.\ (\ref{great property}), we prove that%
\begin{equation}
\left( \mathcal{N}_{rnn^{\prime }}\right) _{\alpha _{1}\cdots \alpha
_{m}\beta _{1}\cdots \beta _{m+1}}^{\mu }=\mathcal{C}_{\left( 1\right)
}\Delta _{\left. {}\right. \alpha _{1}\cdots \alpha _{m}\beta _{1}\cdots
\beta _{m+1}}^{\mu }\;.
\end{equation}%
The coefficient $\mathcal{C}_{\left( 1\right) }$ is obtained from the trace
of $\left( \mathcal{N}_{rnn^{\prime }}\right) _{\beta _{1}\cdots \beta
_{m+1}}^{\mu \alpha _{1}\cdots \alpha _{m}}$, i.e.,%
\begin{align}
\mathcal{C}_{\left( 1\right) }&\equiv \left[ \Delta _{\beta _{1}\cdots \beta
_{m+1}}^{\mu \alpha _{1}\cdots \alpha _{m}}\Delta _{\mu \alpha _{1}\cdots
\alpha _{m}}^{\beta _{1}\cdots \beta _{m+1}}\right] ^{-1}\Delta _{\mu
}^{\left. {}\right. \alpha _{1}\cdots \alpha _{m}\beta _{1}\cdots \beta
_{m+1}}\left( \mathcal{N}_{rnn^{\prime }}\right) _{\alpha _{1}\cdots \alpha
_{m}\beta _{1}\cdots \beta _{m+1}}^{\mu }  \notag \\
& =\frac{1}{\left[ 2\left( m+1\right) +1\right] \nu }\int_{f}E_{\mathbf{k}%
}^{r-1}k_{\mu }  \notag \\
& \times \left( \mathcal{H}_{\mathbf{p}n}^{\left( m\right) }\,\mathcal{H}_{%
\mathbf{p}^{\prime }n^{\prime }}^{(m+1)}\,p_{\left\langle \alpha _{1}\right.
}\cdots p_{\left. \alpha _{m}\right\rangle }\,p^{\prime \left\langle \mu
\right. }p^{\prime \alpha _{1}}\cdots p^{\prime \left. \alpha
_{m}\right\rangle }+\,\mathcal{H}_{\mathbf{p}^{\prime }n}^{\left( m\right)
}\,\mathcal{H}_{\mathbf{p}n^{\prime }}^{(m+1)}\,p_{\left\langle \alpha
_{1}\right. }^{\prime }\cdots p_{\left. \alpha _{m}\right\rangle }^{\prime
}\,p^{\left\langle \mu \right. }p^{\alpha _{1}}\cdots p^{\left. \alpha
_{m}\right\rangle }\right.  \notag \\
& \left. -\mathcal{H}_{\mathbf{k}n}^{\left( m\right) }\,\mathcal{H}_{\mathbf{%
k}^{\prime }n^{\prime }}^{(m+1)}\,k_{\left\langle \alpha _{1}\right. }\cdots
k_{\left. \alpha _{m}\right\rangle }\,\,k^{\prime \left\langle \mu \right.
}k^{\prime \alpha _{1}}\cdots k^{\prime \left. \alpha _{m}\right\rangle }-\,%
\mathcal{H}_{\mathbf{k}^{\prime }n}^{\left( m\right) }\,\mathcal{H}_{\mathbf{%
k}n^{\prime }}^{(m+1)}\,k_{\left\langle \alpha _{1}\right. }^{\prime }\cdots
k_{\left. \alpha _{m}\right\rangle }^{\prime }\,k^{\left\langle \mu \right.
}k^{\alpha _{1}}\cdots k^{\left. \alpha _{m}\right\rangle }\right) .
\end{align}%
Note that $\mathcal{C}_{\left( 1\right) }=\mathcal{C}_{rnn^{\prime
}}^{1\left( m,m+1\right) }$, defined in Eq.\ (\ref{C_lm}) in the main text.
Thus, we obtain $N_{r-1}^{\mu }$ as given in Eq.\ (\ref{Nr_vector}).


\subsection{$\ell = 2$}

For terms with $\ell =2$, two solutions are possible: $m^{\prime
}=m=0,1,\ldots ,$ and $m^{\prime }=m+2=2,3,\ldots\;$. Therefore, two
different type (iii) tensors can be constructed, leading to%
\begin{eqnarray}
\left( \mathcal{N}_{rnn^{\prime }}\right) _{\alpha _{1}\cdots \alpha
_{m}\beta _{1}\cdots \beta _{m^{\prime }}}^{\mu \nu } &=&\delta
_{m+2,m^{\prime }}\mathcal{C}_{\left( 2\right) }\Delta _{(\beta _{1}}^{\mu
}\Delta _{\beta _{2}}^{\nu }\Delta _{\beta _{3}\alpha _{1}}\cdots \Delta
_{\beta _{m+2}\alpha _{m})}  \notag \\
&+&\delta _{mm^{\prime }}\mathcal{D}_{\left( 2\right) }\Delta _{(\beta
_{1}}^{(\mu }\Delta _{\alpha _{1}}^{\nu )}\Delta _{\beta _{2}\alpha
_{2}}\cdots \Delta _{\beta _{m}\alpha _{m})}  \notag \\
&+&[\mbox{terms of type (i)
and (ii)}].  \label{TensorDerivation}
\end{eqnarray}%
All permutations of the $\alpha $--indices and $\beta $--indices among
themselves are allowed, while permutations of the $\alpha $--indices with
the $\beta $--indices are forbidden. Contracting Eq.\ (\ref{TensorDerivation}%
) with $\Delta _{\mu \nu }^{\mu ^{\prime }\nu ^{\prime }}\Delta _{\alpha
_{1}^{\prime }\cdots \alpha _{m}^{\prime }}^{\alpha _{1}\cdots \alpha
_{m}}\Delta _{\beta _{1}^{\prime }\cdots \beta _{m^{\prime }}^{\prime
}}^{\beta _{1}\cdots \beta _{m^{\prime }}}$ and using Eq.\ (\ref{great
property}), we prove that%
\begin{eqnarray}
\left( \mathcal{N}_{rnn^{\prime }}\right) _{\alpha _{1}\cdots \alpha
_{m}\beta _{1}\cdots \beta _{m^{\prime }}}^{\mu \nu } &=&\delta
_{m+2,m^{\prime }}\mathcal{C}_{\left( 2\right) }\Delta _{\left. {}\right.
\left. {}\right. \alpha _{1}\cdots \alpha _{m}\beta _{1}\cdots \beta
_{m+2}}^{\mu \nu }  \notag \\
&+&\delta _{mm^{\prime }}\mathcal{D}_{\left( 2\right) }\Delta _{\lambda
_{1}\sigma }^{\mu \nu }\Delta _{\left. {}\right. \lambda _{2}\cdots \lambda
_{m}\alpha _{1}\cdots \alpha _{m}}^{\sigma }\Delta _{\beta _{1}\cdots \beta
_{m}}^{\lambda _{1}\cdots \lambda _{m}},
\end{eqnarray}%
with the coefficients $\mathcal{C}_{\left( 2\right) }$ and $\mathcal{D}%
_{\left( 2\right) }$ being obtained from the corresponding trace of $\left( 
\mathcal{N}_{rnn^{\prime }}\right) _{\alpha _{1}\cdots \alpha _{m}\beta
_{1}\cdots \beta _{m^{\prime }}}^{\mu \nu }$ when $m^{\prime }=m+2$ and $%
m^{\prime }=m$, respectively. That is, the coefficient $\mathcal{C}_{\left(
2\right) }$ is given by 
\begin{eqnarray}
\mathcal{C}_{\left( 2\right) } &\equiv &\left[ \Delta _{\beta _{1}\cdots
\beta _{m+2}}^{\mu \nu \alpha _{1}\cdots \alpha _{m}}\Delta _{\mu \nu \alpha
_{1}\cdots \alpha _{m}}^{\beta _{1}\cdots \beta _{m+2}}\right] ^{-1}\Delta
_{\mu \nu }^{\left. {}\right. \left. {}\right. \alpha _{1}\cdots \alpha
_{m}\beta _{1}\cdots \beta _{m+2}}\left( \mathcal{N}_{rnn^{\prime }}\right)
_{\alpha _{1}\cdots \alpha _{m}\beta _{1}\cdots \beta _{m+2}}^{\mu \nu }
\notag \\
&=&\frac{1}{\left[ 2\left( m+\ell \right) +1\right] \nu }\Delta ^{\mu \nu
\alpha _{1}\cdots \alpha _{m}\beta _{1}\cdots \beta _{m+2}}\int_{f}E_{%
\mathbf{k}}^{r-1}k_{\left\langle \mu \right. }k_{\left. \nu \right\rangle } 
\notag \\
&\times &\left( \mathcal{H}_{\mathbf{p}n}^{\left( m\right) }\,\mathcal{H}_{%
\mathbf{p}^{\prime }n^{\prime }}^{(m+2)}\,p_{\left\langle \alpha _{1}\right.
}\cdots p_{\left. \alpha _{m}\right\rangle }\,p^{\prime \left\langle \mu
\right. }p^{\prime \nu }p^{\prime \alpha _{1}}\cdots p^{\prime \left. \alpha
_{m}\right\rangle }+\,\mathcal{H}_{\mathbf{p}^{\prime }n}^{\left( m\right)
}\,\mathcal{H}_{\mathbf{p}n^{\prime }}^{(m+2)}\,p_{\left\langle \alpha
_{1}\right. }^{\prime }\cdots p_{\left. \alpha _{m}\right\rangle }^{\prime
}\,p^{\left\langle \mu \right. }p^{\nu }p^{\alpha _{1}}\cdots p^{\left.
\alpha _{m}\right\rangle }\right.   \notag \\
&&\left. -\mathcal{H}_{\mathbf{k}n}^{\left( m\right) }\,\mathcal{H}_{\mathbf{%
k}^{\prime }n^{\prime }}^{(m+2)}\,k_{\left\langle \alpha _{1}\right. }\cdots
k_{\left. \alpha _{m}\right\rangle }\,k^{\prime \left\langle \mu \right.
}k^{\prime \nu }k^{\prime \alpha _{1}}\cdots k^{\prime \left. \alpha
_{m}\right\rangle }-\,\mathcal{H}_{\mathbf{k}^{\prime }n}^{\left( m\right)
}\,\mathcal{H}_{\mathbf{k}n^{\prime }}^{(m+2)}\,k_{\left\langle \alpha
_{1}\right. }^{\prime }\cdots k_{\left. \alpha _{m}\right\rangle }^{\prime
}\,k^{\left\langle \mu \right. }k^{\nu }k^{\alpha _{1}}\cdots k^{\left.
\alpha _{m}\right\rangle }\right) ,
\end{eqnarray}%
while $\mathcal{D}_{\left( 2\right) }$ is%
\begin{eqnarray}
\mathcal{D}_{\left( 2\right) } &\equiv &\left[ d^{\left( m\right) }\right]
^{-1}\Delta _{\mu \nu }^{\lambda _{1}\sigma }\Delta _{\sigma }^{\left.
{}\right. \lambda _{2}\cdots \lambda _{m}\alpha _{1}\cdots \alpha
_{m}}\Delta _{\lambda _{1}\cdots \lambda _{m}}^{\beta _{1}\cdots \beta
_{m}}\left( \mathcal{N}_{rnn^{\prime }}\right) _{\alpha _{1}\cdots \alpha
_{m}\beta _{1}\cdots \beta _{m}}^{\mu \nu }  \notag \\
&=&\left[ d^{\left( m\right) }\right] ^{-1}\frac{1}{\nu }\int_{f}E_{\mathbf{k%
}}^{r-1}k^{\left\langle \lambda _{1}\right. }k_{\left. \sigma \right\rangle }
\notag \\
&\times &\left( \mathcal{H}_{\mathbf{p}n}^{\left( m\right) }\,\mathcal{H}_{%
\mathbf{p}^{\prime }n^{\prime }}^{(m)}\,p^{\left\langle \sigma \right.
}p^{\lambda _{2}}\cdots p^{\left. \lambda _{m}\right\rangle
}\,p_{\left\langle \lambda _{1}\right. }^{\prime }\cdots p_{\left. \lambda
_{m}\right\rangle }^{\prime }-\mathcal{H}_{\mathbf{k}n}^{\left( m\right) }\,%
\mathcal{H}_{\mathbf{k}^{\prime }n^{\prime }}^{(m)}\,k^{\left\langle \sigma
\right. }k^{\lambda _{2}}\cdots k^{\left. \lambda _{m}\right\rangle
}k_{\left\langle \lambda _{1}\right. }^{\prime }\cdots k_{\left. \lambda
_{m}\right\rangle }^{\prime }\right) ,
\end{eqnarray}%
where we defined $d^{\left( m\right) }=\Delta _{\lambda _{1}\sigma }^{\rho
_{1}\psi }\Delta _{\psi }^{\left. {}\right. \rho _{2}\cdots \rho _{m}\alpha
_{1}\cdots \alpha _{m}}\Delta _{\left. {}\right. \lambda _{2}\cdots \lambda
_{m}\alpha _{1}\cdots \alpha _{m}}^{\sigma }\Delta _{\rho _{1}\cdots \rho
_{m}}^{\lambda _{1}\cdots \lambda _{m}}$. The coefficients $\mathcal{C}%
_{\left( 2\right) }$ and $\mathcal{D}_{\left( 2\right) }$ can be identified
with the coefficients $\mathcal{C}_{rnn^{\prime }}^{2\left( m,m+2\right) }$
and $\mathcal{D}_{rnn^{\prime }}^{2\left( mm\right) }$, respectively,
defined in the main text in Eqs.\ (\ref{C_lm}) and (\ref{DDDDD}),
respectively. Thus,%
\begin{eqnarray}
N_{r-1}^{\mu \nu } &=&\sum_{m=0}^{\infty }\sum_{n=0}^{N_{m}}\sum_{n^{\prime
}=0}^{N_{m^{\prime }}}\delta _{m+2,m^{\prime }}\mathcal{C}_{rnn^{\prime
}}^{2\left( m,m+2\right) }\rho _{n}^{\alpha _{1}\cdots \alpha _{m}}\rho
_{n^{\prime }\alpha _{1}\cdots \alpha _{m}}^{\mu \nu }  \notag \\
&+&\sum_{m=1}^{\infty }\sum_{n=0}^{N_{m}}\sum_{n^{\prime }=0}^{N_{m^{\prime
}}}\delta _{mm^{\prime }}\mathcal{D}_{rnn^{\prime }}^{2\left( m,m\right)
}\rho _{n,\lambda _{2}\cdots \lambda _{m}}^{\left\langle \mu \right. }\rho
_{n^{\prime }}^{\left. \nu \right\rangle \lambda _{2}\cdots \lambda _{m}},
\end{eqnarray}%
which was already presented in the main text in Eq.\ (\ref{Nr_tensor}).
Calculating the trace $d^{\left( m\right) }$ for an arbitrary $m$ can be
very complicated. In this paper, we shall only do it for the cases $m=1,2$,
which are actually needed. It follows that, for $m=1$, 
\begin{equation}
d^{\left( 1\right) }\equiv \Delta _{\lambda _{1}\sigma }^{\rho _{1}\psi
}\Delta _{\psi }^{\alpha _{1}}\Delta _{\alpha _{1}}^{\sigma }\Delta _{\rho
_{1}}^{\lambda _{1}}=\Delta _{\lambda _{1}\sigma }^{\lambda _{1}\sigma }=5,
\end{equation}%
while for $m=2$ one obtains%
\begin{equation}
d^{\left( 2\right) }\equiv \Delta _{\lambda _{1}\sigma }^{\rho _{1}\psi
}\Delta _{\psi }^{\left. {}\right. \rho _{2}\alpha _{1}\alpha _{2}}\Delta
_{\left. {}\right. \lambda _{2}\alpha _{1}\alpha _{2}}^{\sigma }\Delta
_{\rho _{1}\rho _{2}}^{\lambda _{1}\lambda _{2}}=\frac{35}{12}.
\end{equation}

\section{Expansion coefficients}

\label{exp_coefficients}

In this appendix, we construct the polynomials $P_{\mathbf{k}n}^{\left( \ell
\right) }$, see Eq.\ (\ref{P_kn}). For any $\ell \geq 0$, we set%
\begin{equation}
P_{\mathbf{k}0}^{\left( \ell \right) }\equiv a_{00}^{(\ell )}=1,
\end{equation}%
and obtain 
\begin{align}
P_{\mathbf{k}1}^{\left( 0\right) }& = a_{11}^{\left( 0\right) }E_{\mathbf{k}%
}+a_{10}^{\left( 0\right) }, \\
P_{\mathbf{k}1}^{\left( 1\right) }& = a_{11}^{\left( 1\right) }E_{\mathbf{k}%
}+a_{10}^{\left( 1\right) }, \\
P_{\mathbf{k}2}^{\left( 0\right) }& = a_{22}^{\left( 0\right) }E_{\mathbf{k}%
}^{2}+a_{21}^{\left( 0\right) }E_{\mathbf{k}}+a_{20}^{\left( 0\right) }.
\end{align}%
From the orthonormality condition $\int dK\ \omega ^{\left( \ell \right) }P_{%
\mathbf{k}i}^{\left( \ell \right) }P_{\mathbf{k}j}^{(\ell )}=\delta _{ij}$,
it follows that the measure, $\omega ^{\left( \ell \right) }$, and the
normalization constant, $W^{\left( \ell \right) }$, are given in Eqs.\ (\ref%
{w_measure},\ref{W_normalization}). Therefore, using the above equations
together with the orthonormality conditions we obtain 
\begin{align}
\frac{a_{10}^{(0)}}{a_{11}^{(0)}}& =-\frac{J_{10}}{J_{00}},\ \left(
a_{11}^{(0)}\right) ^{2}=\frac{J_{00}^{2}}{D_{10}}, \\
\frac{a_{21}^{\left( 0\right) }}{a_{22}^{\left( 0\right) }}& =\frac{G_{12}}{%
D_{10}},\ \frac{a_{20}^{\left( 0\right) }}{a_{22}^{\left( 0\right) }}=\frac{%
D_{20}}{D_{10}}, \\
\left( a_{22}^{\left( 0\right) }\right) ^{2}& =\frac{J_{00}D_{10}}{%
J_{20}D_{20}+J_{30}G_{12}+J_{40}D_{10}}, \\
\frac{a_{10}^{\left( 1\right) }}{a_{11}^{\left( 1\right) }}& =-\frac{J_{31}}{%
J_{21}},\ \left( a_{11}^{\left( 1\right) }\right) ^{2}=\frac{J_{21}^{2}}{%
D_{31}},
\end{align}%
where the $G_{nm}$ and $D_{nq}$ functions were defined in Eqs.\ (\ref{Gnm}),
(\ref{Dnq}).

The coefficients $\mathcal{H}_{\mathbf{k}n}^{(\ell)}$ are defined in Eq.\ (%
\ref{H_kn}). In the 14--moment approximation we only need $\rho _{0}=-3\Pi
/m^{2}$,$\ \rho _{0}^{\mu }=n^{\mu }$ and $\rho _{0}^{\mu \nu }=\pi ^{\mu
\nu }$ with $N_{0}=2$, $N_{1}=1$, and $N_{2}=0$. Furthermore $\rho _{1}=0$
and $\rho _{2}=0$ due to the matching conditions (\ref{Landau_matching}),
while $\rho _{1}^{\mu }=0$ by the choice (\ref{Landau_flow}) of the local
rest frame. Hence, 
\begin{align}
\mathcal{H}_{\mathbf{k}0}^{\left( 0\right) }& \equiv W^{\left( 0\right)
}\left( a_{00}^{(0)}P_{\mathbf{k}0}^{\left( 0\right) }+a_{10}^{(0)}P_{%
\mathbf{k}1}^{\left( 0\right) }+a_{20}^{(0)}P_{\mathbf{k}2}^{\left( 0\right)
}\right) = A_{00}^{(0)}+A_{10}^{(0)}E_{\mathbf{k}}+A_{20}^{(0)}E_{\mathbf{k}%
}^{2}, \\
\mathcal{H}_{\mathbf{k}0}^{\left( 1\right) }& \equiv W^{\left( 1\right)
}\left( a_{00}^{(1)}P_{\mathbf{k}0}^{\left( 1\right) }+a_{10}^{(1)}P_{%
\mathbf{k}1}^{\left( 1\right) }\right) =A_{00}^{(1)}+A_{10}^{(1)}E_{\mathbf{k%
}}, \\
\mathcal{H}_{\mathbf{k}0}^{\left( 2\right) }& \equiv \frac{W^{\left(
2\right) }}{2}a_{00}^{(2)}P_{\mathbf{k}0}^{\left( 2\right) } =A_{00}^{\left(
2\right) },
\end{align}%
where the $A_{rn}^{(\ell )}$ were introduced in Eq.\ (\ref{A_l_rn}) and, in
the 14--moment approximation, are given by 
\begin{align}
A_{00}^{\left( 0\right) }& \equiv W^{\left( 0\right) }\left[ 1+\left(
a_{10}^{\left( 0\right) }\right) ^{2}+\left( a_{20}^{\left( 0\right)
}\right) ^{2}\right] =\frac{D_{30}}{ J_{20}D_{20}+J_{30}G_{12}+J_{40}D_{10} }%
, \\
A_{10}^{\left( 0\right) }& \equiv W^{\left( 0\right) }\left(
a_{10}^{(0)}a_{11}^{\left( 0\right) }+a_{20}^{(0)}a_{21}^{\left( 0\right)
}\right) =\frac{G_{23}}{J_{20}D_{20}+J_{30}G_{12}+J_{40}D_{10} }, \\
A_{20}^{\left( 0\right) }& \equiv W^{\left( 0\right) }\left(
a_{20}^{(0)}a_{22}^{(0)}\right) =\frac{D_{20}}{
J_{20}D_{20}+J_{30}G_{12}+J_{40}D_{10} },
\end{align}%
and 
\begin{align}
A_{00}^{\left( 1\right) }& \equiv W^{\left( 1\right) }\left[ 1+\left(
a_{10}^{\left( 1\right) }\right) ^{2}\right] =-\frac{J_{41}}{D_{31}}, \\
A_{10}^{\left( 1\right) }& \equiv W^{\left( 1\right) }\left(
a_{10}^{(1)}a_{11}^{\left( 1\right) }\right) =\frac{J_{31}}{D_{31}}, \\
A_{00}^{\left( 2\right) }& \equiv \frac{W^{\left( 2\right) }}{2}=\frac{1}{%
2J_{42}}.
\end{align}

Note that these coefficients closely resemble the ones given in Eqs.\
(108) -- (113) of Ref.\ \cite{Denicol:2012es}.

\section{The coefficients of the collision terms}

\label{2nd_order_coll_int}

In this appendix, we calculate the coefficients of the nonlinear collision
integral in the 14--moment approximation. The nonlinear scalar term $N_{r}$
from Eq.\ (\ref{Nr_scalar}) is expanded with the help of the following
coefficients,%
\begin{align}
\mathcal{C}_{r00}^{0\left( 00\right) }& \equiv \frac{1}{\nu }\int_{f}E_{%
\mathbf{k}}^{r-1}\left( \mathcal{H}_{\mathbf{p}0}^{\left( 0\right) }\mathcal{%
H}_{\mathbf{p}^{\prime }0}^{\left( 0\right) }-\mathcal{H}_{\mathbf{k}%
0}^{\left( 0\right) }\mathcal{H}_{\mathbf{k}^{\prime }0}^{\left( 0\right)
}\right)  \notag \\
& =\frac{1}{\nu }\left( A_{10}^{\left( 0\right) }\right) ^{2}\int_{f}E_{%
\mathbf{k}}^{r-1}\left( E_{\mathbf{p}}E_{\mathbf{p}^{\prime }}-E_{\mathbf{k}%
}E_{\mathbf{k}^{\prime }}\right) +\frac{1}{\nu }\left( A_{20}^{\left(
0\right) }\right) ^{2}\int_{f}E_{\mathbf{k}}^{r-1}\left( E_{\mathbf{p}%
}^{2}E_{\mathbf{p}^{\prime }}^{2}-E_{\mathbf{k}}^{2}E_{\mathbf{k}^{\prime
}}^{2}\right)  \notag \\
& +\frac{1}{\nu }\left( A_{00}^{\left( 0\right) }A_{20}^{\left( 0\right)
}\right) \int_{f}E_{\mathbf{k}}^{r-1}\left( E_{\mathbf{p}}^{2}+E_{\mathbf{p}%
^{\prime }}^{2}-E_{\mathbf{k}}^{2}-E_{\mathbf{k}^{\prime }}^{2}\right) +%
\frac{1}{\nu }\left( A_{10}^{\left( 0\right) }A_{20}^{\left( 0\right)
}\right) \int_{f}E_{\mathbf{k}}^{r-1}\left( E_{\mathbf{p}}E_{\mathbf{p}%
^{\prime }}^{2}+E_{\mathbf{p}^{\prime }}E_{\mathbf{p}}^{2}-E_{\mathbf{k}}E_{%
\mathbf{k}^{\prime }}^{2}-E_{\mathbf{k}^{\prime }}E_{\mathbf{k}}^{2}\right) ,
\end{align}%
where terms proportional to $\left( A_{00}^{\left( 0\right) }\right) ^{2}$
and $A_{00}^{\left( 0\right) }A_{10}^{\left( 0\right) }$ vanish on account
of energy conservation in binary collisions. The above result can be
re-expressed using  the $X_{\left( r\right) }$ and $Y_{i\left(
r\right) }$ tensors given in Eq.\ (\ref{X_4rank_tensor}) and Eqs.\ (\ref%
{Y1_4rank_tens}) -- (\ref{Y5_6rank_tens}). Thus, after some calculation we
obtain 
\begin{align}
\mathcal{C}_{r00}^{0\left( 00\right) }& =\left( A_{10}^{\left( 0\right)
}\right) ^{2}Y_{2\left( r-3\right) }^{\mu \nu \alpha \beta }u_{\mu }u_{\nu
}u_{\alpha }u_{\beta }+\left( A_{20}^{\left( 0\right) }\right)
^{2}Y_{5\left( r-3\right) }^{\mu \nu \alpha \beta \kappa \lambda }u_{\mu
}u_{\nu }u_{\alpha }u_{\beta }u_{\kappa }u_{\lambda }  \notag \\
& +\left( A_{00}^{\left( 0\right) }A_{20}^{\left( 0\right) }\right)
X_{\left( r-3\right) }^{\mu \nu \alpha \beta }u_{\mu }u_{\nu }u_{\alpha
}u_{\beta }+\left( A_{10}^{\left( 0\right) }A_{20}^{\left( 0\right) }\right)
Y_{3\left( r-3\right) }^{\mu \nu \alpha \beta \kappa }u_{\mu }u_{\nu
}u_{\alpha }u_{\beta }u_{\kappa }.
\end{align}%
After this step we still need to evaluate the terms $Y_{2\left( r-3\right)
}^{\mu \nu \alpha \beta }u_{\mu }u_{\nu }u_{\alpha }u_{\beta }$, $Y_{5\left(
r-3\right) }^{\mu \nu \alpha \beta \kappa \lambda }u_{\mu }u_{\nu }u_{\alpha
}u_{\beta }u_{\kappa }u_{\lambda }$ etc. This is relegated to Appendix \ref%
{tens_decompositions}, for example $Y_{2\left( r-3\right) }^{\mu \nu \alpha
\beta }u_{\mu }u_{\nu }u_{\alpha }u_{\beta }=Y_{2\left( r\right) ,1}$ as
shown in Eq.\ (\ref{Y_2r_1}). In a similar fashion we repeat the calculation
for all components and later, in Appendix \ref{tensors_massless_limit}, we
calculate them in the massless limit.

The next coefficient is 
\begin{align}
\mathcal{C}_{r00}^{0\left( 11\right) }& \equiv \frac{1}{3\nu }\int_{f}E_{%
\mathbf{k}}^{r-1}\left( \mathcal{H}_{\mathbf{p}0}^{\left( 1\right) }\mathcal{%
H}_{\mathbf{p}^{\prime }0}^{\left( 1\right) }p^{\left\langle \mu
\right\rangle }p_{\left\langle \mu \right\rangle }^{\prime }-\mathcal{H}_{%
\mathbf{k}0}^{\left( 1\right) }\mathcal{H}_{\mathbf{k}^{\prime }0}^{\left(
1\right) }k^{\left\langle \mu \right\rangle }k_{\left\langle \mu
\right\rangle }^{\prime }\right)  \notag \\
& =\frac{1}{3\nu }\left( A_{00}^{\left( 1\right) }\right) ^{2}\int_{f}E_{%
\mathbf{k}}^{r-1}\left( p^{\left\langle \mu \right\rangle }p_{\left\langle
\mu \right\rangle }^{\prime }-k^{\left\langle \mu \right\rangle
}k_{\left\langle \mu \right\rangle }^{\prime }\right)  \notag \\
& +\frac{1}{3\nu }\left( A_{00}^{\left( 1\right) }A_{10}^{\left( 1\right)
}\right) \int_{f}E_{\mathbf{k}}^{r-1}\left( E_{\mathbf{p}}p^{\left\langle
\mu \right\rangle }p_{\left\langle \mu \right\rangle }^{\prime }+E_{\mathbf{p%
}^{\prime }}p^{\left\langle \mu \right\rangle }p_{\left\langle \mu
\right\rangle }^{\prime }-E_{\mathbf{k}}k^{\left\langle \mu \right\rangle
}k_{\left\langle \mu \right\rangle }^{\prime }-E_{\mathbf{k}^{\prime
}}k^{\left\langle \mu \right\rangle }k_{\left\langle \mu \right\rangle
}^{\prime }\right)  \notag \\
& +\frac{1}{3\nu }\left( A_{10}^{\left( 1\right) }\right) ^{2}\int_{f}E_{%
\mathbf{k}}^{r-1}\left( E_{\mathbf{p}}E_{\mathbf{p}^{\prime
}}p^{\left\langle \mu \right\rangle }p_{\left\langle \mu \right\rangle
}^{\prime }-E_{\mathbf{k}}E_{\mathbf{k}^{\prime }}k^{\left\langle \mu
\right\rangle }k_{\left\langle \mu \right\rangle }^{\prime }\right) ,
\end{align}%
so that 
\begin{equation}
\mathcal{C}_{r00}^{0\left( 11\right) }=\frac{1}{3}\left[ \left(
A_{00}^{\left( 1\right) }\right) ^{2}Y_{2\left( r-3\right) }^{\mu \nu \alpha
\beta }u_{\mu }u_{\nu }\Delta _{\alpha \beta }+\left( A_{00}^{\left(
1\right) }A_{10}^{\left( 1\right) }\right) Y_{3\left( r-3\right) }^{\mu \nu
\alpha \beta \kappa }u_{\mu }u_{\nu }u_{\left( \beta \right. }\Delta
_{\left. \kappa \right) \alpha }+\left( A_{10}^{\left( 1\right) }\right)
^{2}Y_{5\left( r-3\right) }^{\mu \nu \alpha \beta \kappa \lambda }u_{\mu
}u_{\nu }u_{\left( \alpha \right. }\Delta _{\left. \beta \right) \left(
\kappa \right. }u_{\left. \lambda \right) }\right] .
\end{equation}%
The last scalar coefficient is,%
\begin{align}
\mathcal{C}_{r00}^{0\left( 22\right) }& \equiv \frac{1}{5\nu }\int_{f}E_{%
\mathbf{k}}^{r-1}\left( \mathcal{H}_{\mathbf{p}0}^{\left( 2\right) }\mathcal{%
H}_{\mathbf{p}^{\prime }0}^{\left( 2\right) }p^{\left\langle \mu \right.
}p^{\left. \nu \right\rangle }p_{\left\langle \mu \right. }^{\prime
}p_{\left. \nu \right\rangle }^{\prime }-\mathcal{H}_{\mathbf{k}0}^{\left(
2\right) }\mathcal{H}_{\mathbf{k}^{\prime }0}^{\left( 2\right)
}k^{\left\langle \mu \right. }k^{\left. \nu \right\rangle }k_{\left\langle
\mu \right. }^{\prime }k_{\left. \nu \right\rangle }^{\prime }\right)  \notag
\\
& =\frac{1}{5\nu }\left( A_{00}^{\left( 2\right) }\right) ^{2}\int_{f}E_{%
\mathbf{k}}^{r-1}\left( p^{\left\langle \mu \right. }p^{\left. \nu
\right\rangle }p_{\left\langle \mu \right. }^{\prime }p_{\left. \nu
\right\rangle }^{\prime }-k^{\left\langle \mu \right. }k^{\left. \nu
\right\rangle }k_{\left\langle \mu \right. }^{\prime }k_{\left. \nu
\right\rangle }^{\prime }\right) ,
\end{align}%
therefore, 
\begin{equation}
\mathcal{C}_{r00}^{0\left( 22\right) }=\frac{1}{5}\left( A_{00}^{\left(
2\right) }\right) ^{2}Y_{5\left( r-3\right) }^{\mu \nu \alpha \beta \kappa
\lambda }u_{\mu }u_{\nu }\Delta _{\alpha \beta \kappa \lambda }.
\end{equation}

The vector coefficients are given by 
\begin{align}
\mathcal{C}_{r00}^{1\left( 01\right) }& \equiv \frac{1}{3\nu }\int_{f}E_{%
\mathbf{k}}^{r-1}k_{\left\langle \mu \right\rangle }\left( \mathcal{H}_{%
\mathbf{p}0}^{\left( 0\right) }\mathcal{H}_{\mathbf{p}^{\prime }0}^{\left(
1\right) }p^{\prime \left\langle \mu \right\rangle }+\mathcal{H}_{\mathbf{p}%
^{\prime }0}^{\left( 0\right) }\mathcal{H}_{\mathbf{p}0}^{\left( 1\right)
}p^{\left\langle \mu \right\rangle } - \mathcal{H}_{\mathbf{k}0}^{\left(
0\right) }\mathcal{H}_{\mathbf{k}^{\prime }0}^{\left( 1\right) }k^{\prime
\left\langle \mu \right\rangle }-\mathcal{H}_{\mathbf{k}^{\prime }0}^{\left(
0\right) }\mathcal{H}_{\mathbf{k}0}^{\left( 1\right) }k^{\left\langle \mu
\right\rangle }\right)  \notag \\
& =\frac{1}{3\nu }\left( A_{00}^{\left( 0\right) }A_{10}^{\left( 1\right)
}\right) \int_{f}E_{\mathbf{k}}^{r-1}k_{\left\langle \mu \right\rangle
}\left( E_{\mathbf{p}}p^{\left\langle \mu \right\rangle }+E_{\mathbf{p}%
^{\prime }}p^{\prime \left\langle \mu \right\rangle } - E_{\mathbf{k}%
}k^{\left\langle \mu \right\rangle }-E_{\mathbf{k}^{\prime }}k^{\prime
\left\langle \mu \right\rangle }\right)  \notag \\
& +\frac{1}{3\nu }\left( A_{10}^{\left( 0\right) }A_{00}^{\left( 1\right)
}\right) \int_{f}E_{\mathbf{k}}^{r-1}k_{\left\langle \mu \right\rangle
}\left( E_{\mathbf{p}}p^{\prime \left\langle \mu \right\rangle }+E_{\mathbf{p%
}^{\prime }}p^{\left\langle \mu \right\rangle } - E_{\mathbf{k}}k^{\prime
\left\langle \mu \right\rangle }-E_{\mathbf{k}^{\prime }}k^{\left\langle \mu
\right\rangle }\right)  \notag \\
& +\frac{1}{3\nu }\left( A_{10}^{\left( 0\right) }A_{10}^{\left( 1\right)
}\right) \int_{f}E_{\mathbf{k}}^{r-1}k_{\left\langle \mu \right\rangle
}\left(E_{\mathbf{p}}E_{\mathbf{p}^{\prime }}p^{\prime \left\langle \mu
\right\rangle }+E_{\mathbf{p}^{\prime }}E_{\mathbf{p}}p^{\left\langle \mu
\right\rangle } - E_{\mathbf{k}}E_{\mathbf{k}^{\prime }}k^{\prime
\left\langle \mu \right\rangle }-E_{\mathbf{k}^{\prime }}E_{\mathbf{k}%
}k^{\left\langle \mu \right\rangle }\right)  \notag \\
& +\frac{1}{3\nu }\left( A_{20}^{\left( 0\right) }A_{00}^{\left( 1\right)
}\right) \int_{f}E_{\mathbf{k}}^{r-1}k_{\left\langle \mu \right\rangle
}\left( E_{\mathbf{p}^{\prime }}^{2}p^{\left\langle \mu \right\rangle }+E_{%
\mathbf{p}}^{2}p^{\prime \left\langle \mu \right\rangle } - E_{\mathbf{k}%
^{\prime }}^{2}k^{\left\langle \mu \right\rangle }-E_{\mathbf{k}%
}^{2}k^{\prime \left\langle \mu \right\rangle }\right)  \notag \\
& +\frac{1}{3\nu }\left( A_{20}^{\left( 0\right) }A_{10}^{\left( 1\right)
}\right) \int_{f}E_{\mathbf{k}}^{r-1}k_{\left\langle \mu \right\rangle
}\left( E_{\mathbf{p}}^{2}E_{\mathbf{p}^{\prime }}p^{\prime \left\langle \mu
\right\rangle }+E_{\mathbf{p}^{\prime }}^{2}E_{\mathbf{p}}p^{\left\langle
\mu \right\rangle } - E_{\mathbf{k}}^{2}E_{\mathbf{k}^{\prime }}k^{\prime
\left\langle \mu \right\rangle }-E_{\mathbf{k}^{\prime }}^{2}E_{\mathbf{k}%
}k^{\left\langle \mu \right\rangle }\right) ,
\end{align}%
hence%
\begin{align}
\mathcal{C}_{r00}^{1\left( 01\right) }& =\frac{1}{3}\left[ \left(
A_{00}^{\left( 0\right) }A_{10}^{\left( 1\right) }\right) X_{\left(
r-2\right) }^{\mu \nu \alpha \beta }u_{\left( \mu \right. }\Delta _{\left.
\nu \right) \left( \alpha \right. }u_{\left. \beta \right) }+\left(
A_{10}^{\left( 0\right) }A_{00}^{\left( 1\right) }\right) Y_{1\left(
r-2\right) }^{\mu \nu \alpha \beta }u_{\left( \mu \right. }\Delta _{\left.
\nu \right) \left( \alpha \right. }u_{\left. \beta \right) }\right.  \notag
\\
& \left. +\left( A_{10}^{\left( 0\right) }A_{10}^{\left( 1\right) }\right)
Y_{3\left(r-2\right) }^{\mu \nu \alpha \beta \kappa }u_{\left( \mu \right. }\Delta
_{\left. \nu \right) \left( \beta \right. }u_{\left. \kappa \right)
}u_{\alpha }+\left(
A_{20}^{\left( 0\right) }A_{00}^{\left( 1\right) }\right)
Y_{3\left( r-2\right) }^{\mu \nu \alpha \beta \kappa }\Delta _{\alpha \left(
\mu \right. }u_{\left. \nu \right) }u_{\beta }u_{\kappa }\right.  \notag \\
& \left. +\left( A_{20}^{\left( 0\right) }A_{10}^{\left( 1\right) }\right)
Y_{4\left( r-2\right) }^{\mu \nu \alpha \beta \kappa \lambda }u_{\left( \mu
\right. }\Delta _{\left. \nu \right) \left( \kappa \right. }u_{\left.
\lambda \right) }u_{\alpha }u_{\beta }\right] .
\end{align}

The second term is 
\begin{align}
\mathcal{C}_{r00}^{1\left( 12\right) }& \equiv \frac{1}{5\nu }\int_{f}E_{%
\mathbf{k}}^{r-1}k_{\left\langle \mu \right\rangle }\left( \mathcal{H}_{%
\mathbf{p}0}^{\left( 1\right) }\mathcal{H}_{\mathbf{p}^{\prime }0}^{\left(
2\right) }p_{\left\langle \nu \right\rangle }p^{\prime \left\langle \mu
\right. }p^{\prime \left. \nu \right\rangle }+\mathcal{H}_{\mathbf{p}%
^{\prime }0}^{\left( 1\right) }\mathcal{H}_{\mathbf{p}0}^{\left( 2\right)
}p_{\left\langle \nu \right\rangle }^{\prime }p^{\left\langle \mu \right.
}p^{\left. \nu \right\rangle }-\mathcal{H}_{\mathbf{k}0}^{\left( 1\right) }%
\mathcal{H}_{\mathbf{k}^{\prime }0}^{\left( 2\right) }k_{\left\langle \nu
\right\rangle }k^{\prime \left\langle \mu \right. }k^{\prime \left. \nu
\right\rangle }-\mathcal{H}_{\mathbf{k}^{\prime }0}^{\left( 1\right) }%
\mathcal{H}_{\mathbf{k}0}^{\left( 2\right) }k_{\left\langle \nu
\right\rangle }^{\prime }k^{\left\langle \mu \right. }k^{\left. \nu
\right\rangle }\right)  \notag \\
& =\frac{1}{5\nu }\left( A_{00}^{\left( 1\right) }A_{00}^{\left( 2\right)
}\right) \int_{f}E_{\mathbf{k}}^{r-1}k_{\left\langle \mu \right\rangle
}\left( p_{\left\langle \nu \right\rangle }p^{\prime \left\langle \mu
\right. }p^{\prime \left. \nu \right\rangle }+p_{\left\langle \nu
\right\rangle }^{\prime }p^{\left\langle \mu \right. }p^{\left. \nu
\right\rangle }-k_{\left\langle \nu \right\rangle }k^{\prime \left\langle
\mu \right. }k^{\prime \left. \nu \right\rangle }-k_{\left\langle \nu
\right\rangle }^{\prime }k^{\left\langle \mu \right. }k^{\left. \nu
\right\rangle }\right)  \notag \\
& +\frac{1}{5\nu }\left( A_{10}^{\left( 1\right) }A_{00}^{\left( 2\right)
}\right) \int_{f}E_{\mathbf{k}}^{r-1}k_{\left\langle \mu \right\rangle
}\left( E_{\mathbf{p}}p_{\left\langle \nu \right\rangle }p^{\prime
\left\langle \mu \right. }p^{\prime \left. \nu \right\rangle }+E_{\mathbf{p}%
^{\prime }}p_{\left\langle \nu \right\rangle }^{\prime }p^{\left\langle \mu
\right. }p^{\left. \nu \right\rangle }-E_{\mathbf{k}}k_{\left\langle \nu
\right\rangle }k^{\prime \left\langle \mu \right. }k^{\prime \left. \nu
\right\rangle }-E_{\mathbf{k}^{\prime }}k_{\left\langle \nu \right\rangle
}^{\prime }k^{\left\langle \mu \right. }k^{\left. \nu \right\rangle }\right)
,
\end{align}%
hence%
\begin{equation}
\mathcal{C}_{r00}^{1\left( 12\right) }=\frac{1}{5}A_{00}^{\left( 2\right) }%
\left[ A_{00}^{\left( 1\right) }Y_{3\left( r-2\right) }^{\mu \nu \alpha
\beta \kappa }u_{\left( \mu \right. }\Delta _{\left. \nu \right) \alpha
\beta \kappa }+\frac{1}{2}A_{10}^{\left( 1\right) }Y_{4\left( r-2\right)
}^{\mu \nu \alpha \beta \kappa \lambda }\left( u_{\left( \mu \right. }\Delta
_{\left. \nu \right) \alpha \kappa \lambda }u_{\beta }+u_{\left( \mu \right.
}\Delta _{\left. \nu \right) \beta \kappa \lambda }u_{\alpha }\right) \right]
.
\end{equation}

The first tensor coefficient is given by 
\begin{align}
\mathcal{C}_{r00}^{2\left( 02\right) }& \equiv \frac{1}{5\nu }\int_{f}E_{%
\mathbf{k}}^{r-1}k_{\left\langle \mu \right. }k_{\left. \nu \right\rangle
}\left( \mathcal{H}_{\mathbf{p}0}^{\left( 0\right) }\mathcal{H}_{\mathbf{p}%
^{\prime }0}^{\left( 2\right) }p^{\prime \left\langle \mu \right. }p^{\prime
\left. \nu \right\rangle }+\mathcal{H}_{\mathbf{p}^{\prime }0}^{\left(
0\right) }\mathcal{H}_{\mathbf{p}0}^{\left( 2\right) }p^{\left\langle \mu
\right. }p^{\left. \nu \right\rangle } - \mathcal{H}_{\mathbf{k}0}^{\left(
0\right) }\mathcal{H}_{\mathbf{k}^{\prime }0}^{\left( 2\right) }k^{\prime
\left\langle \mu \right. }k^{\prime \left. \nu \right\rangle }-\mathcal{H}_{%
\mathbf{k}^{\prime }0}^{\left( 0\right) }\mathcal{H}_{\mathbf{k}0}^{\left(
2\right) }k^{\left\langle \mu \right. }k^{\left. \nu \right\rangle }\right) 
\notag \\
& =\frac{1}{5\nu }\left( A_{00}^{\left( 0\right) }A_{00}^{\left( 2\right)
}\right) \int_{f}E_{\mathbf{k}}^{r-1}k_{\left\langle \mu \right. }k_{\left.
\nu \right\rangle }\left( p^{\prime \left\langle \mu \right. }p^{\prime
\left. \nu \right\rangle }+p^{\left\langle \mu \right. }p^{\left. \nu
\right\rangle }- k^{\prime \left\langle \mu \right. }k^{\prime \left. \nu
\right\rangle }-k^{\left\langle \mu \right. }k^{\left. \nu \right\rangle
}\right)  \notag \\
& +\frac{1}{5\nu }\left( A_{10}^{\left( 0\right) }A_{00}^{\left( 2\right)
}\right) \int_{f}E_{\mathbf{k}}^{r-1}k_{\left\langle \mu \right. }k_{\left.
\nu \right\rangle }\left( E_{\mathbf{p}}p^{\prime \left\langle \mu \right.
}p^{\prime \left. \nu \right\rangle }+E_{\mathbf{p}^{\prime
}}p^{\left\langle \mu \right. }p^{\left. \nu \right\rangle } - E_{\mathbf{k}%
}k^{\prime \left\langle \mu \right. }k^{\prime \left. \nu \right\rangle }-E_{%
\mathbf{k}^{\prime }}k^{\left\langle \mu \right. }k^{\left. \nu
\right\rangle }\right)  \notag \\
& +\frac{1}{5\nu }\left( A_{20}^{\left( 0\right) }A_{00}^{\left( 2\right)
}\right) \int_{f}E_{\mathbf{k}}^{r-1}k_{\left\langle \mu \right. }k_{\left.
\nu \right\rangle }\left( E_{\mathbf{p}}^{2}p^{\prime \left\langle \mu
\right. }p^{\prime \left. \nu \right\rangle }+E_{\mathbf{p}^{\prime
}}^{2}p^{\left\langle \mu \right. }p^{\left. \nu \right\rangle } - E_{%
\mathbf{k}}^{2}k^{\prime \left\langle \mu \right. }k^{\prime \left. \nu
\right\rangle }-E_{\mathbf{k}^{\prime }}^{2}k^{\left\langle \mu \right.
}k^{\left. \nu \right\rangle }\right) ,
\end{align}%
and so,%
\begin{equation}
\mathcal{C}_{r00}^{2\left( 0,2\right) }=\frac{1}{5}A_{00}^{\left( 2\right) }%
\left[ A_{00}^{\left( 0\right) }X_{\left( r-1\right) }^{\mu \nu \alpha \beta
}\Delta _{\mu \nu \alpha \beta }+A_{10}^{\left( 0\right) }Y_{3\left(
r-1\right) }^{\mu \nu \alpha \beta \kappa }u_{\alpha }\Delta _{\mu \nu \beta
\kappa }+A_{20}^{\left( 0\right) }Y_{4\left( r-1\right) }^{\mu \nu \alpha
\beta \kappa \lambda }u_{\alpha }u_{\beta }\Delta _{\mu \nu \kappa \lambda }%
\right] .
\end{equation}%
The $\mathcal{D}_{r00}^{2\left( 11\right) }$ term is 
\begin{align}
\mathcal{D}_{r00}^{2\left( 11\right) }& \equiv \frac{1}{5\nu }\int_{f}E_{%
\mathbf{k}}^{r-1}k_{\left\langle \mu \right. }k_{\left. \nu \right\rangle
}\left(\mathcal{H}_{\mathbf{p}0}^{\left( 1\right) }\mathcal{H}_{\mathbf{p}%
^{\prime }0}^{\left( 1\right) }p^{\left\langle \mu \right\rangle }p^{\prime
\left\langle \nu \right\rangle }- \mathcal{H}_{\mathbf{k}0}^{\left( 1\right)
}\mathcal{H}_{\mathbf{k}^{\prime }0}^{\left( 1\right) }k^{\left\langle \mu
\right\rangle }k^{\prime \left\langle \nu \right\rangle }\right)  \notag \\
& =\frac{1}{5\nu }\left( A_{00}^{\left( 1\right) }\right) ^{2}\int_{f}E_{%
\mathbf{k}}^{r-1}k_{\left\langle \mu \right. }k_{\left. \nu \right\rangle
}\left(p^{\left\langle \mu \right\rangle }p^{\prime \left\langle \nu
\right\rangle }-k^{\left\langle \mu \right\rangle }k^{\prime \left\langle
\nu \right\rangle }\right)  \notag \\
& +\frac{1}{5\nu }\left( A_{00}^{\left( 1\right) }A_{10}^{\left( 1\right)
}\right) \int_{f}E_{\mathbf{k}}^{r-1}k_{\left\langle \mu \right. }k_{\left.
\nu \right\rangle }\left( E_{\mathbf{p}^{\prime }}p^{\left\langle \mu
\right\rangle }p^{\prime \left\langle \nu \right\rangle }+E_{\mathbf{p}%
}p^{\left\langle \mu \right\rangle }p^{\prime \left\langle \nu \right\rangle
} - E_{\mathbf{k}^{\prime }}k^{\left\langle \mu \right\rangle }k^{\prime
\left\langle \nu \right\rangle }-E_{\mathbf{k}}k^{\left\langle \mu
\right\rangle }k^{\prime \left\langle \nu \right\rangle }\right)  \notag \\
& +\frac{1}{5\nu }\left( A_{10}^{\left( 1\right) }\right) ^{2}\int_{f}E_{%
\mathbf{k}}^{r-1}k_{\left\langle \mu \right. }k_{\left. \nu \right\rangle
}\left(E_{\mathbf{p}}E_{\mathbf{p}^{\prime }}p^{\left\langle \mu
\right\rangle }p^{\prime \left\langle \nu \right\rangle }-E_{\mathbf{k}}E_{%
\mathbf{k}^{\prime }}k^{\left\langle \mu \right\rangle }k^{\prime
\left\langle \nu \right\rangle }\right) ,
\end{align}%
hence, 
\begin{align}
\mathcal{D}_{r00}^{2\left( 11\right) }& =\frac{1}{5}\left[ \left(
A_{00}^{\left( 1\right) }\right) ^{2}Y_{2\left( r-1\right) }^{\mu \nu \alpha
\beta }\Delta _{\mu \nu \alpha \beta }+\left( A_{00}^{\left( 1\right)
}A_{10}^{\left( 1\right) }\right) Y_{3\left( r-1\right) }^{\mu \nu \alpha
\beta \kappa }\Delta _{\mu \nu \alpha \left( \beta \right. }u_{\left. \kappa
\right) }\right.  \notag \\
& \left. +\frac{1}{2}\left( A_{10}^{\left( 1\right) }\right) ^{2}Y_{5\left(
r-1\right) }^{\mu \nu \alpha \beta \kappa \lambda }\left( \Delta _{\mu \nu
\alpha \left( \kappa \right. }u_{\left. \lambda \right) }u_{\beta }+\Delta
_{\mu \nu \beta \left( \kappa \right. }u_{\left. \lambda \right) }u_{\alpha
}\right) \right] ,
\end{align}%
The last coefficient we need is 
\begin{align}
\mathcal{D}_{r00}^{2\left( 22\right) }& \equiv\frac{12}{35\nu }\int_{f}E_{%
\mathbf{k}}^{r-1}k_{\left\langle \mu \right. }k_{\left. \nu \right\rangle
}\left( \mathcal{H}_{\mathbf{p}0}^{\left( 2\right) }\mathcal{H}_{\mathbf{p}%
^{\prime }0}^{\left( 2\right) }p^{\left\langle \mu \right. }p_{\left.
\lambda \right\rangle }p^{\prime \left\langle \nu \right. }p^{\prime \left.
\lambda \right\rangle }-\mathcal{H}_{\mathbf{k}0}^{\left( 2\right) }\mathcal{%
H}_{\mathbf{k}^{\prime }0}^{\left( 2\right) }k^{\left\langle \mu \right.
}k_{\left. \lambda \right\rangle }k^{\prime \left\langle \nu \right.
}k^{\prime \left. \lambda \right\rangle }\right)  \notag \\
& =\frac{12}{35\nu }\left( A_{00}^{\left( 2\right) }\right) ^{2}\int_{f}E_{%
\mathbf{k}}^{r-1}k_{\left\langle \mu \right. }k_{\left. \nu \right\rangle
}\left( p^{\left\langle \mu \right. }p_{\left. \lambda \right\rangle
}p^{\prime \left\langle \nu \right. }p^{\prime \left. \lambda \right\rangle
}-k^{\left\langle \mu \right. }k_{\left. \lambda \right\rangle }k^{\prime
\left\langle \nu \right. }k^{\prime \left. \lambda \right\rangle }\right) ,
\end{align}%
and so,%
\begin{equation}
\mathcal{D}_{r00}^{2\left( 22\right) }=\frac{12}{35}\left( A_{00}^{\left(
2\right) }\right) ^{2}Y_{5\left( r-1\right) }^{\mu \nu \alpha \beta \kappa
\lambda }\left[ \frac{1}{2}\left( \Delta _{\mu \nu \mu _{1}\kappa }\Delta
_{\alpha \beta \lambda }^{\mu _{1}}+\Delta _{\mu \nu \mu _{1}\lambda }\Delta
_{\alpha \beta \kappa }^{\mu _{1}}\right) -\frac{1}{3}\Delta _{\mu \nu
\alpha \beta }\Delta _{\kappa \lambda }\right] .
\end{equation}

\section{Tensor decompositions}

\label{tens_decompositions}

In this appendix, we discuss the decompositions of all collision tensors
required in the 14--moment approximation. The collision tensor$\ X_{\left(
r\right) }^{\mu \nu \alpha \beta }=X_{\left( r\right) }^{\left( \mu \nu
\right) \left( \alpha \beta \right) }$\ from Eq.\ (\ref{X_4rank_tensor}) is
symmetric upon the interchange of indices $\left( \mu ,\nu \right) $ and $%
\left( \alpha ,\beta \right) $, and it is also traceless in the latter
indices, $X_{\left( r\right) }^{\mu \nu \alpha \beta }g_{\alpha \beta }=0$,
which follows from the mass-shell condition, $k^{\mu }k_{\mu }\equiv p^{\mu
}p_{\mu }=m^{2}$. Using these properties one can show that $X_{\left(
r\right) }^{\mu \nu \alpha \beta }$ is a spatially isotropic tensor which
can be constructed using the 4-velocity $u^{\mu }$, the projector $%
\Delta ^{\mu \nu }$, and different scalar coefficients $x_{ij}$. The most
general decomposition of $X_{\left( r\right) }^{\mu \nu \alpha \beta }$
which is symmetric upon the interchange of indices $\left( \mu ,\nu \right) $
and $\left( \alpha ,\beta \right) $ is 
\begin{align}
X^{\left( \mu \nu \right) \left( \alpha \beta \right) }& \equiv x_{10}u^{\mu
}u^{\nu }u^{\alpha }u^{\beta }+x_{11}u^{\mu }u^{\nu }\Delta ^{\alpha \beta
}+x_{21}u^{\alpha }u^{\beta }\Delta ^{\mu \nu }+4x_{31}u^{\left( \mu \right.
}\Delta ^{\left. \nu \right) \left( \alpha \right. }u^{\left. \beta \right)
} +x_{12}\Delta ^{\mu \nu }\Delta ^{\alpha \beta }+2x_{22}\Delta ^{\mu
\left( \alpha \right. }\Delta ^{\left. \beta \right) \nu }.
\end{align}%
The indices of the scalar coefficients $x_{ij}$ are chosen such that the
second index ($j$) denotes the number of projection tensors belonging to the
respective coefficient while the first index ($i$) counts the number of such
coefficients. For example $x_{10}$ is the coefficient without any projection
tensor, while $x_{12}$ is the first coefficient which contains two
elementary projection tensors.

From the tracelessness relation, $X_{\left( r\right) }^{\mu \nu \alpha \beta
}g_{\alpha \beta }=0$, follows that we only have four independent
coefficients, since 
\begin{equation}
x_{11}=-\frac{x_{10}}{3},\ x_{22}=-\frac{1}{2} \left(x_{21}+ 3x_{12} \right).
\end{equation}%
Thus we obtain 
\begin{equation}
X_{\left( r\right) }^{\mu \nu \alpha \beta }\equiv \left( x_{10}u^{\mu
}u^{\nu }+x_{21}\Delta ^{\mu \nu }\right) \left( u^{\alpha }u^{\beta }-\frac{%
1}{3}\Delta ^{\alpha \beta }\right) +4x_{31}u^{(\mu }\Delta ^{\nu )(\alpha
}u^{\beta )}-\left( x_{21}+3x_{12}\right) \Delta ^{\mu \nu \alpha \beta }.
\end{equation}%
Introducing the notation $X_{\left( r\right) ,1}=x_{10}$,$\ X_{\left(
r\right) ,2}=x_{21}$,$\ X_{\left( r\right) ,3}=x_{31}$, and$\ X_{\left(
r\right) ,4}=2x_{22}$, one can easily show that these coefficients are the
result of the following contractions, 
\begin{align}
X_{\left( r\right) ,1}& \equiv X_{\left( r\right) }^{\mu \nu \alpha \beta
}u_{\mu }u_{\nu }u_{\alpha }u_{\beta }=-X_{\left( r\right) }^{\mu \nu \alpha
\beta }u_{\mu }u_{\nu }\Delta _{\alpha \beta },  \label{X1_tens} \\
X_{\left( r\right) ,2}& \equiv \frac{1}{3}X_{\left( r\right) }^{\mu \nu
\alpha \beta }\Delta _{\mu \nu }u_{\alpha }u_{\beta }=-\frac{1}{3}X_{\left(
r\right) }^{\mu \nu \alpha \beta }\Delta _{\mu \nu }\Delta _{\alpha \beta },
\label{X2_tens} \\
X_{\left( r\right) ,3}& = \frac{1}{3}X_{\left( r\right) }^{\mu \nu \alpha
\beta }u_{\left( \mu \right. }\Delta _{\left. \nu \right) \left( \alpha
\right. }u_{\left. \beta \right) },  \label{X3_tens} \\
X_{\left( r\right) ,4}& = \frac{1}{5}X_{\left( r\right) }^{\mu \nu \alpha
\beta }\Delta _{\mu \nu \alpha \beta }.  \label{X4_tens}
\end{align}

Similarly as shown here, the other rank-4 collision tensors $Y_{1\left(
r\right) }^{\mu \nu \alpha \beta }$ and $Y_{2\left( r\right) }^{\mu \nu
\alpha \beta }$ from Eqs.\ (\ref{Y1_4rank_tens}), (\ref{Y2_4rank_tens}) can
be decomposed taking into account the symmetry properties $Y_{1\left(
r\right) }^{\mu \nu \alpha \beta }=Y_{1\left( r\right) }^{\left( \mu \nu
\right) \left( \alpha \beta \right) }$, $Y_{2\left( r\right) }^{\mu \nu
\alpha \beta }=Y_{2\left( r\right) }^{\left( \mu \nu \right) \alpha \beta }$
and the tracelessness relations, $Y_{1\left( r\right) }^{\mu \nu \alpha
\beta }g_{\alpha \beta }=0$, $Y_{2\left( r\right) }^{\mu \nu \alpha \beta
}g_{\alpha \beta }=0$. Here we only list the coefficients which are needed
for later calculations, 
\begin{align}
Y_{1\left( r\right) ,3}& =\frac{1}{3}Y_{1\left( r\right) }^{\mu \nu \alpha
\beta }u_{\left( \mu \right. }\Delta _{\left. \nu \right) \left( \alpha
\right. }u_{\left. \beta \right) }, \\
Y_{2\left( r\right) ,1}& \equiv Y_{2\left( r\right) }^{\mu \nu \alpha \beta
}u_{\mu }u_{\nu }u_{\alpha }u_{\beta }=-Y_{2\left( r\right) }^{\mu \nu
\alpha \beta }u_{\mu }u_{\nu }\Delta _{\alpha \beta },  \label{Y_2r_1} \\
Y_{2\left( r\right) ,4}& =\frac{1}{5}Y_{2\left( r\right) }^{\mu \nu \alpha
\beta }\Delta _{\mu \nu \alpha \beta }.
\end{align}

In the following we will decompose the rank-5 collision tensor $Y_{3\left(
r\right) }^{\mu \nu \alpha \beta \kappa }=Y_{3\left( r\right) }^{\left( \mu
\nu \right) \alpha \left( \beta \kappa \right) }$ from Eq.\ (\ref%
{Y3_5rank_tens}), 
\begin{align}
Y_{3}^{\left( \mu \nu \right) \alpha \left( \beta \kappa \right) }&
=y_{10}u^{\mu }u^{\nu }u^{\beta }u^{\kappa }u^{\alpha }+y_{11}u^{\mu }u^{\nu
}\Delta ^{\beta \kappa }u^{\alpha }+y_{21}u^{\beta }u^{\kappa }\Delta ^{\mu
\nu }u^{\alpha }+4y_{31}u^{\left( \mu \right. }\Delta ^{\left. \nu \right)
\left( \beta \right. }u^{\left. \kappa \right) }u^{\alpha }  \notag \\
& +2y_{41}u^{\mu }u^{\nu }u^{\left( \beta \right. }\Delta ^{\left. \kappa
\right) \alpha }+2y_{51}u^{\beta }u^{\kappa }u^{\left( \mu \right. }\Delta
^{\left. \nu \right) \alpha }+y_{12}\Delta ^{\mu \nu }\Delta ^{\beta \kappa
}u^{\alpha }+2y_{22}\Delta ^{\mu \left( \beta \right. }\Delta ^{\left.
\kappa \right) \nu }u^{\alpha }  \notag \\
& +2y_{32}\Delta ^{\mu \nu }u^{\left( \beta \right. }\Delta ^{\left. \kappa
\right) \alpha }+2y_{42}\Delta ^{\beta \kappa }u^{\left( \mu \right. }\Delta
^{\left. \nu \right) \alpha }+4y_{52}\Delta ^{\alpha \left( \mu \right.
}\Delta ^{\left. \nu \right) \left( \beta \right. }u^{\kappa
)}+4y_{62}\Delta ^{\alpha \left( \beta \right. }\Delta ^{\left. \kappa
\right) \left( \mu \right. }u^{\left. \nu \right) }\,.
\end{align}%
Using the tracelessness relation $Y_{3}^{\mu \nu \alpha \beta \kappa
}g_{\beta \kappa }=0$, 
\begin{equation}
y_{11}=-\frac{y_{10}}{3},\ y_{22}=-\frac{1}{2} \left(y_{21}+3y_{12}\right),\
y_{51}=-3y_{42}-2y_{62},
\end{equation}%
thus the number of unknown coefficients is nine. We can express them
similarly as in the previous cases, however, here we only list the ones
which will be used later 
\begin{align}
Y_{3\left( r\right) ,1}& \equiv y_{10}=Y_{3\left( r\right) }^{\mu \nu \alpha
\beta \kappa }u_{\mu }u_{\nu }u_{\alpha }u_{\beta }u_{\kappa }, \\
Y_{3\left( r\right) ,3}& \equiv y_{31}=\frac{1}{3}Y_{3\left( r\right) }^{\mu
\nu \alpha \beta \kappa }u_{\left( \mu \right. }\Delta _{\left. \nu \right)
\left( \beta \right. }u_{\left. \kappa \right) }u_{\alpha }, \\
Y_{3\left( r\right) ,4}& \equiv 2y_{22}=\frac{1}{5}Y_{3\left( r\right)
}^{\mu \nu \alpha \beta \kappa }\Delta _{\mu \nu \beta \kappa }u_{\alpha },
\\
Y_{3\left( r\right) ,5}& \equiv y_{41}=\frac{1}{3}Y_{3\left( r\right) }^{\mu
\nu \alpha \beta \kappa }u_{\mu }u_{\nu }u_{\left( \beta \right. }\Delta
_{\left. \kappa \right) \alpha }, \\
Y_{3\left( r\right) ,6}& \equiv y_{42}=-\frac{1}{15}Y_{3\left( r\right)
}^{\mu \nu \alpha \beta \kappa }\left[ \Delta _{\alpha \left( \beta \right.
}\Delta _{\left. \kappa \right) \left( \mu \right. }u_{\left. \nu \right)
}+2u_{\beta }u_{\kappa }u_{\left( \mu \right. }\Delta _{\left. \nu \right)
\alpha }\right] , \\
Y_{3\left( r\right) ,7}& \equiv y_{52}=\frac{1}{30}Y_{3\left( r\right)
}^{\mu \nu \alpha \beta \kappa }\left[ 3\Delta _{\alpha \left( \mu \right.
}\Delta _{\left. \nu \right) \left( \beta \right. }u_{\left. \kappa \right)
}-\Delta _{\mu \nu }u_{\left( \beta \right. }\Delta _{\left. \kappa \right)
\alpha }\right] =\frac{1}{10}Y_{3\left( r\right) }^{\mu \nu \alpha \beta
\kappa }\Delta _{\mu \nu \alpha \left( \beta \right. }u_{\left. \kappa
\right) }, \\
Y_{3\left( r\right) ,8}& \equiv y_{62}=\frac{1}{30}Y_{3\left( r\right)
}^{\mu \nu \alpha \beta \kappa }\left[ 3\Delta _{\alpha \left( \beta \right.
}\Delta _{\left. \kappa \right) \left( \mu \right. }u_{\left. \nu \right)
}+u_{\beta }u_{\kappa }u_{\left( \mu \right. }\Delta _{\left. \nu \right)
\alpha }\right] .
\end{align}
Note that in Eq.\ (\ref{C101}) we made use of 
\begin{equation}
\ y_{51}\equiv \frac{1}{3}Y_{3\left( r\right) }^{\mu \nu \alpha \beta \kappa
}u_{\left( \mu \right. }\Delta _{\left. \nu \right) \alpha }u_{\beta
}u_{\kappa }=- 3Y_{3\left( r\right) ,6}-2Y_{3\left( r\right) ,8}.
\end{equation}

The last two tensors we need are $Y_{4\left( r\right) }^{\mu \nu \alpha
\beta \kappa \lambda }\equiv Y_{4\left( r\right) }^{\left( \mu \nu \right)
\left( \alpha \beta \right) \left( \kappa \lambda \right) }=Y_{4\left(
r\right) }^{\left( \mu \nu \right) \left( \kappa \lambda \right) \left(
\alpha \beta \right) }$ and $Y_{5\left( r\right) }^{\mu \nu \alpha \beta
\kappa \lambda }=Y_{5\left( r\right) }^{\left( \mu \nu \right) \left( \alpha
\beta \right) \left( \kappa \lambda \right) }$, where the trace relations
are for example\ $Y_{4\left( r\right) }^{\mu \nu \alpha \beta \kappa \lambda
}g_{\kappa \lambda }=m^{2}X_{\left( r\right) }^{\mu \nu \alpha \beta }$, $%
Y_{4\left( r\right) }^{\mu \nu \alpha \beta \kappa \lambda }g_{\beta \lambda
}=m^{2}Y_{1\left( r\right) }^{\mu \nu \alpha \kappa }$,$\ $and\ $Y_{5\left(
r\right) }^{\mu \nu \alpha \beta \kappa \lambda }g_{\beta \lambda
}=m^{2}Y_{2\left( r\right) }^{\mu \nu \alpha \kappa }$, as well as $%
Y_{4\left( r\right) }^{\mu \nu \alpha \beta \kappa \lambda }g_{\kappa
\lambda }g_{\alpha \beta }\equiv Y_{5\left( r\right) }^{\mu \nu \alpha \beta
\kappa \lambda }g_{\kappa \lambda }g_{\alpha \beta }=0$. Although both
rank-6 tensor are very similar, there are important differences due to
symmetry, and so we write down both decompositions, 
\begin{align}
Y_{4\left( r\right) }^{\left( \mu \nu \right) \left( \alpha \beta \right)
\left( \kappa \lambda \right) }& =w_{10}u^{\mu }u^{\nu }u^{\alpha }u^{\beta
}u^{\kappa }u^{\lambda }+w_{11}u^{\mu }u^{\nu }\left[ \Delta ^{\alpha \beta
}u^{\kappa }u^{\lambda }+\Delta ^{\kappa \lambda }u^{\alpha }u^{\beta }%
\right] +w_{21}u^{\alpha }u^{\beta }\Delta ^{\mu \nu }u^{\kappa }u^{\lambda }
\notag \\
& +4w_{31}\left[ u^{\left( \mu \right. }\Delta ^{\left. \nu \right) \left(
\alpha \right. }u^{\left. \beta \right) }u^{\kappa }u^{\lambda }+u^{\left(
\mu \right. }\Delta ^{\left. \nu \right) \left( \kappa \right. }u^{\left.
\lambda \right) }u^{\alpha }u^{\beta }\right] +4w_{41}u^{\mu }u^{\nu
}u^{\left( \alpha \right. }\Delta ^{\left. \beta \right) \left( \kappa
\right. }u^{\left. \lambda \right) }  \notag \\
& +w_{12}\Delta ^{\mu \nu }\left[ \Delta ^{\alpha \beta }u^{\kappa
}u^{\lambda }+\Delta ^{\kappa \lambda }u^{\alpha }u^{\beta }\right] +2w_{22}%
\left[ \Delta ^{\mu \left( \alpha \right. }\Delta ^{\left. \beta \right) \nu
}u^{\kappa }u^{\lambda }+\Delta ^{\mu \left( \kappa \right. }\Delta ^{\left.
\lambda \right) \nu }u^{\alpha }u^{\beta }\right]  \notag \\
& +4w_{32}\Delta ^{\mu \nu }u^{\left( \alpha \right. }\Delta ^{\left. \beta
\right) \left( \kappa \right. }u^{\left. \lambda \right) }+4w_{42}\left[
\Delta ^{\alpha \beta }u^{\left( \mu \right. }\Delta ^{\left. \nu \right)
\left( \kappa \right. }u^{\left. \lambda \right) }+\Delta ^{\kappa \lambda
}u^{\left( \mu \right. }\Delta ^{\left. \nu \right) \left( \alpha \right.
}u^{\left. \beta \right) }\right]  \notag \\
& +8w_{52}u^{\left( \kappa \right. }\Delta ^{\left. \lambda \right) \left(
\mu \right. }\Delta ^{\left. \nu \right) \left( \alpha \right. }u^{\left.
\beta \right) }+8w_{62}\left[ u^{\left( \mu \right. }\Delta ^{\left. \nu
\right) \left( \alpha \right. }\Delta ^{\left. \beta \right) \left( \kappa
\right. }u^{\left. \lambda \right) }+u^{\left( \mu \right. }\Delta ^{\left.
\nu \right) \left( \kappa \right. }\Delta ^{\left. \lambda \right) \left(
\alpha \right. }u^{\left. \beta \right) }\right]  \notag \\
& +2w_{72}u^{\mu }u^{\nu }\Delta ^{\kappa \left( \alpha \right. }\Delta
^{\left. \beta \right) \lambda }+w_{82}u^{\mu }u^{\nu }\Delta ^{\alpha \beta
}\Delta ^{\kappa \lambda }  \notag \\
& +w_{13}\Delta ^{\mu \nu }\Delta ^{\alpha \beta }\Delta ^{\kappa \lambda
}+8w_{23}\Delta ^{\left( \kappa \right. \left( \mu \right. }\Delta ^{\left.
\nu \right) \left( \alpha \right. }\Delta ^{\left. \beta \right) \left.
\lambda \right) }  \notag \\
& +2w_{33}\left[ \Delta ^{\alpha \beta }\Delta ^{\mu \left( \kappa \right.
}\Delta ^{\left. \lambda \right) \nu }+\Delta ^{\kappa \lambda }\Delta ^{\mu
\left( \alpha \right. }\Delta ^{\left. \beta \right) \nu }\right]
+2w_{43}\Delta ^{\mu \nu }\Delta ^{\alpha \left( \kappa \right. }\Delta
^{\left. \lambda \right) \beta },
\end{align}%
where $\Delta ^{\left( \kappa \right. \left( \mu \right. }\Delta ^{\left.
\nu \right) \left( \alpha \right. }\Delta ^{\left. \beta \right) \left.
\lambda \right) }=\frac{1}{2}\left( \Delta ^{\kappa \left( \mu \right.
}\Delta ^{\left. \nu \right) \left( \alpha \right. }\Delta ^{\left. \beta
\right) \lambda }+\Delta ^{\lambda \left( \mu \right. }\Delta ^{\left. \nu
\right) \left( \alpha \right. }\Delta ^{\left. \beta \right) \kappa }\right) 
$\ and%
\begin{align}
Y_{5\left( r\right) }^{\left( \mu \nu \right) \left( \alpha \beta \right)
\left( \kappa \lambda \right) }& =z_{10}u^{\mu }u^{\nu }u^{\alpha }u^{\beta
}u^{\kappa }u^{\lambda }+z_{11}u^{\mu }u^{\nu }\Delta ^{\alpha \beta
}u^{\kappa }u^{\lambda }+z_{11}^{\prime }u^{\mu }u^{\nu }\Delta ^{\kappa
\lambda }u^{\alpha }u^{\beta }+z_{21}u^{\alpha }u^{\beta }\Delta ^{\mu \nu
}u^{\kappa }u^{\lambda }  \notag \\
& +4z_{31}u^{\left( \mu \right. }\Delta ^{\left. \nu \right) \left( \alpha
\right. }u^{\left. \beta \right) }u^{\kappa }u^{\lambda }+4z_{31}^{\prime
}u^{\left( \mu \right. }\Delta ^{\left. \nu \right) \left( \kappa \right.
}u^{\left. \lambda \right) }u^{\alpha }u^{\beta }+4z_{41}u^{\mu }u^{\nu
}u^{\left( \alpha \right. }\Delta ^{\left. \beta \right) \left( \kappa
\right. }u^{\left. \lambda \right) }  \notag \\
& +z_{12}\Delta ^{\mu \nu }\Delta ^{\alpha \beta }u^{\kappa }u^{\lambda
}+z_{12}^{\prime }\Delta ^{\mu \nu }\Delta ^{\kappa \lambda }u^{\alpha
}u^{\beta }+2z_{22}\Delta ^{\mu \left( \alpha \right. }\Delta ^{\left. \beta
\right) \nu }u^{\kappa }u^{\lambda }+2z_{22}^{\prime }\Delta ^{\mu \left(
\kappa \right. }\Delta ^{\left. \lambda \right) \nu }u^{\alpha }u^{\beta } 
\notag \\
& +4z_{32}\Delta ^{\mu \nu }u^{\left( \alpha \right. }\Delta ^{\left. \beta
\right) \left( \kappa \right. }u^{\left. \lambda \right) }+4z_{42}\Delta
^{\alpha \beta }u^{\left( \mu \right. }\Delta ^{\left. \nu \right) \left(
\kappa \right. }u^{\left. \lambda \right) }+4z_{42}^{\prime }\Delta ^{\kappa
\lambda }u^{\left( \mu \right. }\Delta ^{\left. \nu \right) \left( \alpha
\right. }u^{\left. \beta \right) }  \notag \\
& +8z_{52}u^{\left( \kappa \right. }\Delta ^{\left. \lambda \right) \left(
\mu \right. }\Delta ^{\left. \nu \right) \left( \alpha \right. }u^{\left.
\beta \right) }+8z_{62}u^{\left( \mu \right. }\Delta ^{\left. \nu \right)
\left( \alpha \right. }\Delta ^{\left. \beta \right) \left( \kappa \right.
}u^{\left. \lambda \right) }+8z_{62}^{\prime }u^{\left( \mu \right. }\Delta
^{\left. \nu \right) \left( \kappa \right. }\Delta ^{\left. \lambda \right)
\left( \alpha \right. }u^{\left. \beta \right) }  \notag \\
& +2z_{72}u^{\mu }u^{\nu }\Delta ^{\kappa \left( \alpha \right. }\Delta
^{\left. \beta \right) \lambda }+z_{82}u^{\mu }u^{\nu }\Delta ^{\alpha \beta
}\Delta ^{\kappa \lambda }  \notag \\
& +z_{13}\Delta ^{\mu \nu }\Delta ^{\alpha \beta }\Delta ^{\kappa \lambda
}+8z_{23}\Delta ^{\left( \kappa \right. \left( \mu \right. }\Delta ^{\left.
\nu \right) \left( \alpha \right. }\Delta ^{\left. \beta \right) \left.
\lambda \right) }+2z_{33}\Delta ^{\alpha \beta }\Delta ^{\mu \left( \kappa
\right. }\Delta ^{\left. \lambda \right) \nu }  \notag \\
& +2z_{33}^{\prime }\Delta ^{\kappa \lambda }\Delta ^{\mu \left( \alpha
\right. }\Delta ^{\left. \beta \right) \nu }+2z_{43}\Delta ^{\mu \nu }\Delta
^{\alpha \left( \kappa \right. }\Delta ^{\left. \lambda \right) \beta }.
\end{align}%
There are only a few coefficients which we will need to recall later, these
are 
\begin{align}
Y_{4\left( r\right) ,3}& \equiv w_{31}=\frac{1}{3}Y_{4\left( r\right) }^{\mu
\nu \alpha \beta \kappa \lambda }u_{\left( \mu \right. }\Delta _{\left. \nu
\right) \left( \alpha \right. }u_{\left. \beta \right) }u_{\kappa
}u_{\lambda } =\frac{1}{3}Y_{4\left( r\right) }^{\mu \nu \alpha \beta \kappa
\lambda }u_{\left( \mu \right. }\Delta _{\left. \nu \right) \left( \kappa
\right. }u_{\left. \lambda \right) }u_{\alpha }u_{\beta }, \\
Y_{4\left( r\right) ,4}& \equiv 2w_{22}=\frac{1}{5}Y_{4\left( r\right)
}^{\mu \nu \alpha \beta \kappa \lambda }\Delta _{\mu \nu \alpha \beta
}u_{\kappa }u_{\lambda }, \\
Y_{4\left( r\right) ,8}& \equiv w_{62}=\frac{1}{30}Y_{4\left( r\right)
}^{\mu \nu \alpha \beta \kappa \lambda }\left[ 3u_{\left( \mu \right.
}\Delta _{\left. \nu \right) \left( \kappa \right. }\Delta _{\left. \lambda
\right) \left( \alpha \right. }u_{\left. \beta \right) }-\Delta _{\kappa
\lambda }u_{\left( \mu \right. }\Delta _{\left. \nu \right) \left( \alpha
\right. }u_{\left. \beta \right) }\right] ,  \notag \\
& =\frac{1}{30}Y_{4\left( r\right) }^{\mu \nu \alpha \beta \kappa \lambda }%
\left[ 3u_{\left( \mu \right. }\Delta _{\left. \nu \right) \left( \alpha
\right. }\Delta _{\left. \beta \right) \left( \kappa \right. }u_{\left.
\lambda \right) }-\Delta _{\alpha \beta }u_{\left( \mu \right. }\Delta
_{\left. \nu \right) \left( \kappa \right. }u_{\left. \lambda \right) }%
\right] ,
\end{align}%
and 
\begin{align}
Y_{5\left( r\right) ,1}& \equiv z_{10}=Y_{5\left( r\right) }^{\mu \nu \alpha
\beta \kappa \lambda }u_{\mu }u_{\nu }u_{\alpha }u_{\beta }u_{\kappa
}u_{\lambda }, \\
Y_{5\left( r\right) ,5}& \equiv z_{41}=\frac{1}{3}Y_{5\left( r\right) }^{\mu
\nu \alpha \beta \kappa \lambda }u_{\mu }u_{\nu }u_{\left( \alpha \right.
}\Delta _{\left. \beta \right) \left( \kappa \right. }u_{\left. \lambda
\right) }, \\
Y_{5\left( r\right) ,7}& \equiv z_{52}=\frac{1}{30}Y_{5\left( r\right)
}^{\mu \nu \alpha \beta \kappa \lambda }\left[ 3u_{\left( \kappa \right.
}\Delta _{\left. \lambda \right) \left( \mu \right. }\Delta _{\left. \nu
\right) \left( \alpha \right. }u_{\left. \beta \right) }-\Delta _{\mu \nu
}u_{\left( \alpha \right. }\Delta _{\left. \beta \right) \left( \kappa
\right. }u_{\left. \lambda \right) }\right] , \\
Y_{5\left( r\right) ,9}& \equiv z_{72}=\frac{1}{30}Y_{5\left( r\right)
}^{\mu \nu \alpha \beta \kappa \lambda }\left[ 3\Delta _{\kappa \left(
\alpha \right. }\Delta _{\left. \beta \right) \lambda }u_{\mu }u_{\nu
}-\Delta _{\alpha \beta }\Delta _{\kappa \lambda }u_{\mu }u_{\nu }\right] =%
\frac{1}{10}Y_{5\left( r\right) }^{\mu \nu \alpha \beta \kappa \lambda
}u_{\mu }u_{\nu }\Delta _{\alpha \beta \kappa \lambda }, \\
Y_{5\left( r\right) ,11}& \equiv z_{23}=\frac{1}{210}Y_{5\left( r\right)
}^{\mu \nu \alpha \beta \kappa \lambda }\left[ 2\Delta _{\mu \nu }\Delta
_{\alpha \beta }\Delta _{\kappa \lambda }+9\Delta _{\left( \kappa \right.
\left( \mu \right. }\Delta _{\left. \nu \right) \left( \alpha \right.
}\Delta _{\left. \beta \right) \left. \lambda \right) }\right]  \notag \\
& \ \ \ \ \ \ \ \ \ \ -\frac{1}{70}Y_{5\left( r\right) }^{\mu \nu \alpha
\beta \kappa \lambda }\left[ \Delta _{\alpha \beta }\Delta _{\mu \left(
\kappa \right. }\Delta _{\left. \lambda \right) \nu }+\Delta _{\mu \nu
}\Delta _{\alpha \left( \kappa \right. }\Delta _{\left. \lambda \right)
\beta }+\Delta _{\kappa \lambda }\Delta _{\mu \left( \alpha \right. }\Delta
_{\left. \beta \right) \nu }\right] .
\end{align}

It is easy to realize that in case the particles have the same mass, $%
Y_{4\left( r\right) }^{\mu \nu \alpha \beta \kappa \lambda }g_{\kappa
\lambda }=m^{2}X_{\left( r\right) }^{\mu \nu \alpha \beta }$, $Y_{4\left(
r\right) }^{\mu \nu \alpha \beta \kappa \lambda }g_{\beta \lambda
}=m^{2}Y_{1\left( r\right) }^{\mu \nu \alpha \kappa }$, also\ $Y_{5\left(
r\right) }^{\mu \nu \alpha \beta \kappa \lambda }g_{\beta \lambda
}=Y_{2\left( r\right) }^{\mu \nu \alpha \kappa }$, and so on. This means
that we only have rank-5 and -6 tensors to evaluate, $Y_{3\left( r\right)
}^{\mu \nu \alpha \beta \kappa }$, $Y_{4\left( r\right) }^{\mu \nu \alpha
\beta \kappa \lambda }$ $Y_{5\left( r\right) }^{\mu \nu \alpha \beta \kappa
\lambda }$. However, since it is much simpler to decompose and project
rank-4 tensors and then use them to check the results through the trace
relations for the rank-6 tensors, we will keep using all of them.


\section{Collision integrals in the Boltzmann limit}

\label{Coll_BL}

In this appendix we calculate the previously defined collision tensors Eqs.\
(\ref{X_4rank_tensor}) -- (\ref{Y5_6rank_tens}) in the massless Boltzmann
limit. These tensors are based on the collision integral $C\left[ f\right] $
defined in Eq.\ (\ref{COLL_INT}), where the Lorentz-invariant transition
rate, $W_{\mathbf{kk}\prime \rightarrow \mathbf{pp}\prime }$, may only
depend on the following collision invariant 
\begin{equation}
s\equiv \left( k^{\mu }+k^{\prime \mu }\right) ^{2}=\left( p^{\mu
}+p^{\prime \mu }\right) ^{2}.
\end{equation}%
In the center of mass (CM) frame, where the sum of 3-momenta vanishes $%
\mathbf{k+k}^{\prime }=\mathbf{p+p}^{\prime }=0$ and so $k^{0}=k^{\prime
0}=p^{0}=p^{\prime 0}$, the scattering angle $\theta _{s}$ is given with the
help of another collision invariant, $t\equiv \left( k^{\mu }-p^{\mu
}\right) ^{2}=\left( k^{\prime \mu }-p^{\prime \mu }\right) ^{2}$, such that%
\begin{equation}
\cos \theta _{s}\equiv 1+\frac{2t}{s-4m^{2}}=\frac{\left( k^{\mu }-k^{\prime
\mu }\right) \left( p_{\mu }-p_{\mu }^{\prime }\right) }{\left( k^{\mu
}-k^{\prime \mu }\right) ^{2}}.
\end{equation}%
The above defined collision invariants together with $u\equiv \left( k^{\mu
}-p^{\prime \mu }\right) ^{2}=\left( k^{\prime \mu }-p^{\mu }\right) ^{2}$
are the so-called Mandelstam variables, satisfying $s+t+u=4m^{2}$, where $m$
is the mass of the particles.

Therefore, the total momentum involved in a binary collision$\ P_{T}^{\mu
}\equiv k^{\mu }+k^{\prime \mu }=$ $p^{\mu }+p^{\prime \mu }$ is related to
the total energy $P_{T}^{\mu }=\left( \sqrt{s},0,0,0\right) $, where $%
s\equiv 2\left( m^{2}+k^{\mu }k_{\mu }^{\prime }\right) =\left(
2k^{0}\right) ^{2}$, in the CM frame. The Lorentz-invariant transition rate
is defined as 
\begin{equation}
g^{2}W_{\mathbf{kk}\prime \rightarrow \mathbf{pp}\prime }=\left( 2\pi
\right) ^{6}s\sigma \left( s,\theta _{s}\right) \delta ^{4}\left( k^{\mu
}+k^{\prime \mu }-p^{\mu }-p^{\prime \mu }\right) .
\end{equation}%
The quantity $\sigma \left( s,\theta _{CM}\right) $ is the differential
cross section and $\delta ^{4}(k^{\mu }+k^{\prime \mu }-p^{\mu }-p^{\prime
\mu })$ enforces energy and momentum conservation in binary collisions.
Furthermore, we also define the total cross section $\sigma _{T}\left(
s\right) \ $as the integral of the differential cross section over the solid
angle, $d\Omega =\int_{0}^{2\pi }d\varphi \int_{0}^{\pi }\sin \theta
_{s}d\theta _{s}$, 
\begin{equation}
\sigma _{T}\left( s\right) =\frac{1}{\nu }\int_{0}^{\pi }d\Omega \ \sigma
\left( s,\theta _{s}\right) .
\end{equation}

For later purposes, let us define the rank-$n$ tensor 
\begin{align}
\Theta ^{\mu _{1}\cdots \mu _{n}}& \equiv \frac{1}{\nu }\int \frac{d^{3}%
\mathbf{p}}{p^{0}}\frac{d^{3}\mathbf{p}^{\prime }}{p^{\prime 0}}s\sigma
\left( s,\theta _{s}\right) p^{\mu _{1}}\cdots p^{\mu _{n}}\delta ^{4}\left(
k^{\mu }+k^{\prime \mu }-p^{\mu }-p^{\prime \mu }\right) ,  \notag \\
& =\sum_{q=0}^{\left[ n/2\right] }\left( -1\right) ^{q} b_{nq}\mathcal{B}%
_{nq}\Delta _{P_{T}}^{\left( \mu _{1}\mu _{2}\right. }\cdots \Delta
_{P_{T}}^{\mu _{2q-1}\mu _{_{2q}}}\,P_{T}^{\mu _{2q+1}}\cdots P_{T}^{\left.
\mu _{n}\right) },  \label{Theta_1_n}
\end{align}%
where we assumed that the differential cross section is isotropic, i.e., $%
\sigma \left( s,\theta _{s}\right) =\sigma \left( s\right) $.\ The $b_{nq}$
coefficients are given in Eq.\ (\ref{a_nq}) and 
\begin{align}
\mathcal{B}_{nq}& \equiv \frac{\left( -1\right) ^{q}}{\left( 2q+1\right) !!}%
\frac{1}{\nu }\int \frac{d^{3}\mathbf{p}}{p^{0}}\frac{d^{3}\mathbf{p}%
^{\prime }}{p^{\prime 0}}s^{-\left( n-2q\right) /2}\sigma \left( s,\theta
_{s}\right) \left( P_{T}^{\mu }p_{\mu }\right) ^{n-2q}\left( \Delta
_{P_{T}}^{\alpha \beta }p_{\alpha }p_{\beta }\right) ^{q},  \notag \\
& =\frac{1}{\left( 2q+1\right) !!}\frac{\sigma _{T}\left( s\right) }{2^{n+1}}%
s^{1/2}\left( s-4m^{2}\right) ^{\left( 2q+1\right) /2}.
\end{align}%
Here we defined the projection orthogonal to the total momentum, $\Delta
_{P}^{\alpha \beta }=g^{\mu \nu }-P_{T}^{\mu }P_{T}^{\nu }/s$, hence the
particle 4-momentum can be decomposed as $p^{\mu }=P_{T}^{\mu }\left(
P_{T}^{\alpha }p_{\alpha }\right) /s+p_{\alpha }\Delta _{P_{T}}^{\mu \alpha
} $ and thus $\Delta _{P_{T}}^{\alpha \beta }p_{\alpha }p_{\beta }=m^{2}-s/4$%
. Similarly as in the previous integral we define a slightly different rank-$%
n$ tensor $\Gamma _{(n,m)}^{\mu _{1}\cdots \mu _{n-m}\nu _{n-m+1}\cdots \nu
_{n}}$ 
\begin{align}
\Gamma _{(n,m)}^{\mu _{1}\cdots \mu _{n-m}\nu _{n-m+1}\cdots \nu _{n}}&
\equiv \frac{1}{\nu }\int \frac{d^{3}\mathbf{p}}{p^{0}}\frac{d^{3}\mathbf{p}%
^{\prime }}{p^{\prime 0}}s\sigma \left( s,\theta _{CM}\right) p^{\mu
_{1}}\cdots p^{\mu _{n-m}}p^{\prime \nu _{n-m+1}}\cdots p^{\prime \nu
_{n}}\delta ^{4}\left( k^{\mu }+k^{\prime \mu }-p^{\mu }-p^{\prime \mu
}\right) ,  \notag \\
& =\left( -1\right) ^{n-m}\Theta ^{\mu _{1}\cdots \mu _{n-m}\nu
_{n-m+1}\cdots \nu _{n}}+\left( -1\right) ^{n-m-1}P_{T}^{\nu _{n}}\Theta
^{\mu _{1}\cdots \mu _{n-m}\nu _{n-m+1}\cdots \nu _{n-1}}  \notag \\
& +\cdots +\ P_{T}^{_{\nu _{n-m+1}}}\cdots P_{T}^{\nu _{n}}\Theta ^{\mu
_{1}\cdots \mu _{n-m}},
\end{align}%
where we replaced $p^{\prime \nu }=P_{T}^{\nu }-p^{\nu }$ to obtain the
above result.

\section{Coefficients in the massless limit}

\label{tensors_massless_limit}

Using the results from Appendix \ref{Coll_BL} we can express the collision
integrals from Eqs.\ (\ref{X_4rank_tensor}) -- (\ref{Y5_6rank_tens}) as a
function of particle momenta and the metric tensor. As an example, here we
discuss the $X_{\left( r\right) }^{\mu \nu \alpha \beta }$ tensor, which we
separate into gain $\mathcal{G}_{\left( r\right) }^{\mu \nu \alpha \beta }$%
and loss $\mathcal{L}_{\left( r\right) }^{\mu \nu \alpha \beta }\ $parts as $%
X_{\left( r\right) }^{\mu \nu \alpha \beta }=\mathcal{G}_{\left( r\right)
}^{\mu \nu \alpha \beta }-\mathcal{L}_{\left( r\right) }^{\mu \nu \alpha
\beta }$, where%
\begin{align}
\mathcal{G}_{\left( r\right) }^{\mu \nu \alpha \beta }& \equiv \frac{1}{\nu }%
\int dKdK^{\prime }dPdP^{\prime }f_{0\mathbf{k}}f_{0\mathbf{k}\prime }W_{%
\mathbf{kk}\prime \rightarrow \mathbf{pp}\prime }E_{\mathbf{k}}^{r}k^{\mu
}k^{\nu }\left( p^{\alpha }p^{\beta }+p^{\prime \alpha }p^{\prime \beta
}\right) ,  \notag \\
& =2\int dKdK^{\prime }f_{0\mathbf{k}}f_{0\mathbf{k}\prime }E_{\mathbf{k}%
}^{r}k^{\mu }k^{\nu }\Theta ^{\alpha \beta },
\end{align}%
and 
\begin{align}
\mathcal{L}_{\left( r\right) }^{\mu \nu \alpha \beta }& \equiv \frac{1}{\nu }%
\int dKdK^{\prime }dPdP^{\prime }f_{0\mathbf{k}}f_{0\mathbf{k}\prime }W_{%
\mathbf{kk}\prime \rightarrow \mathbf{pp}\prime }E_{\mathbf{k}}^{r}k^{\mu
}k^{\nu }\left( k^{\alpha }k^{\beta }+k^{\prime \alpha }k^{\prime \beta
}\right) ,  \notag \\
& =\int dKdK^{\prime }f_{0\mathbf{k}}f_{0\mathbf{k}^{\prime }}E_{\mathbf{k}%
}^{r}k^{\mu }k^{\nu }\left( k^{\alpha }k^{\beta }+k^{\prime \alpha
}k^{\prime \beta }\right) \Theta .
\end{align}%
Here $\Theta \equiv \frac{\sigma _{T}}{2}\sqrt{s\left( s-4m^{2}\right) }%
=\sigma _{T}\sqrt{\left( k^{\mu }k_{\mu }^{\prime }\right) ^{2}-m^{4}}$ is
the invariant flux and $\Theta ^{\mu \nu }=\frac{\Theta }{4}\left[
P_{T}^{\mu }P_{T}^{\nu }-\frac{1}{3}\left( s-4m^{2}\right) \Delta
_{P_{T}}^{\mu \nu }\right] $, cf.\ Eq.\ (\ref{Theta_1_n}). The invariant flux%
$\ $can be expressed using the relative velocity between particles, $\Theta
=v_{rel}k^{\mu }k_{\mu }^{\prime }$ where $v_{rel}=\left( 1-\mathbf{v\cdot v}%
^{\prime }\right) ^{-1}\sqrt{\left( \mathbf{v}-\mathbf{v}^{\prime }\right)
^{2}-\left( \mathbf{v\times v}^{\prime }\right) ^{2}}$ and $\mathbf{v}%
=k^{\mu }/k^{0}$, $\mathbf{v}^{\prime }=k^{\prime \mu }/k^{\prime 0}$. Hence
the total cross section is $\sigma _{T}=v_{rel}\sigma _{T}\left( s\right) $.
However, in the massless limit, $m\rightarrow 0$, the relative velocity is $%
v_{rel}=1$.

From now on we will work in the massless limit $m\rightarrow 0$. For example
the solution for $\mathcal{G}_{\left( r\right) }^{\mu \nu \alpha \beta }$
can be written formally as $\mathcal{G}_{\left( r\right) }^{\mu \nu \alpha
\beta }=\int \left( A_{X}-B_{X}/s\right) P_{T}^{\mu }P_{T}^{\nu
}+B_{X}g^{\mu \nu }$. After finding the coefficients $A_{X}$ and $B_{X}$ we
replace $P_{T}^{\mu }=k^{\mu }+k^{\prime \mu }$. Thus using the
above definitions and relations we obtain 
\begin{eqnarray}
\mathcal{G}_{\left( r\right) }^{\mu \nu \alpha \beta } &=&\frac{2\sigma _{T}%
}{3}I_{\left( r+5\right) }^{\mu \nu \alpha \beta \kappa }I_{\left( 1\right)
\kappa }+\frac{4\sigma _{T}}{3}I_{\left( r+4\right) }^{\mu \nu \kappa \left(
\alpha \right. }I_{\left( 2\right) \kappa }^{\left. \beta \right) }+\frac{%
2\sigma _{T}}{3}I_{\left( r+3\right) }^{\mu \nu \kappa }I_{\left( 3\right)
\kappa }^{\alpha \beta }-\frac{\sigma _{T}}{3}g^{\alpha \beta }I_{\left(
r+4\right) }^{\mu \nu \kappa \lambda }I_{\left( 2\right) \kappa \lambda }, \\
\mathcal{L}_{\left( r\right) }^{\mu \nu \alpha \beta } &=&\sigma
_{T}I_{\left( r+5\right) }^{\mu \nu \alpha \beta \kappa }I_{\left( 1\right)
\kappa }+\sigma _{T}I_{\left( r+3\right) }^{\mu \nu \kappa }I_{\left(
3\right) \kappa }^{\alpha \beta },
\end{eqnarray}%
and hence,%
\begin{equation}
X_{\left( r\right) }^{\mu \nu \alpha \beta }=-\frac{\sigma _{T}}{3}I_{\left(
r+5\right) }^{\mu \nu \alpha \beta \kappa }I_{\left( 1\right) \kappa }+\frac{%
4\sigma _{T}}{3}I_{\left( r+4\right) }^{\mu \nu \kappa \left( \alpha \right.
}I_{\left( 2\right) \kappa }^{\left. \beta \right) }-\frac{\sigma _{T}}{3}%
I_{\left( r+3\right) }^{\mu \nu \kappa }I_{\left( 3\right) \kappa }^{\alpha
\beta }-\frac{\sigma _{T}}{3}g^{\alpha \beta }I_{\left( r+4\right) }^{\mu
\nu \kappa \lambda }I_{\left( 2\right) \kappa \lambda },
\end{equation}%
where the $I_{\left( r+n\right) }^{\mu _{1}\cdots \mu _{n}}$ tensors were
defined in Eq.\ (\ref{I_n_moment}).

Similarly, we obtain for the next rank-4 collision tensor 
\begin{equation}
Y_{1\left( r\right) }^{\mu \nu \alpha \beta }=-X_{\left( r\right) }^{\mu \nu
\alpha \beta }.
\end{equation}%
The last rank-4 collision tensor we need is 
\begin{equation}
Y_{2\left( r\right) }^{\mu \nu \alpha \beta }=\frac{\sigma _{T}}{6}I_{\left(
r+5\right) }^{\mu \nu \alpha \beta \kappa }I_{\left( 1\right) \kappa }+\frac{%
\sigma _{T}}{3}I_{\left( r+4\right) }^{\mu \nu \kappa \left( \alpha \right.
}I_{\left( 2\right) \kappa }^{\left. \beta \right) }+\frac{\sigma _{T}}{6}%
I_{\left( r+3\right) }^{\mu \nu \kappa }I_{\left( 3\right) \kappa }^{\alpha
\beta }-\sigma _{T}I_{\left( r+4\right) }^{\mu \nu \kappa \alpha }I_{\left(
2\right) \kappa }^{\beta }+\frac{\sigma _{T}}{6}g^{\alpha \beta }I_{\left(
r+4\right) }^{\mu \nu \kappa \lambda }I_{\left( 2\right) \kappa \lambda }.
\end{equation}%
The only rank-5 tensor we have is 
\begin{align}
Y_{3\left( r\right) }^{\left( \mu \nu \right) \alpha \left( \beta \kappa
\right) }& =\frac{\sigma _{T}}{6}\left[ I_{\left( r+6\right) }^{\mu \nu
\alpha \beta \kappa \lambda }I_{\left( 1\right) \lambda }+2I_{\left(
r+5\right) }^{\mu \nu \lambda \alpha \left( \beta \right. }I_{\left(
2\right) \lambda }^{\left. \kappa \right) }+I_{\left( r+5\right) }^{\mu \nu
\lambda \kappa \beta }I_{\left( 2\right) \lambda }^{\alpha }\right] +\frac{%
\sigma _{T}}{6}\left[ 2I_{\left( r+4\right) }^{\mu \nu \lambda \left( \beta
\right. }I_{\left( 3\right) \lambda }^{\left. \kappa \right) \alpha
}+I_{\left( r+4\right) }^{\mu \nu \lambda \alpha }I_{\left( 3\right) \lambda
}^{\beta \kappa }+I_{\left( r+3\right) }^{\mu \nu \lambda }I_{\left(
4\right) \lambda }^{\alpha \beta \kappa }\right]  \notag \\
& +\frac{\sigma _{T}}{6}2g^{\alpha \left( \beta \right. }\left[ I_{\left(
r+5\right) }^{\left. \kappa \right) \mu \nu \lambda \sigma }I_{\left(
2\right) \lambda \sigma }+I_{\left( 3\right) \lambda \sigma }^{\left. \kappa
\right) }I_{\left( r+4\right) }^{\mu \nu \lambda \sigma }\right] -\frac{%
\sigma _{T}}{6}g^{\beta \kappa }\left[ I_{\left( r+5\right) }^{\alpha \mu
\nu \lambda \sigma }I_{\left( 2\right) \lambda \sigma }+I_{\left( 3\right)
\lambda \sigma }^{\alpha }I_{\left( r+4\right) }^{\mu \nu \lambda \sigma }%
\right]  \notag \\
& -\sigma _{T}\left[ I_{\left( r+4\right) }^{\mu \nu \alpha \lambda
}I_{\left( 3\right) \lambda }^{\beta \kappa }+I_{\left( r+5\right) }^{\mu
\nu \beta \kappa \lambda }I_{\left( 2\right) \lambda }^{\alpha }\right] ,
\end{align}%
where the solution $Y_{3\left( r\right) }^{\left( \mu \nu \right)
\alpha \left( \beta \kappa \right) }\sim \int \left(
A_{3}-2B_{3}/s-C_{3}/s\right) P_{T}^{\alpha }P_{T}^{\beta }P_{T}^{\kappa
}+2B_{3}g^{\alpha (\beta }P_{T}^{\kappa )}+C_{3}g^{\beta \kappa
}P_{T}^{\alpha }$.

The rank-6 tensors are%
\begin{align}
Y_{4\left( r\right) }^{\mu \nu \alpha \beta \kappa \lambda }& =\frac{\sigma
_{T}}{15}\left[ I_{\left( r+7\right) }^{\mu \nu \alpha \beta \kappa \lambda
\sigma }I_{\left( 1\right) \sigma }+4I_{\left( r+6\right) }^{\mu \nu \sigma
\left( \alpha \beta \kappa \right. }I_{\left( 2\right) \sigma }^{\left.
\lambda \right) }+6I_{\left( r+5\right) }^{\mu \nu \sigma \left( \alpha
\beta \right. }I_{\left( 3\right) \sigma }^{\left. \kappa \lambda \right)
}+4I_{\left( r+4\right) }^{\mu \nu \sigma \left( \alpha \right. }I_{\left(
4\right) \sigma }^{\left. \beta \kappa \lambda \right) }+I_{\left(
r+3\right) }^{\mu \nu \sigma }I_{\left( 5\right) \sigma }^{\alpha \beta
\kappa \lambda }\right]  \notag \\
& +\frac{\sigma _{T}}{15}2g^{\beta \left( \kappa \right. }\left[ I_{\left(
r+6\right) }^{\left. \lambda \right) \alpha \mu \nu \sigma \rho }I_{\left(
2\right) \sigma \rho }+I_{\left( 3\right) \sigma \rho }^{\left. \lambda
\right) }I_{\left( r+5\right) }^{\alpha \mu \nu \sigma \rho }+I_{\left(
r+5\right) }^{\left. \lambda \right) \mu \nu \sigma \rho }I_{\left( 3\right)
\sigma \rho }^{\alpha }+I_{\left( 4\right) \sigma \rho }^{\left. \lambda
\right) \alpha }I_{\left( r+4\right) }^{\mu \nu \sigma \rho }\right]  \notag
\\
& +\frac{\sigma _{T}}{15}2g^{\alpha \left( \kappa \right. }\left[ I_{\left(
r+6\right) }^{\left. \lambda \right) \beta \mu \nu \sigma \rho }I_{\left(
2\right) \sigma \rho }+I_{\left( 3\right) \sigma \rho }^{\left. \lambda
\right) }I_{\left( r+5\right) }^{\beta \mu \nu \sigma \rho }+I_{\left(
r+5\right) }^{\left. \lambda \right) \mu \nu \sigma \rho }I_{\left( 3\right)
\sigma \rho }^{\beta }+I_{\left( 4\right) \sigma \rho }^{\left. \lambda
\right) \beta }I_{\left( r+4\right) }^{\mu \nu \sigma \rho }\right]  \notag
\\
& +\frac{\sigma _{T}}{30}\left[ 2g^{\alpha \left( \kappa \right. }g^{\left.
\lambda \right) \beta }+g^{\alpha \beta }g^{\kappa \lambda }\right]
I_{\left( r+5\right) }^{\mu \nu \sigma \rho \tau }I_{\left( 3\right) \sigma
\rho \tau }  \notag \\
& -\frac{\sigma _{T}}{10}g^{\alpha \beta }\left[ I_{\left( r+6\right)
}^{\kappa \lambda \mu \nu \sigma \rho }I_{\left( 2\right) \sigma \rho
}+I_{\left( 3\right) \sigma \rho }^{\lambda }I_{\left( r+5\right) }^{\kappa
\mu \nu \sigma \rho }+I_{\left( 3\right) \sigma \rho }^{\kappa }I_{\left(
r+5\right) }^{\lambda \mu \nu \sigma \rho }+I_{\left( 4\right) \sigma \rho
}^{\kappa \lambda }I_{\left( r+4\right) }^{\mu \nu \sigma \rho }\right] 
\notag \\
& -\frac{\sigma _{T}}{10}g^{\kappa \lambda }\left[ I_{\left( r+6\right)
}^{\alpha \beta \mu \nu \sigma \rho }I_{\left( 2\right) \sigma \rho
}+I_{\left( 3\right) \sigma \rho }^{\beta }I_{\left( r+5\right) }^{\alpha
\mu \nu \sigma \rho }+I_{\left( 3\right) \sigma \rho }^{\alpha }I_{\left(
r+5\right) }^{\beta \mu \nu \sigma \rho }+I_{\left( 4\right) \sigma \rho
}^{\alpha \beta }I_{\left( r+4\right) }^{\mu \nu \sigma \rho }\right]  \notag
\\
& -\sigma _{T}I_{\left( r+5\right) }^{\mu \nu \alpha \beta \sigma }I_{\left(
3\right) \sigma }^{\kappa \lambda }-\sigma _{T}I_{\left( r+5\right) }^{\mu
\nu \kappa \lambda \sigma }I_{\left( 3\right) \sigma }^{\alpha \beta },
\end{align}%
and%
\begin{equation}
Y_{5\left( r\right) }^{\mu \nu \alpha \beta \kappa \lambda }=\frac{1}{2}%
\left[ Y_{4\left( r\right) }^{\mu \nu \alpha \beta \kappa \lambda }-\sigma
_{T}I_{\left( r+5\right) }^{\mu \nu \alpha \beta \sigma }I_{\left( 3\right)
\sigma }^{\kappa \lambda }+\sigma _{T}I_{\left( r+5\right) }^{\mu \nu \kappa
\lambda \sigma }I_{\left( 3\right) \sigma }^{\alpha \beta }\right] .
\end{equation}%
Here the solution has the form $Y^{\mu \nu \alpha \beta \kappa \lambda }\sim
\int A_{Y}P_{T}^{\alpha }P_{T}^{\beta }P_{T}^{\kappa }P_{T}^{\lambda
}+2B_{Y}\left( P_{T}^{\alpha }g^{\beta (\kappa }P^{\lambda )}+P_{T}^{\beta
}g^{\alpha (\kappa }P^{\lambda )}\right) +2C_{Y}g^{\alpha (\kappa
}g^{\lambda )\beta }+D_{Y}g^{\alpha \beta }P_{T}^{\lambda }P_{T}^{\kappa
}+E_{Y}g^{\kappa \lambda }P_{T}^{\alpha }P_{T}^{\beta }+F_{Y}g^{\alpha \beta
}g^{\kappa \lambda }$.

In order to calculate the above collision tensors, we need the following $%
I_{\left( r+n\right) }^{\mu _{1}\cdots \mu _{n}}$ tensors from Eq.\ (\ref%
{I_n_moment}),%
\begin{align}
I_{\left( r+1\right) }^{\mu }& =I_{r+1,0}u^{\mu }, \\
I_{\left( r+2\right) }^{\mu \nu }& =I_{r+2,0}u^{\mu }u^{\nu
}-I_{r+2,1}\Delta ^{\mu \nu }, \\
I_{\left( r+3\right) }^{\mu \nu \alpha }& =I_{r+3,0}u^{\mu }u^{\nu
}u^{\alpha }-3I_{r+3,1}u^{\left( \mu \right. }\Delta ^{\left. \nu \alpha
\right) }\,, \\
I_{\left( r+4\right) }^{\mu \nu \alpha \beta }& =I_{r+4,0}u^{\mu }u^{\nu
}u^{\alpha }u^{\beta }-6I_{r+4,1}u^{\left( \mu \right. }u^{\nu }\Delta
^{\left. \alpha \beta \right) }+3I_{r+4,2}\Delta ^{\left( \mu \nu \right.
}\Delta ^{\left. \alpha \beta \right) }, \\
I_{\left( r+5\right) }^{\mu \nu \alpha \beta \kappa }& =I_{r+5,0}u^{\mu
}u^{\nu }u^{\alpha }u^{\beta }u^{\kappa }-10I_{r+5,1}u^{\left( \mu \right.
}u^{\nu }u^{\alpha }\Delta ^{\left. \beta \kappa \right)
}+15I_{r+5,2}u^{\left( \mu \right. }\Delta ^{\nu \alpha }\Delta ^{\left.
\beta \kappa \right) }, \\
I_{\left( r+6\right) }^{\mu \nu \alpha \beta \kappa \lambda }&
=I_{r+6,0}u^{\mu }u^{\nu }u^{\alpha }u^{\beta }u^{\kappa }u^{\lambda
}-15I_{r+6,1}u^{\left( \mu \right. }u^{\nu }u^{\alpha }u^{\beta }\Delta
^{\left. \kappa \lambda \right) }  \notag \\
& +45I_{r+6,2}u^{\left( \mu \right. }u^{\nu }\Delta ^{\alpha \beta }\Delta
^{\left. \kappa \lambda \right) }-15I_{r+6,3}\Delta ^{\left( \mu \nu \right.
}\Delta ^{\alpha \beta }\Delta ^{\left. \kappa \lambda \right) }, \\
I_{\left( r+7\right) }^{\mu \nu \alpha \beta \kappa \lambda \tau }&
=I_{r+7,0}u^{\mu }u^{\nu }u^{\alpha }u^{\beta }u^{\kappa }u^{\lambda
}u^{\tau }-21I_{r+7,1}u^{\left( \mu \right. }u^{\nu }u^{\alpha }u^{\beta
}u^{\kappa }\Delta ^{\left. \lambda \tau \right) }  \notag \\
& +105I_{r+7,2}u^{\left( \mu \right. }u^{\nu }u^{\alpha }\Delta ^{\beta
\kappa }\Delta ^{\left. \lambda \tau \right) }-105I_{r+7,3}\Delta ^{\left(
\mu \nu \right. }\Delta ^{\alpha \beta }\Delta ^{\kappa \lambda }u^{\left.
\tau \right) }.
\end{align}%
In the massless Boltzmann limit, we use the following formula for the
thermodynamic integrals, 
\begin{equation}
I_{n+r,q}\left( \alpha _{0},\beta _{0}\right) \underset{m\rightarrow 0}{=}%
\frac{P_{0}\left( r+n+1\right) !}{2\beta _{0}^{r+n-2}(2q+1)!!},
\end{equation}%
where $P_{0}=ge^{\alpha _{0}}\beta _{0}^{-4}/\pi ^{2}$, and we extensively
use the recursion relation (\ref{I_nq_recursive}) in the massless limit,
i.e., $I_{r+n,q}=\left( 2q+3\right) I_{r+n,q+1}$.

The coefficients for the linear part of the collision integral are%
\begin{align}
X_{\left( r\right) ,1}& \equiv x_{10}=-\frac{\sigma _{T}}{3}\left[
I_{r+5,0}I_{1,0}+I_{r+3,0}I_{3,0}-24I_{r+4,1}I_{2,1}\right] =-\frac{\sigma
_{T}P_{0}^{2}\left( r+4\right) !}{6\beta _{0}^{r+2}}\left( r^{2}+3r+2\right)
, \\
X_{\left( r\right) ,3}& \equiv x_{31}=\frac{\sigma _{T}}{3}\left[
I_{r+5,1}I_{1,0}-I_{r+3,1}I_{3,1}-4I_{r+4,1}I_{2,1}\right] =\frac{\sigma
_{T}P_{0}^{2}\left( r+4\right) !}{18\beta _{0}^{r+2}}\left(
r^{2}+7r+6\right) , \\
X_{\left( r\right) ,4}& \equiv 2x_{22}=-\frac{2\sigma _{T}}{3}\left[
I_{r+5,2}I_{1,0}+4I_{r+4,2}I_{2,1}\right] =-\frac{\sigma _{T}P_{0}^{2}\left(
r+5\right) !}{45\beta _{0}^{r+2}}\left( r+10\right) .
\end{align}%
The other coefficients resulting from the rank-4 tensor are 
\begin{align}
Y_{1\left( r\right) ,3}& =-X_{\left( r\right) ,3}, \\
Y_{2\left( r\right) ,1}& =-\frac{1}{2}X_{\left( r\right) ,1}, \\
Y_{2\left( r\right) ,4}& =-\frac{1}{2}X_{\left( r\right) ,4}.
\end{align}%
The coefficients of the rank-5 tensor are%
\begin{align}
Y_{3\left( r\right) ,1}& \equiv y_{10}=\frac{\sigma _{T}}{6}\left[
I_{r+6,0}I_{1,0}-15I_{r+5,1}I_{2,1}-15I_{r+4,1}I_{3,1}+I_{r+3,0}I_{4,0}%
\right]  \notag \\
& =\frac{\sigma _{T}P_{0}^{2}\left( r+4\right) !}{12\beta _{0}^{r+3}}\left(
r+1\right) \left( r+2\right) \left( r+10\right) , \\
Y_{3\left( r\right) ,3}& \equiv y_{31}=-\frac{\sigma _{T}}{6}\left[
I_{r+6,1}I_{1,0}-9I_{r+5,1}I_{2,1}+5I_{r+4,1}I_{3,1}-I_{r+3,1}I_{4,1}\right]
\notag \\
& =-\frac{\sigma _{T}P_{0}^{2}\left( r+4\right) !}{36\beta _{0}^{r+3}}\left(
r+1\right) \left( r^{2}+8r+20\right) , \\
Y_{3\left( r\right) ,4}& \equiv 2y_{22}=\frac{\sigma _{T}}{3}\left[
I_{r+6,2}I_{1,0}-17I_{r+5,2}I_{2,1}-2I_{r+4,2}I_{3,1}\right]  \notag \\
& =\frac{\sigma _{T}P_{0}^{2}\left( r+5\right) !}{90\beta _{0}^{r+3}}\left(
r^{2}-4r-68\right) , \\
Y_{3\left( r\right) ,5}& \equiv y_{41}=-\frac{1}{3}Y_{3\left( r\right) ,1},
\\
Y_{3\left( r\right) ,6}& \equiv y_{42}=\frac{\sigma _{T}}{30}\left[
I_{r+6,1}I_{1,0}+23I_{r+5,1}I_{2,1}-35I_{r+4,1}I_{3,1}-I_{r+3,1}I_{4,1}%
\right]  \notag \\
& =\frac{\sigma _{T}P_{0}^{2}\left( r+4\right) !}{180\beta _{0}^{r+3}}\left(
r+1\right) \left( r^{2}+40r+180\right) , \\
Y_{3\left( r\right) ,7}& \equiv y_{52}=\frac{\sigma _{T}}{30}\left[
I_{r+6,1}I_{1,0}+7I_{r+5,1}I_{2,1}+4I_{r+4,1}I_{3,1}\right]  \notag \\
& =\frac{\sigma _{T}P_{0}^{2}\left( r+5\right) !}{180\beta _{0}^{r+3}}\left(
r+10\right) ^{2}. \\
Y_{3\left( r\right) ,8}& \equiv y_{62}=\frac{\sigma _{T}}{30}\left[
I_{r+6,1}I_{1,0}-17I_{r+5,1}I_{2,1}+15I_{r+4,1}I_{3,1}-I_{r+3,1}I_{4,1}%
\right]  \notag \\
& =\frac{\sigma _{T}P_{0}^{2}\left( r+4\right) !}{180\beta _{0}^{r+3}}\left(
r+1\right) \left( r^{2}-20\right) .
\end{align}

The coefficients from the rank-6 tensors are%
\begin{align}
Y_{4\left( r\right) ,3}& \equiv w_{31}=-\frac{\sigma _{T}}{15}\left[
I_{r+7,1}I_{1,0}+10I_{r+6,1}I_{2,1}-23I_{r+5,1}I_{3,1}-I_{r+4,1}I_{4,1}-I_{r+3,1}I_{5,1}%
\right]  \notag \\
& =-\frac{\sigma _{T}P_{0}^{2}\left( r+4\right) !}{90\beta _{0}^{r+4}}\left(
r+1\right) \left( r^{3}+35r^{2}+304r+800\right) , \\
Y_{4\left( r\right) ,4}& \equiv 2w_{22}=\frac{\sigma _{T}}{75}\left[
10I_{r+7,2}I_{1,0}-20I_{r+6,2}I_{2,1}-400I_{r+5,2}I_{3,1}-26I_{r+4,2}I_{4,1}%
\right]  \notag \\
& =\frac{\sigma _{T}P_{0}^{2}\left( r+5\right) !}{225\beta _{0}^{r+4}}\left(
r+19\right) \left( r^{2}-40\right) , \\
Y_{4\left( r\right) ,8}& \equiv w_{62}=\frac{\sigma _{T}}{150}\left[
10I_{r+7,2}I_{1,0}-40I_{r+6,1}I_{2,1} +
14I_{r+5,1}I_{3,1}+28I_{r+4,1}I_{4,1}-2I_{r+3,1}I_{5,1}\right]  \notag \\
& =\frac{\sigma _{T}P_{0}^{2}\left( r+4\right) !}{450\beta _{0}^{r+4}}\left(
r+1\right) \left( r^{3}+5r^{2}-86r-400\right) ,
\end{align}%
and 
\begin{align}
Y_{5\left( r\right) ,1}& \equiv z_{10}=\frac{\sigma _{T}}{30}\left[
I_{r+7,0}I_{1,0}+48I_{r+6,1}I_{2,1}-165I_{r+5,1}I_{3,1}+48I_{r+4,1}I_{4,1}+9I_{r+3,1}I_{5,1 }%
\right]  \notag \\
& =\frac{\sigma _{T}P_{0}^{2}\left( r+4\right) !}{60\beta _{0}^{r+4}}\left(
r+1\right) \left( r+2\right) \left( r^{2}+39r+200\right) , \\
Y_{5\left( r\right) ,5}& \equiv z_{41}=-\frac{\sigma _{T}}{30}\left[
I_{r+7,1}I_{1,0}-4I_{r+6,1}I_{2,1}-5I_{r+5,1}I_{3,1}-4I_{r+4,1}I_{4,1}+3I_{r+3,1}I_{5,1}%
\right]  \notag \\
& =-\frac{\sigma _{T}P_{0}^{2}\left( r+4\right) !}{180\beta _{0}^{r+4}}%
\left( r+1\right) \left( r+2\right) \left( r^{2}+19r+100\right) , \\
Y_{5\left( r\right) ,7}& \equiv z_{52}=\frac{\sigma _{T}}{150}\left[
5I_{r+7,2}I_{1,0}+40I_{r+6,2}I_{2,1}+125I_{r+5,2}I_{3,1}-8I_{r+4,2}I_{4,1}%
\right]  \notag \\
& =\frac{\sigma _{T}P_{0}^{2}\left( r+5\right) !}{900\beta _{0}^{r+4}}\left(
r^{3}+29r^{2}+350r+1240\right) , \\
Y_{5\left( r\right) ,9}& \equiv z_{72}=\frac{\sigma _{T}}{150}\left[
5I_{r+7,2}I_{1,0}-44I_{r+6,1}I_{2,1}+95I_{r+5,1}I_{3,1}-44I_{r+4,1}I_{4,1}+3I_{r+3,1}I_{5,1}%
\right]  \notag \\
& =\frac{\sigma _{T}P_{0}^{2}\left( r+4\right) !}{900\beta _{0}^{r+4}}
\left(r^{4}-18r^{3}-161r^{2}-342r-200\right) , \\
Y_{5\left( r\right) ,11}& \equiv z_{23}=-\frac{\sigma _{T}}{210}\left[
I_{r+7,2}I_{1,0}-32I_{r+6,2}I_{2,1}+35I_{r+5,2}I_{3,1}-28I_{r+4,2}I_{4,2}%
\right]  \notag \\
& =-\frac{\sigma _{T}P_{0}^{2}\left( r+5\right) !}{6300\beta _{0}^{r+4}}%
\left( r^{3}-11r^{2}-130r-280\right) .
\end{align}%
We remark that to crosscheck some of our calculations found in this appendix
we made use of the symbolic computer algebra software Cadabra by K.\ Peeters 
\cite{Peeters:2006kp,Peeters:2007wn}, as well as of the Mathematica package
FeynCalc developed and maintained by Rolf Mertig and Frederik Orellana \cite%
{Mertig:1990an}.

\section{Coefficients of the $\mathcal{K}$ tensors}

\label{Kcoeff}

In this appendix, we list the coefficients of the tensors $\mathcal{K},%
\mathcal{K}^{\mu }$, and $\mathcal{K}^{\mu \nu }$ in Eqs.\ (\ref{Ks}), in
the notation of Ref.\ \cite{Denicol:2012cn}. They can be derived following
the derivation of the equations of motion for the dissipative quantities
presented in that reference. For the sake of brevity, they were not
explicitly given in Ref.\ \cite{Denicol:2012cn}, so we decided to list them
here for further reference. Note, however, that in the 14-moment
approximation, where $N_{0}=2$, $N_{1}=1$, and $N_{2}=0$, they vanish
identically. 
\begin{align}
\tilde{\zeta}_{1}& =\sum_{r=3}^{N_{0}}\tau _{0r}^{(0)}\left( \zeta
_{r}-\Omega _{r0}^{(0)}\zeta \right) , \\
\tilde{\zeta}_{2}& =-\frac{2m^{2}}{3}\sum_{r=1}^{N_{0}-2}\tau
_{0,r+2}^{(0)} (r+1)\left( \eta _{r}-\Omega _{r0}^{(2)}\eta \right) -\tilde{\zeta}%
_{1},
\end{align}%
\begin{align}
\tilde{\zeta}_{3}& = \sum_{r=3}^{N_{0}}\tau _{0r}^{(0)}\bar{F}_{r}(\alpha
_{0},\beta _{0})+\frac{1}{3}\sum_{r=3}^{N_{0}}\tau _{0r}^{(0)}(r+1)\left(
\zeta _{r}-\Omega _{r0}^{(0)}\zeta \right) -\frac{m^{2}}{3}%
\sum_{r=3}^{N_{0}-2}\tau _{0,r+2}^{(0)}(r+1)\left( \zeta _{r}-\Omega
_{r0}^{(0)}\zeta \right) ,  \label{zeta3} \\
\tilde{\zeta}_{4}& =\frac{m^{2}}{3}\sum_{r=2}^{N_{0}-1}\tau
_{0,r+1}^{(0)}\left( \frac{\partial }{\partial \alpha _{0}}+\frac{n_{0}}{%
\varepsilon _{0}+P_{0}}\frac{\partial }{\partial \beta _{0}}\right) \left(
\kappa _{r}-\Omega _{r0}^{(1)}\kappa \right) , \\
\tilde{\zeta}_{5}& =-\frac{2\left( \varepsilon _{0}+P_{0}\right) +\beta
_{0}J_{30}}{(\varepsilon _{0}+P_{0})^{3}}\,\tilde{\zeta}_{1}, \\
\tilde{\zeta}_{6}& =-\frac{\left( \varepsilon _{0}+P_{0}\right)
J_{20}-n_{0}J_{30}}{(\varepsilon _{0}+P_{0})^{3}}\,\tilde{\zeta}_{1}-\frac{%
m^{2}}{3}\frac{1}{\varepsilon _{0}+P_{0}}\sum_{r=2}^{N_{0}-1}\tau
_{0,r+1}^{(0)}\left[ (r+1)\left( \kappa _{r}-\Omega _{r0}^{(1)}\kappa
\right) +\frac{\partial \left( \kappa _{r}-\Omega _{r0}^{(1)}\kappa \right) 
}{\partial \ln \beta _{0}}\right] , \\
\tilde{\zeta}_{7}& =\frac{m^{2}}{3}\sum_{r=2}^{N_{0}-1}\tau
_{0,r+1}^{(0)}\left( \kappa _{r}-\Omega _{r0}^{(1)}\kappa \right) , \\
\tilde{\zeta}_{8}& =\frac{\tilde{\zeta}_{1}}{\varepsilon _{0}+P_{0}}.
\end{align}%
where $\zeta =\zeta ^{0}$, $\kappa =\kappa ^{0}$, $\eta =\eta ^{0}$ in
consistency with Eqs.\ (\ref{Final1}) -- (\ref{Final3}).

In Eq.\ (\ref{zeta3}), we defined the function $\bar{F}_{r}(\alpha
_{0},\beta _{0})$ through the relation 
\begin{equation}
D\left( \zeta _{r}-\Omega _{r0}^{(0)}\zeta \right) =\bar{F}_{r}(\alpha
_{0},\beta _{0})\theta .
\end{equation}%
Furthermore, 
\begin{align}
\tilde{\kappa}_{1}& =-2\sum_{r=1}^{N_{1}-1}\tau _{0,r+1}^{(1)}\left( \frac{%
\partial }{\partial \alpha _{0}}+\frac{n_{0}}{\varepsilon _{0}+P_{0}}\frac{%
\partial }{\partial \beta _{0}}\right) \left( \eta _{r}-\Omega
_{r0}^{(2)}\eta \right)  \notag \\
& +\frac{2m^{2}}{5}\sum_{r=2}^{N_{1}-2}\tau _{0,r+2}^{(1)}(r+1)\left( \kappa
_{r}-\Omega _{r0}^{(1)}\kappa \right) -\frac{2}{5}\sum_{r=2}^{N_{1}}\tau
_{0r}^{(1)}(r-1)\left( \kappa _{r}-\Omega _{r0}^{(1)}\kappa \right) , \\
\tilde{\kappa}_{2}& =\frac{2}{\varepsilon _{0}+P_{0}}\sum_{r=1}^{N_{1}-1}%
\tau _{0,r+1}^{(1)}\left[ (r+1)\left( \eta _{r}-\Omega _{r0}^{(2)}\eta
\right) +\frac{\partial \left( \eta _{r}-\Omega _{r0}^{(2)}\eta \right) }{%
\partial \ln \beta _{0}}\right] , \\
\tilde{\kappa}_{3}& =-\sum_{r=2}^{N_{1}}\tau _{0r}^{(1)}\left[ \bar{G}%
_{r}(\alpha _{0},\beta _{0})+\left( \frac{r+2}{3}+\frac{\partial \mathcal{H}%
(\alpha _{0},\beta _{0})}{\partial \alpha _{0}}+\frac{n_{0}}{\varepsilon
_{0}+P_{0}}\frac{\partial \mathcal{H}(\alpha _{0},\beta _{0})}{\partial
\beta _{0}}\right) \left( \kappa _{r}-\Omega _{r0}^{(1)}\kappa \right) %
\right]  \notag \\
& +\frac{m^{2}}{3}\sum_{r=2}^{N_{1}-2}\tau _{0,r+2}^{(1)}(r+1)\left( \kappa
_{r}-\Omega _{r0}^{(1)}\kappa \right) -\sum_{r=3}^{N_{1}-1}\tau
_{0,r+1}^{(1)}\left( \frac{\partial }{\partial \alpha _{0}}+\frac{n_{0}}{%
\varepsilon _{0}+P_{0}}\frac{\partial }{\partial \beta _{0}}\right) \left(
\zeta _{r}-\Omega _{r0}^{(0)}\zeta \right)  \notag \\
& +\frac{1}{m^{2}}\sum_{r=3}^{N_{1}+1}\tau _{0,r-1}^{(1)}\left( \frac{%
\partial }{\partial \alpha _{0}}+\frac{n_{0}}{\varepsilon _{0}+P_{0}}\frac{%
\partial }{\partial \beta _{0}}\right) \left( \zeta _{r}-\Omega
_{r0}^{(0)}\zeta \right) , \\
\tilde{\kappa}_{4}& =\frac{1}{\varepsilon _{0}+P_{0}}\left\{ \left[
\mathcal{H}(\alpha _{0},\beta _{0}) + \frac{\partial \mathcal{H}(\alpha _{0},\beta _{0}) %
}{\partial \ln \beta _{0}} \right] \sum_{r=2}^{N_{1}}\tau _{0r}^{(1)}\left(
\kappa _{r}-\Omega _{r0}^{(1)}\kappa \right) +\sum_{r=3}^{N_{1}-1}\tau
_{0,r+1}^{(1)}\left( r+1+\frac{\partial }{\partial \ln \beta _{0}}\right)
\left( \zeta _{r}-\Omega _{r0}^{(0)}\zeta \right) \right.  \notag \\
& -\left. \frac{1}{m^{2}}\sum_{r=3}^{N_{1}+1}\tau _{0,r-1}^{(1)}\left( r+2+%
\frac{\partial }{\partial \ln \beta _{0}}\right) \left( \zeta _{r}-\Omega
_{r0}^{(0)}\zeta \right) \right\} , \\
\tilde{\kappa}_{5}& =2\sum_{r=2}^{N_{1}}\tau _{0r}^{(1)}\left( \kappa
_{r}-\Omega _{r0}^{(1)}\kappa \right) , \\
\tilde{\kappa}_{6}& =-2\sum_{r=1}^{N_{1}-1}\tau _{0,r+1}^{(1)}\left( \eta
_{r}-\Omega _{r0}^{(2)}\eta \right) , \\
\tilde{\kappa}_{7}& =-\mathcal{H}(\alpha _{0},\beta _{0})\frac{\tilde{\kappa}%
_{5}}{2}-\sum_{r=3}^{N_{1}-1}\tau _{0,r+1}^{(1)}\left( \zeta _{r}-\Omega
_{r0}^{(0)}\zeta \right) +\frac{1}{m^{2}}\sum_{r=3}^{N_{1}+1}\tau
_{0,r-1}^{(1)}\left( \zeta _{r}-\Omega _{r0}^{(0)}\zeta \right) .
\end{align}%
Here, we defined the function $\bar{G}_{r}(\alpha _{0},\beta _{0})$ through
the relation 
\begin{equation}
D\left( \kappa _{r}-\Omega _{r0}^{(1)}\kappa \right) =\bar{G}_{r}(\alpha
_{0},\beta _{0})\theta .
\end{equation}%
Moreover, 
\begin{equation}
\mathcal{H}(\alpha _{0},\beta _{0})=\frac{(\varepsilon
_{0}+P_{0})J_{20}-n_{0}J_{30}}{D_{20}}.
\end{equation}%
Finally, 
\begin{align}
\tilde{\eta}_{1}& =2\sum_{r=1}^{N_{2}}\tau _{0r}^{(2)}\left( \eta
_{r}-\Omega _{r0}^{(2)}\eta \right) , \\
\tilde{\eta}_{2}& =2\left\{ -\sum_{r=1}^{N_{2}}\tau _{0r}^{(2)}\left[ \bar{H}%
_{r}(\alpha _{0},\beta _{0})+\frac{1}{3}(r+2)\left( \eta _{r}-\Omega
_{r0}^{(2)}\eta \right) \right] \right.  \notag \\
& +\frac{m^{2}}{3}\sum_{r-1}^{N_{2}-2}\tau _{0,r+2}^{(2)}(r+1)\left( \eta
_{r}-\Omega _{r0}^{(2)}\eta \right) +\frac{m^{2}}{5}\sum_{r=3}^{N_{2}-2}\tau
_{0,r+2}^{(2)}(r+1)\left( \zeta _{r}-\Omega _{r0}^{(0)}\zeta \right)  \notag
\\
& \left. -\frac{1}{5}\sum_{r=3}^{N_{2}}\tau _{0r}^{(2)}(2r+3)\left( \zeta
_{r}-\Omega _{r0}^{(0)}\zeta \right) +\frac{1}{5m^{2}}\sum_{r=3}^{N_{2}+2}%
\tau _{0,r-2}^{(2)}(r+2)\left( \zeta _{r}-\Omega _{r0}^{(0)}\zeta \right)
\right\} , \\
\tilde{\eta}_{3}& =\frac{2}{7}\left[ -\sum_{r=1}^{N_{2}}\tau
_{0r}^{(2)}(4r+3)\left( \eta _{r}-\Omega _{r0}^{(2)}\eta \right)
+4m^{2}\sum_{r=1}^{N_{2}-2}\tau _{0,r+2}^{(2)}(r+1)\left( \eta _{r}-\Omega
_{r0}^{(2)}\eta \right) \right] , \\
\tilde{\eta}_{4}& =2\tilde{\eta}_{1}, \\
\tilde{\eta}_{5}& =\frac{2}{5}\left[ \sum_{r=2}^{N_{2}+1}\tau
_{0,r-1}^{(2)}\left( \frac{\partial }{\partial \alpha _{0}}+\frac{n_{0}}{%
\varepsilon _{0}+P_{0}}\frac{\partial }{\partial \beta _{0}}\right) \left(
\kappa _{r}-\Omega _{r0}^{(1)}\kappa \right) -m^{2}\sum_{r=2}^{N_{2}-1}\tau
_{0,r+1}^{(2)}\left( \frac{\partial }{\partial \alpha _{0}}+\frac{n_{0}}{%
\varepsilon _{0}+P_{0}}\frac{\partial }{\partial \beta _{0}}\right) \left(
\kappa _{r}-\Omega _{r0}^{(1)}\kappa \right) \right] , \\
\tilde{\eta}_{6}& =\frac{2(\varepsilon _{0}+P_{0})+\beta _{0}J_{30}}{%
(\varepsilon _{0}+P_{0})^{3}}\tilde{\eta}_{1}, \\
\tilde{\eta}_{7}& =\frac{2}{5(\varepsilon _{0}+P_{0})}\left\{ \frac{5}{2}%
\frac{(\varepsilon _{0}+P_{0})J_{20}-n_{0}J_{30}}{(\varepsilon
_{0}+P_{0})^{2}}\tilde{\eta}_{1}-\sum_{r=2}^{N_{2}+1}\tau
_{0,r-1}^{(2)}\left( r+4+\frac{\partial }{\partial \ln \beta _{0}}\right)
\left( \kappa _{r}-\Omega _{r0}^{(1)}\kappa \right) \right.  \notag \\
& \left. +m^{2}\sum_{r=2}^{N_{2}-1}\tau _{0,r+1}^{(2)}\left( r+1+\frac{%
\partial }{\partial \ln \beta _{0}}\right) \left( \kappa _{r}-\Omega
_{r0}^{(1)}\kappa \right) \right\} , \\
\tilde{\eta}_{8}& =\frac{2}{5}\left[ \sum_{r=2}^{N_{2}+1}\tau
_{0,r-1}^{(2)}\left( \kappa _{r}-\Omega _{r0}^{(1)}\kappa \right)
-m^{2}\sum_{r=2}^{N_{2}-1}\tau _{0,r+1}^{(2)}\left( \kappa _{r}-\Omega
_{r0}^{(1)}\kappa \right) \right] , \\
\tilde{\eta}_{9}& =-\frac{\tilde{\eta}_{1}}{\varepsilon _{0}+P_{0}}.
\end{align}%
Here, we defined the function $\bar{H}_{r}(\alpha _{0},\beta _{0})$ through
the relation 
\begin{equation}
D\left( \eta _{r}-\Omega _{r0}^{(2)}\eta \right) =\bar{H}_{r}(\alpha
_{0},\beta _{0})\theta .
\end{equation}

\end{document}